\documentclass[a4paper, print,12pt,oneside]{book}
\usepackage[font=small,format=plain,labelfont=bf,up,textfont=it,up]{caption}
\usepackage{amsmath} 		
\usepackage{amssymb} 		
\usepackage{pst-node}
\usepackage{graphicx}		
\usepackage{datetime}
\usepackage[hidelinks]{hyperref}
\usepackage{amsfonts}
\usepackage{supertabular}
\usepackage{color}
\usepackage{booktabs}
\usepackage{enumitem}
\usepackage{lscape}
\usepackage[english]{babel}
\usepackage[toc]{glossaries}
\usepackage{fancyref}
\usepackage{longtable} 
\usepackage[titletoc,toc,page]{appendix}
\usepackage{setspace}

\usepackage{tocloft} 
\usepackage[square,numbers,sort&compress]{natbib}
\usepackage{vmargin}
\usepackage[toc,page]{appendix}
\usepackage{morefloats}
\usepackage{epstopdf}
\usepackage{mathrsfs} 	
\usepackage{epigraph}   
\usepackage{pdflscape}  
\usepackage{multirow}		

\usepackage{color}  
\usepackage[normalem]{ulem}
\include{definitions}

\usepackage{fancyhdr}

\usepackage{titlesec}
\titleformat{\chapter}[display]   
{\normalfont\huge\bfseries}{\chaptertitlename\ \thechapter}{10pt}{\huge}   
\titlespacing*{\chapter}{0pt}{-40pt}{32pt}

\usepackage[T1]{fontenc}
\usepackage{mathptmx}

\setmarginsrb
{4cm}   
{3cm}   
{2cm}   
{2cm}   
{3pt}    
{0.3in}  
{3pt}     
{0.3in}   

\begin{document}
  \frontmatter
%
\begin{titlepage}
\ifpdf
    \graphicspath{{0/figures/PNG/}{0/figures/PDF/}{0/figures/}}
\else
    \graphicspath{{0/figures/EPS/}{0/figures/}}
\fi
\begin{center}
 \Huge{Mathematical modelling to investigate a \textit{Wolbachia} intervention to reduce dengue transmission}
\vfill
\vfill
\vfill
\vfill
\vfill
\vfill
 \Large{\textbf{Meksianis Zadrak Ndii}}\\ 
\normalsize{S. Si (Bachelor of Science) (Universitas Nusa Cendana, Indonesia)} \\
\normalsize{M.Sc (The Australian National University, Australia)}\\
\vfill
\vfill
\vfill
\vfill
\vfill
\vfill
\vfill
\vfill
\vfill
\vfill
\vfill
\vfill
\vfill
\vfill
\vfill
\includegraphics[scale=0.75]{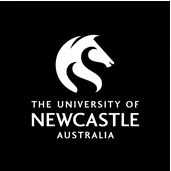}

 \large A thesis submitted for the degree of \\
         \textbf{Doctor of Philosophy}\\
         at The University of Newcastle, Australia\\
         \Large 2015 \\
\vfill
\vfill
\end{center}

\end{titlepage}

\chapter*{Statement of originality}\label{declaration}
\thispagestyle{empty}
The thesis contains no material which has been accepted for the award of any other degree
or diploma in any university or other tertiary institution and, to the best of my knowledge and
belief, contains no material previously published or written by another person, except where
due reference has been made in the text. I give consent to the final version of my thesis
being made available worldwide when deposited in the University's Digital Repository,
subject to the provisions of the Copyright Act 1968. Unless an Embargo has been approved for a determined period.


\vspace{1in}

\hfill\hfill
Callaghan, August 2015\\
\vspace{1in}
\hfill\hfill
Meksianis Zadrak Ndii\\

\hspace*{\fill}

\pagestyle{myheadings}

\pagestyle{empty}
\vspace*{\fill}
\vfill
\begin{center}
\textit{\large{I dedicate this work to my wife, Bertha, and our daugther, Geavanna.}}
\end{center}



\vfill\vfill\vfill

\chapter*{Acknowledgements}\label{acknowledgements}
\addcontentsline{toc}{chapter}{Acknowledgements}

I would like to take this opportunity to sincerely thank numerous people for their contributions and support throughout my studies.  I realise that one or two sentences are not enough to fully acknowledge their contributions.

It was not easy to make the transition from simply being a course taker to becoming an independent researcher. This transition would not have been possible without the help and guidance of three inspiring, dedicated and approachable supervisors: Dr.~Roslyn~I. Hickson, Dr. David Allingham,  and Dr. Kathryn Glass.  My PhD studies would have taken much longer and been much more difficult without your help. To Roslyn, thank you for our regular meetings and discussions in the early phase of my PhD (the first two years), which have equipped me to conduct research. Thank you all for your tireless supervision, expertise, wisdom, and encouragement.  Our weekly skype meetings triggered  discussions, generated ideas, and of course made  the completion of this thesis possible. Thank you for your constructive feedback, suggestions of directions for my research, and proofreading of thesis and paper drafts, which helped to improve my English writing skills. I have learned responsibility from you all. I am fortunate to have worked with you, and have enjoyed this challenging journey. The contribution of the late Prof. Geoffry N. Mercer to this work is also acknowledged. He and Roslyn were the first people to introduce me to this field, and together with David and Kathryn have helped me through my PhD journey. I would also like to thank Dr. Teresa Bates, who, together with all my supervisors, greatly helped me to improve my English writing.

I would like to thank my wife, Bertha, for her love and support. Thanks for looking after our daughter, Gea, without me. I hope this is the last time I will be away from home for a long time. I thank Gea for her smile and laugh, which motivated me to work harder and complete this work. I would also like to express my very great appreciation to my parents, the Ndii and Djahi families for their support and prayers.

I acknowledge the University of Newcastle for awarding me a PhD scholarship and the Faculty of Science and Information Technology for awarding me an RHD conference scholarship. I would also like to acknowledge ANZIAM for partial funding through the CSIRO/ ANZIAM student support scheme for my travel to the MISG and ANZIAM conferences. I acknowledge Universitas Nusa Cendana for allowing me to pursue my further study.

Thanks to my friends from Indonesia in Australia:  Astija Surya,   Elisa Sesa, Junaidi, Muhammad Ilyas, Novi Bong, Sri Hastuti and Tadulako's group, Stefa Yuwiko, and Yodi Christiani who have made me feel at home.  I would also like to thank David and Tricia Mileham for their caring.  

Above all, I believe that God has been working through these wonderful people to make His purpose in my life fulfilled, as is stated in Jeremiah 29:11: \textit{``For I know the thoughts that I think toward you, saith the Lord, thoughts of peace, and not of evil, to give you hope in your latter end''}. I pray for all the best for your careers, and future endeavours in your lives.

\chapter*{Abstract}\label{abstract}

\addcontentsline{toc}{chapter}{Abstract}

The introduction of \textit{Wolbachia}-carrying mosquitoes into the population has recently been proposed as an alternative strategy against dengue. Although laboratory experiments have shown that the \textit{Wolbachia} bacterium can reduce the levels of dengue virus in mosquitoes, it is also important to assess the performance of \textit{Wolbachia} in reducing the incidence of dengue in human populations.

In this thesis, deterministic mathematical models of human and mosquito populations in which either one or two dengue serotypes circulate are developed.  We adapt these models to enable the investigation of dengue disease dynamics in the absence and presence of \textit{Wolbachia} in order to assess the performance of \textit{Wolbachia} as a strategy to reduce human dengue incidence.

When studying the situation in which a single dengue serotype is present in the population, we consider  scenarios where dengue is introduced into the human population once and multiple times. We find that when mosquitoes infected with the \textit{Wolbachia} strain \textit{WMel}, which reduces the mosquito lifespan by at most 10\%, are released into the population, the \textit{Wolbachia}-carrying mosquitoes  persist. The ranges of the reproductive  and  death rates for \textit{Wolbachia}-carrying mosquitoes which allow mosquitoes carrying \textit{Wolbachia} to persist in competition with non-\textit{Wolbachia} carrying mosquitoes are also found. Furthermore, the transmission probability, the biting rate and the average death rate are the  parameters exerting the most influence on the cumulative number of infectious individuals in the population.   An analysis of the basic reproduction number, $\mathcal{R}_0$, for the model considering the absence and presence of \textit{Wolbachia}-carrying mosquitoes shows that  the presence of \textit{Wolbachia}-carrying mosquitoes reduces the number of days for which $\mathcal{R}_0>1$. When multiple introductions of dengue are considered, it is found that the presence of \textit{Wolbachia} reduces the potential lengths  of  the seasons in which epidemics are likely to occur.   The strength of seasonality also affects the reduction in dengue incidence caused by the introduction of \textit{Wolbachia}:  if seasonality is strong, then there are some seasons when mosquitoes have longer life spans and more individuals are infected in each outbreak so that \textit{Wolbachia} becomes less effective in reducing dengue incidence.

Our two-serotype dengue models are used to investigate dengue serotypes with symmetric and asymmetric characteristics.  For serotypes with symmetric characteristics, we investigate the performance of \textit{Wolbachia} in reducing dengue incidence under different disease introduction scenarios, and find that a difference in the disease introduction scenario does not affect the performance of \textit{Wolbachia} in reducing dengue incidence. Furthermore, the transmission probability is a more influential parameter regulating dengue dynamics than antibody-dependent enhancement. When dengue serotype characteristics differ (asymmetry), the more transmissible dengue serotype will dominate the primary infection, while the other serotype will dominate the secondary infections.   The number of secondary infections caused by the more transmissible serotype can still be reduced by the introduction of \textit{Wolbachia}-carrying mosquitoes, but the proportional reduction in dengue cases is not as high.

Our findings suggest that \textit{Wolbachia} intervention can be used as an effective alternative strategy against dengue. \textit{Wolbachia} should reduce the number of primary dengue cases in areas with moderate transmission levels, and can provide an even greater reduction in the number of secondary cases.  Given the higher risk of severe outcomes in secondary cases, \textit{Wolbachia} has great potential for improving public health.


\chapter*{Publications, Talks and Awards}\label{paperstalk}
\addcontentsline{toc}{chapter}{Publications, Talks and Awards}
\textbf{The following is a list of papers and manuscripts that I have written during my PhD candidature}

\begin{itemize}[label={}] 
  \item \textbf{Ndii, M.Z.}, Hickson, R.I., Mercer, G.N. 2012. Modelling the introduction of \textit{Wolbachia} into \textit{Aedes aegypti} to reduce dengue transmission. \textit{The ANZIAM Journal}. 53(3), 213--227. \url{http://dx.doi.org/10.1017/S1446181112000132}. (This paper was written and published during my PhD candidature, but was based on the work undertaken during my Masters study at the ANU. The work is therefore not included in this thesis.)

  \item  \textbf{Ndii, M.Z.}, Hickson, R.I., Allingham, D., Mercer, G.N. 2015. Modelling the transmission dynamics of dengue in the presence of \textit{Wolbachia}. \textit{Mathematical Biosciences}. 262,156--166. \url{http://dx.doi.org/10.1016/j.mbs.2014.12.011}. (This paper is based on the work in Chapters 3 and 4.)

  \item \textbf{Ndii, M.Z.}, Allingham, D., Hickson, R.I. Glass, K. The effect of \textit{Wolbachia} on dengue transmission dynamics when dengue is repeatedly introduced. This paper has been submitted for publication. (This paper is based on the work in Chapters 3 and 4.)

  \item \textbf{Ndii, M.Z.}, Allingham, D., Hickson, R.I. Glass, K.The effect of \textit{Wolbachia} on dengue dynamics in the presence of two serotypes of dengue: symmetric and asymmetric  epidemiological characteristics. This paper has been submitted for publication. (This paper is based on the work in Part 2; Chapters 6, 7, and 8.)

\end{itemize}

\textbf{The following is a list of the talks I have given during my PhD candidature}
\begin{itemize}[label={}]
\item \textbf{Ndii, M.Z.}, Allingham D.,  Hickson, R.I., Glass, K. The effects of \textit{Wolbachia} on dengue outbreaks when dengue is repeatedly introduced. \textit{51st Applied Mathematics Conference ANZIAM, 2015}, Gold Coast.

\item \textbf{Ndii, M.Z.}., Hickson, R.I., Allingham D., Mercer, G.N., Glass, K. Introduction of \textit{Wolbachia} into \textit{Aedes aegypti} significantly reduces human dengue cases, \textit{Mathematical Biology Conference, Joint meeting Society for Mathematical Biology and Japanese Mathematical Biology 2014}. 28 July--1 Agustus 2014. Osaka, Japan.

\item \textbf{Ndii, M.Z.}., Hickson, R.I., Allingham D., Mercer, G.N. The effect of \textit{Wolbachia} on dengue transmission.  \textit{50th Applied Mathematics Conference, ANZIAM 2014}, Rotorua, New Zealand.

\item \textbf{Ndii, M.Z.}, Hickson, R.I., Allingham D., Mercer, G.N. An analysis of a seasonal dengue model with the presence of \textit{Wolbachia}. \textit{Joint ACT/NSW ANZIAM mini meeting 2013}, Sydney.

\item  \textbf{Ndii, M.Z.}, Hickson, R.I., Allingham D., Mercer, G.N. An analysis of a seasonal dengue model with the presence of  \textit{Wolbachia}. \textit{Australian Mathematical Society Workshop on Infectious Disease Modelling, September  2013.} Newcastle.

\item \textbf{Ndii, M.Z.}, Hickson, R.I., Allingham D., Mercer, G.N. Dengue and mosquitoes: can we stop transmission?. \textit{49th Applied Mathematics Conference ANZIAM, 2013}, Newcastle.

\item \textbf{Ndii, M.Z.}, Hickson, R.I., Mercer, G.N. \textit{CARMA Seminar, 2012}. The University of Newcastle, Australia.
\end{itemize}

\textbf{The following supports were received during my candidature}
\begin{itemize}
\item Faculty of Science and Information Technology RHD Conference Scholarship to attend the 2014 Annual Meeting of the Society for Mathematical Biology
and Japanese Society for Mathematical Biology, 28 July-1 August 2014, Osaka, Japan.
\item   Funding  from  the  CSIRO/ANZIAM  student  support  scheme  to  attend the \textit{50th Applied Mathematics Conference, ANZIAM}, 2014, New Zealand.
\item   Funding  from  the  CSIRO/ANZIAM  student  support  scheme  to  attend  \textit{the Mathematics in Industry Study Group} 2013 in Brisbane, and the  \textit{49th Applied Mathematics Conference ANZIAM}, 2013,  Newcastle.
\end{itemize}

\tableofcontents
\pagestyle{myheadings}
\listoffigures\addcontentsline{toc}{chapter}{List of Figures} 
\clearpage
\listoftables\addcontentsline{toc}{chapter}{List of Tables}

\mainmatter

\chapter{Introduction}\label{chap:intro}

\section{Background and Research Motivation}

Dengue is a vector-borne disease which is transmitted by mosquitoes.
Approximately two thirds of the world's population is living in dengue-endemic regions, with around 390 million individuals infected annually~\cite{Bhatt2013}.

There are four serotypes of dengue: DEN1, DEN2, DEN3, and DEN4. Individuals  infected with one of the serotypes obtain life long immunity to that serotype, but only temporary immunity to the other serotypes. When infected with a second serotype, individuals are at greater risk of developing severe forms of dengue such as Dengue Haemmorhagic Fever (DHF) and Dengue Shock Syndrome (DSS). Interactions between dengue serotypes may affect dengue transmission dynamics because of differences in epidemiological characteristics, and  disease severity~\cite{Vaughn2000, Althouse2014, Barraquer2014, Tricou2011, Nishiura2007, Ferguson1999a} between serotypes.

Fluctuations in mosquito populations  due to climatic factors such as temperature and rainfall results in seasonality in dengue transmission dynamics. In areas and regions where dengue is not endemic, but mosquitoes reside, dengue outbreaks are triggered by  imported cases~\cite{Knope2013,Lin2010}.   These are individuals who have acquired dengue overseas and brought it to the non-endemic regions.  After mosquitoes are exposed to dengue, a certain period of time is required for the dengue viruses to replicate in the mosquito's body so that the mosquito can transmit the disease. This period is known as the extrinsic incubation period (EIP). 

A number of strategies for controlling dengue have been implemented, but they are difficult to sustain~\cite{Grisales2013, Montella2007,Ritchie2002,Ooi2006,Freitas2014}, particularly in the developing world. For example, the use of insecticides has become largely ineffective as mosquitoes have developed resistance to the chemicals~\cite{Grisales2013, Montella2007,Freitas2014}. Therefore, new control strategies are necessary.

One strategy that has been recently proposed is to infect mosquito populations with the \textit{Wolbachia} bacterium~\cite{Bian2010, McMeniman2009, Moreira2009, Hofman2011, Walker2011, Frentiu2014}. There are two mechanisms by which \textit{Wolbachia} can work to control dengue transmission. First, \textit{Wolbachia} impairs the ability of the dengue virus to replicate inside the mosquito~\cite{Bian2010, Moreira2009, Hofman2011, Walker2011, Frentiu2014}, particularly in salivary gland~\cite{Walker2011}. Consequently, the ability of  \textit{Wolbachia}-carrying mosquitoes to transmit dengue viruses can be reduced. Second, \textit{Wolbachia} reduces the mosquito's lifespan~\cite{Yeap2011, Walker2011}. As a result, the mosquito's lifespan may be lower than the extrinsic incubation period (EIP) so the transmission will not occur. However, in order for this intervention to be successful, it is necessary for  \textit{Wolbachia}-carrying mosquitoes to be established and persist in the field.

Studies utilising mathematical models have suggested that \textit{Wolbachia}-carrying mosquitoes are likely to persist in the field~\cite{Ndii2012, Hancock2011,Hancock2012,Chan2013}.  This has been confirmed by field experiments~\cite{Hoffmann2014,Hofman2011,Walker2011}. An open question that then arises is ``to what extent can \textit{Wolbachia} reduce dengue transmission?''.  This question forms the central focus of this thesis.

To measure the effectiveness of \textit{Wolbachia} in reducing dengue at the population level is challenging since many factors such as  seasonality, importation of dengue  and dengue serotype characteristics affect dengue transmission dynamics. An efficient way to measure the effectiveness of \textit{Wolbachia}  is by the use of a mathematical model. A mathematical model allows us to determine the underlying causes of outbreaks and to understand the effectiveness of the control strategies that have been implemented. To date, few  studies have  used mathematical models to understand the effect of \textit{Wolbachia} in reducing dengue transmission, and hence our models will provide new insight on the effectiveness of \textit{Wolbachia} in reducing dengue at a population level.  Further, we shall investigate the effects of seasonality, dengue importation, and dengue serotype characteristics, which have received relatively little attention.

We are particularly interested in quantifying the effectiveness of \textit{Wolbachia}-carrying mosquitoes on the reduction of dengue incidence once their population has been established.  As a result, we shall assume that \textit{Wolbachia}-carrying mosquitoes persist and are established in the population, and shall not investigate the spread of mosquitoes carrying \textit{Wolbachia} into the population in this thesis.

Throughout this thesis, the terminology ``dengue/disease introduction'' refers to dengue/ disease importation. For clarity, two things should be noted.  When studying the effects of \textit{Wolbachia} in the presence of a single dengue serotype, single and multiple dengue introductions mean that we introduce a single dengue serotype once and multiple times, respectively. Second, when studying the effects of \textit{Wolbachia} in the presence of two dengue serotypes, dengue cases are introduced multiple times, with different dengue serotypes. The term,  ``dengue introduction scenarios'' is used to refer to the manner in which a particular dengue serotype is introduced into the population.

\section{Research Aims}\label{ResearchAims}
The main question addressed in this thesis is  ``to what extent can \textit{Wolbachia} reduce dengue transmission?''. This has been approached by formulating novel mathematical models in the absence and presence of \textit{Wolbachia}-carrying mosquitoes for single and two dengue serotypes. The specific aims of this thesis are to:

\begin{enumerate}[wide=0pt,labelwidth = 1.3333em, labelsep = 0.3333em, leftmargin = \dimexpr\labelwidth + \labelsep\relax ]
\item Determine the level of reduction in dengue incidence caused by the presence of \textit{Wolbachia}-carrying mosquitoes.
\item Explore the effects of seasonality and other important parameters on dengue transmission dynamics and the persistence of \textit{Wolbachia}-carrying mosquitoes.
\item Determine the effects of dengue introduction scenarios on the performance of \textit{Wolbachia} in reducing dengue incidence.
\item Investigate the effects of two dengue serotypes and dengue serotype characteristics on the performance of \textit{Wolbachia} in reducing dengue incidence.
\end{enumerate}

\section{Contributions}
The novel contributions of this thesis are as follows.
\begin{enumerate}[wide=0pt,labelwidth = 1.3333em, labelsep = 0.3333em, leftmargin = \dimexpr\labelwidth + \labelsep\relax ]
\item Deterministic mathematical models of scenarios incorporating one and two dengue  serotypes in the absence and presence of \textit{Wolbachia}-carrying mosquitoes are produced, incorporating sinusoidally forced death rates for the mosquitoes. These models take into account important biological features of dengue and \textit{Wolbachia}. The models are used to study the effectiveness of \textit{Wolbachia} in reducing the incidence of dengue at a population level. Furthermore, these models serve as a baseline for further investigation of  \textit{Wolbachia} interventions. To the best of our knowledge, the two serotype dengue model in the presence of \textit{Wolbachia} is the first population level model to investigate the effectiveness of \textit{Wolbachia} in the presence of more than one dengue serotype.

\item The important parameters of the models that drive dengue epidemics, are determined using sensitivity analysis through the standard combination of Latin Hypercube Sampling and Partial Rank Correlation Coefficient (LHS/PRCC)~\cite{Blower1994,Marino2008,Wu2013}, where the parameters relative significances are ranked. These findings will be useful to guide future data collections to better inform future models.

\item The ranges of biologically realistic parameter values, which determine the persistence of \textit{Wolbachia}-carrying mosquitoes  and dengue epidemics, are calculated.  These findings confirm the potential \textit{Wolbachia} strains that can persist in the field, and also the maximum benefits of \textit{Wolbachia} for the different parameter values.

\item The level of reduction in dengue incidence due to \textit{Wolbachia} intervention is determined by comparing the relative difference of the outbreak sizes in the absence and presence of \textit{Wolbachia}-carrying mosquitoes. We estimate the level of reduction in dengue incidence due to \textit{Wolbachia} in scenarios where  one and  two dengue serotypes are present. In the single serotype scenario, the levels of dengue incidence reduction after single and multiple introductions of dengue are determined. In the presence of  two dengue serotypes, the levels of dengue incidence reduction for serotypes with symmetric and asymmetric epidemiological characteristics are determined.   
\end{enumerate}

\section{Outline of the Thesis}
In the next chapter, we present a review of literature related to this thesis, and then in the subsequent six chapters, we present the original contributions of this thesis.  These contributions are divided into two parts corresponding to the presence of one (Chapters~~\ref{chap:singlemathmodel}-\ref{chap:multipleintro})  and two  (Chapters~~\ref{chap:modelmultiplestrains}--\ref{chap:asymmetry}) dengue serotypes in the population. The  final chapter discusses the conclusions drawn and future directions for research.  A brief overview of the contents of each chapter is given below.

\noindent
\textbf{Chapter~\ref{chap:literaturereview}: Literature Review}\\
A review of literature is presented in this chapter. Chapter \ref{chap:literaturereview} contains reviews of relevant literature from two main viewpoints: biological and mathematical reviews. The biological literature reviewed presents background information about dengue, dengue vectors and \textit{Wolbachia}. The mathematical reviews provide background on the deterministic SIR model, vector-borne disease modelling, seasonality, the basic reproduction number, and introduce models specific for \textit{Wolbachia} population dynamics  and models for dengue epidemiology in the absence and presence of \textit{Wolbachia}-carrying mosquitoes. 

\noindent
\textbf{Chapter~\ref{chap:singlemathmodel}: Mathematical Modelling of a Single Dengue Serotype}\\
This chapter contains the full derivation of our single serotype dengue models in the absence and  presence of \textit{Wolbachia}-carrying mosquitoes. The assumptions and important features of these models are described. The models are an extension of previously published models by Ndii~\textit{et al.}~\cite{Ndii2012}\footnote{This work was conducted during my Masters study at the ANU} and introduce a human population in which dengue is present as well as seasonally varying death rates for adult mosquitoes. These models are then used to study the effect of single and multiple dengue introductions, which are presented in Chapter~\ref{chap:singleintro} and Chapter~\ref{chap:multipleintro}, respectively. The expressions for the basic and effective reproduction numbers for these models are also derived.

\noindent
\textbf{Chapter~\ref{chap:singleintro}: Dengue Transmission Dynamics for a Single Introduction Event}\\
In this chapter, the performance of \textit{Wolbachia} in reducing dengue incidence after a single introduction of dengue cases is investigated. The parameter estimation is also presented. We justify the parameter values chosen and explore the parameter spaces of several parameters. A global sensitivity analysis for the model is also conducted, and we compare the reproduction numbers obtained in the absence and presence of \textit{Wolbachia}-carrying mosquitoes. 

\noindent
\textbf{Chapter~\ref{chap:multipleintro}: Dengue Transmission Dynamics for Multiple Introductions}\\
In this chapter, we investigate multiple introductions of dengue cases over weekly, year-round and three month periods, and assess the performance of \textit{Wolbachia} in reducing dengue in these scenarios.  We then  investigate the potential for an outbreak, and the effects of the transmission rate and seasonality on \textit{Wolbachia} performance in reducing dengue. We also compare the effective reproduction number and our simulations where dengue cases are introduced for one week of the year.

\noindent
\textbf{Chapter~\ref{chap:modelmultiplestrains}: Mathematical Modelling of Two Dengue Serotypes}\\
After studying the performance of \textit{Wolbachia} in reducing dengue incidence when only a single dengue serotype circulates, we turn our attention to the performance of \textit{Wolbachia} in reducing dengue in the presence of two dengue serotypes. We begin by developing two serotype dengue models in the absence and presence of \textit{Wolbachia}-carrying mosquitoes.   These models take into account  antibody-dependent enhancement (ADE) and temporary immunity. As  a reminder for readers, we re-describe the parameter values used in our investigation.  The models are then used to study the effects of the symmetric and asymmetric epidemiological characteristics of dengue serotypes which are presented in Chapters~\ref{chap:symmetry} and ~\ref{chap:asymmetry}, respectively.

\noindent
\textbf{Chapter~\ref{chap:symmetry}: Two Serotypes with Symmetric Epidemiological Characteristics}\\
In this chapter, the symmetric epidemiological characteristics of dengue serotypes under different disease introduction scenarios are investigated. Two disease introduction scenarios are considered. First, individuals carrying dengue serotype~1 are introduced into the population at weekly intervals for one year, and then individuals carrying another dengue serotype are introduced into the population at weekly intervals for another year. Second, individuals carrying dengue serotype~1 are introduced into the population at weekly intervals over a period of two years, and then individuals carrying another dengue serotype are introduced into the population at weekly intervals over a further two year period.  We  explore the effects of the antibody-dependent enhancement and the transmission probability on \textit{Wolbachia} performance in reducing dengue incidence in these two scenarios.    

\noindent
\textbf{Chapter~\ref{chap:asymmetry}: Two Serotypes with Asymmetric Epidemiological Characteristics.}\\
In this chapter, we study the performance of \textit{Wolbachia} in reducing dengue incidence when the epidemiological characteristics of dengue serotypes differ.  As in Chapter~\ref{chap:symmetry}, we explore the effects of the antibody-dependent enhancement and the transmission probability on \textit{Wolbachia} performance in reducing dengue incidence. 
\enlargethispage{\baselineskip}

\noindent
\textbf{Chapter~\ref{chap:Conclusions}: Conclusions and Future Directions}\\
In this chapter, we summarise the original results discovered in this thesis, present our conclusions and suggest several possible future directions for research.


\chapter[Literature Review]{Literature Review}\label{chap:literaturereview}

\ifpdf
    \graphicspath{{2/figures/PNG/}{2/figures/PDF/}{2/figures/}}
\else
    \graphicspath{{2/figures/EPS/}{2/figures/}}
\fi

The review of literature  presented in this chapter is divided into two parts: a  review of biological literature and a review of mathematical literature. The biological literature review introduces dengue (Section~\ref{dengue}), the dengue vector (Section~\ref{denguevector}), and  \textit{Wolbachia} (Section~\ref{WolbachiaReview}). The mathematical literature review explains the deterministic SIR model (Section~\ref{SIRReview}), vector-borne disease modelling (Section~\ref{VectorBorneModel}), the inclusion of seasonality in compartmental models (Section~\ref{SeasonalityReview}), the basic reproduction number (Section~\ref{R0Review}), and introduces a model for \textit{Wolbachia} population dynamics (Section~\ref{WolbachiaModelReview}), a model for dengue in the absence (Section~\ref{denguemodelReview}) and presence of \textit{Wolbachia} (Section~\ref{modeldenguepresenceReview}), and  the Latin Hypercube Sampling (LHS) and  Partial Rank Correlation Coefficient (PRCC) methods (Section~\ref{SensitivityReview}).

When material from this review is required in later chapters, we shall reference the appropriate sections so that the interested reader can review the details. 

\section{Biological Review}\label{Biological aspects}
This section presents background information about dengue, dengue vectors and \textit{Wolbachia}. 
\subsection{Dengue}\label{dengue}

Dengue is a vector borne disease that is transmitted by mosquitoes, and has attracted public health concern worldwide. Over 40\% of the world's population are living in dengue-endemic regions and approximately 390 million individuals are infected with dengue annually~\cite{Bhatt2013}. Infections with the dengue virus can lead to dengue fever (DF), dengue haemorrhagic fever (DHF) and dengue shock syndrome (DSS), the last two of which are the most severe forms of dengue.  The fatality rates for DHF and DSS can be as high as 20\%, although this can be reduced to less than  1\% if the proper treatment is accessible~\cite{WHO}, which is generally in developed countries.

\indent
The number of dengue cases is increasing worldwide~\cite{NNDSS2013,Karyanti2014,Mia2013}. For example, in Australia,  there has been an increasing trend in the number of dengue cases over the last two decades, an increase from around 17 in 1991 to over 1000 in 2015 \cite{NNDSS2013}. In dengue endemic countries there are also increasing numbers of dengue cases.  In Indonesia, the incidence of dengue has been increasing since 1968~\cite{Karyanti2014}. In Malaysia, the number of dengue cases increased by 14\% annually from  2000--2010~\cite{Mia2013}.

In  regions or countries where dengue is not endemic,  outbreaks of the disease are triggered by imported cases. These are individuals who have acquired dengue overseas. In Australia, local outbreaks are generally caused by imported cases, with an estimated 1132 overseas-acquired cases of dengue entering Australia in 2010~\cite{Knope2013}. In Taiwan, dengue outbreaks begin by importation of dengue.  The disease  then spreads until transmission ends in winter due to the cold weather. The same pattern is repeated every year~\cite{Lin2010}. These examples highlight the effect of dengue importation on disease transmission dynamics.

\indent
There are four serotypes of dengue viruses: DEN1, DEN2, DEN3 and DEN4. Individuals obtain lifelong immunity to a given dengue serotype after infection with it,  and have a short-term antibody response to the other serotypes, which lasts for around 2--9 months~\cite{Wearing2006}. If they are subsequently infected by a different serotype, they are likely to contract DHF or DSS.  Patients with a severe form of dengue show peak viraemia levels (levels of the virus in the blood) 100--1000 times higher than DF patients~\cite{Vaughn2000}.

DHF is associated with the antibody dependent enhancement (ADE) of dengue virus replications. The ADE mechanism is as follows. An individual who is primarily infected by a particular serotype will produce neutralising antibodies to that serotype, but when the same individual is secondarily infected with other serotypes, the pre-existing antibodies to the previous infection do not neutralise, but rather enhance, the replication of the secondary virus~\cite{Whitehead2007, Murphy2011}. ADE may result in higher transmissibility rates of dengue from humans to mosquitoes.

There is variability in the epidemiological characteristics of the four dengue serotypes, and the severity of the disease varies between serotypes~\cite{Vaughn2000, Althouse2014, Barraquer2014, Tricou2011, Nishiura2007, Ferguson1999a}. Nishiura and Halstead~\cite{Nishiura2007} found that DEN1 infection causes more severe symptoms than DEN4, while Tricou \textit{et al.}~\cite{Tricou2011} found that infections with DEN1 result in higher viraemia levels than infections with DEN2.  Estimates of the basic reproduction numbers from serological data showed  relatively little difference between dengue serotypes~\cite{Ferguson1999a,Barraquer2014}. 

The distribution and the dominance of dengue serotypes in a given outbreak varies over time. In Thailand, all four dengue serotypes circulated during the period 2000-2010, with DEN3 being dominant during the periods 2000--2002 and 2008--2010, whereas DEN4 was most common between 2003 and 2008 as reported by Limkittikul \textit{et al}.~\cite{Limkittikul2014}. In Malaysia, DEN1 was dominant during 2004 and 2005, but was overtaken by DEN2, which was dominant during 2006 and 2007.  Although all four dengue serotypes were circulating in 2014, none of them was dominant~\cite{Mia2013}. In Taiwan, DEN2 dominated during 2002 and 2003, DEN3 dominated during 2005--2006, and DEN1 dominated in 2004 and 2007~\cite{Lin2010}. More information about dengue epidemics in several countries can be found in the literature (see, for example, ~\cite{Villar2015, Bravo2014, Azou2014, Dantes2014}).

 In general, the characteristics of DF infection are high fever with a temperature of  approximately $40^\circ C$, severe headache, pain behind the eyes, muscle and joint pains, nausea, vomiting, swollen glands or rash.  These symptoms last for approximately 2--7 days. People with a severe form of dengue exhibit all DF symptoms, together with plasma leakage, fluid accumulation, respiratory distress, severe bleeding, or organ impairment, symptoms which may lead to death if proper treatment is not administered~\cite{WHO}.

The dynamics of dengue transmission are very seasonal due to fluctuations in the dengue vector caused by climatic factors such as temperature and rainfall. Furthermore, there are differences between regions or countries in the periods when epidemics occur. For example, in Thailand, the number of cases peaks between May and September each year as reported by Limkittikul \textit{et al.}~\cite{Limkittikul2014}. In Taiwan, the peak is around October and then incidence diminishes around December~\cite{Lin2010}. 

There are several important aspects to be considered when studying dengue dynamics at a population level: importation of dengue, dengue serotype characteristics and the seasonal factors which affect the mosquito population dynamics.

\subsection{Dengue Vectors}\label{denguevector}
Two vectors that transmit dengue are \textit{Aedes albocpictus} and \textit{Aedes aegypti}.   The latter is considered to be the main vector.  \textit{Aedes albocpictus} mostly lives in rural areas and large water containers. \textit{Aedes aegypti} is highly anthropophilic, preferring to feed on humans only, and lives in urban and semi urban areas which are densely populated. \textit{Aedes aegypti} moves only short distances of  around 300 $m$ during its lifetime~\cite{Harrington2005}.

\begin{figure}
\centering
\includegraphics[scale=1]{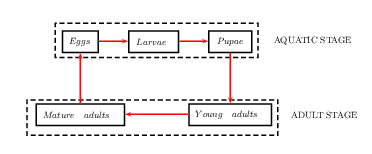} 
\caption[The lifecycle of mosquitoes]{ The lifecycle of \textit{Aedes aegypti}: eggs progress to larvae, then pupae and become mature. The aquatic stage encompasses eggs, larvae, and pupae. The adult stage comprises young and mature mosquitoes.}
\label{LifeCycle}		
\end{figure}

\begin{figure}[h]
\centering
\includegraphics[scale=1]{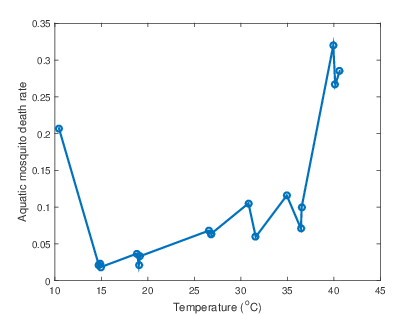}
  \caption[The aquatic mosquito death rate for different temperatures]{Plot of the aquatic mosquito death rate for different temperatures as given in Yang \textit{et al~\cite{Yang2011,Yang2009}.}}
    \label{AquaticDeathRate}		
\end{figure}

\begin{figure}[h!]
\centering
\includegraphics[scale=1]{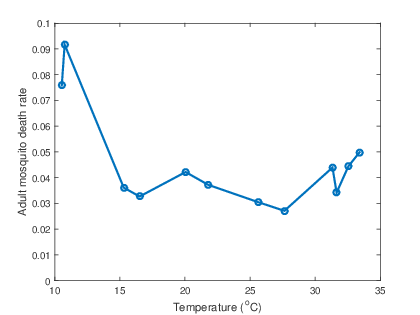}
  \caption[The adult mosquito death rate for different temperatures]{Plot of the adult mosquito death rate for different temperatures as given in Yang \textit{et al.}~\cite{Yang2011,Yang2009}. }
    \label{AdultDeathRate}
\end{figure}

\indent
Figure \ref{LifeCycle} shows the \textit{Aedes aegypti} lifecycle, which consists of eggs, larvae, pupae, young adult and adult \textit{Aedes aegypti}. In modelling, eggs, larvae and pupae are commonly grouped into one category because only adult mosquitoes can transmit dengue and the dengue virus cannot influence the aquatic stage of development. The mosquito life cycle is highly dependent on climatic conditions such as temperature and rainfall.  Figures~\ref{AquaticDeathRate}--\ref{AdultDeathRate} show the plots of the aquatic death rate and the adult mosquito death rate against temperature given by Yang \textit{et al.}~\cite{Yang2011,Yang2009}. These figures show that the aquatic death rate and the adult mosquito death rate
vary as the temperature changes. Figure~\ref{AquaticDeathRate} shows that the aquatic death rate is high when the temperature is less than approximately $13^\circ C$ or greater than $35^\circ C$. The adult death rate is very high when the temperature is below $15^\circ C$ (see Figure \ref{AdultDeathRate}).

\noindent

\indent
When feeding on an infected human, \textit{Aedes aegypti} ingests viraemic blood.  After a period of time known as the Extrinsic Incubation Period (EIP), the salivary glands of the mosquito become infected.  During the EIP the viruses replicate and are internally disseminated within the mosquito.  The EIP varies depending on temperature, but on average lasts for one to two weeks~\cite{Watts1987, Chan2012}. Infected mosquitoes are able to transmit dengue if they live longer than the EIP. When an infected mosquito bites a susceptible human, the human becomes exposed, but not yet infectious, for approximately 4--7 days and then becomes infectious for the following 3--7 days. If susceptible mosquito subsequently feeds on the infectious human, the cycle continues.

\indent
To date, no vaccines against dengue are commercially available. Although human behavioural changes can  be helpful in reducing dengue transmission,   most interventions against dengue focus on vector control.  A number of strategies have been implemented to control vector populations, but they are difficult to sustain~\cite{Grisales2013, Montella2007,Ritchie2002,Ooi2006,Freitas2014}, particularly in the developing world. Insecticides, for example, become less effective as mosquitoes develop resistance to the chemicals~\cite{Grisales2013, Montella2007,Freitas2014}. The removal of mosquito breeding sites is another intervention that has  been carried out, but this must be repeated often, as has been found in Cairns and Singapore~\cite{Ritchie2002,Ooi2006}. A promising strategy against dengue is the \textit{Wolbachia} intervention~\cite{Bian2010, McMeniman2009, Moreira2009, Hofman2011, Walker2011, Frentiu2014}. Details about  \textit{Wolbachia} are presented in the following sub-section.

\subsection{\textit{Wolbachia}}\label{WolbachiaReview}
This section discusses the \textit{Wolbachia} bacterium, its characteristics, and effects on mosquitoes and the dengue virus.

\textit{Wolbachia} is a genus of bacteria which has infected around 66\% of all different species of insects~\cite{Hilgenboecker2008}, but does not naturally infect \textit{Aedes aegypti}, the main vector transmitting dengue. Infecting \textit{Aedes aegypti} with \textit{Wolbachia} has been found to reduce dengue viral load in mosquitoes~\cite{Bian2010, Moreira2009, Hofman2011, Walker2011, Frentiu2014, Ye2015}. Consequently, it has been proposed as  a strategy to reduce the spread of dengue.

\indent
\textit{Wolbachia} are maternally inherited intracellular bacteria that manipulate the reproduction of a diverse range of arthropod hosts. One form of reproductive manipulation caused by \textit{Wolbachia} infection is known as cytoplasmic incompatibility (CI)~\cite{Stouthamer1999, Weren1997, Hofman2011, Walker2011}.  CI gives a female \textit{Aedes aegypti} mosquito a reproductive advantage, the details of which are given below.
\begin{enumerate}[wide=0pt,labelwidth = 1.3333em, labelsep = 0.3333em, leftmargin = \dimexpr\labelwidth + \labelsep\relax ]
\item \textit{Wolbachia}-carrying female mosquitoes produce viable offspring when mating with either \textit{Wolbachia}-carrying or non-\textit{Wolbachia} male mosquitoes, and immature mosquitoes mature to non \textit{Wolbachia} and \textit{Wolbachia}-carrying mosquitoes.
\item Non-\textit{Wolbachia} female mosquitoes can only reproduce successfully with non-\textit{Wolbachia} male mosquitoes, and immature mosquitoes mature to non-\textit{Wolbachia} mosquitoes only.
\item  If non-\textit{Wolbachia} female mosquitoes mate with \textit{Wolbachia}-carrying males, they cannot produce offspring successfully, although an embryo is formed, further blocking reproduction~\cite{Walker2011}.
\end{enumerate}

\indent
Laboratory experiments have been conducted to assess the effects of \textit{Wolbachia} on \textit{Aedes aegypti}.  Two \textit{Wolbachia} strains were used in these experiments: \textit{WMelPop} and \textit{WMel} \cite{Hofman2011, Turley2013, Walker2011,Turley2014}. \textit{WMelPop} halves the mosquito lifespan~\cite{Yeap2011, Walker2011} and reduces egg viability~\cite{McMeniman2010}. \textit{WMelPop} has been found to reduce blood feeding success~\cite{Turley2009}. Furthermore, a fecundity cost of up to 56\% has been found in \textit{WMelPop}-carrying mosquitoes~\cite{McMeniman2011,Walker2011}. By contrast, \textit{WMel} reduces the mosquito lifespan by approximately 10\%~\cite{Walker2011}, does not influence the time development of \textit{Aedes aegypti}  and has no significant effect on egg viability after oviposition~\cite{Walker2011}. Both strains of \textit{Wolbachia} have the ability to reduce the levels of dengue viruses in the salivary glands of \textit{Aedes aegypti}~\cite{Moreira2009, Walker2011}.

If \textit{Wolbachia} causes a marked reduction in the mosquito lifespan, multiple introductions of \textit{Wolbachia}-carrying mosquitoes may be required to ensure \textit{Wolbachia} persistence in the wild. The elapsed time between introductions  and the number of released mosquitoes determines the likelihood of persistence~\cite{Hancock2011}.  The release of more males than females can also  lead to \textit{Wolbachia} persistence~\cite{Hancock2011a, Zheng2014}.

Results from field experiments show that \textit{Wolbachia}-carrying mosquitoes can persist~\cite{Hoffmann2014,Hofman2011}. These results suggest that the population of \textit{Wolbachia}-carrying mosquitoes can stabilise over time, and hence, potentially reduce dengue transmission. Furthermore, the proportion of \textit{Wolbachia}-carrying mosquitoes in the field has been found to be around 90\% in comparison to non-\textit{Wolbachia} mosquitoes~\cite{Hofman2011, Hoffmann2014}. Moreover, the maternal transmission of \textit{Wolbachia} may be perfect or near to perfect, as seen in field trials~\cite{Hoffmann2014}. Lower rates of infection with dengue viruses of serotypes 1--3 in mosquitoes carrying \textit{Wolbachia} have also been found.

In summary, there are several important characteristics of \textit{Wolbachia} that can affect the population dynamics of mosquitoes, namely cytoplasmic incompatibility (CI), the mosquito death rate and the maternal transmission. Dengue-related characteristics such as reduced biting rates, and reduced levels of dengue virus in mosquitoes are also key to the potential reduction in dengue transmission.

\section{Mathematical Review}\label{Deterministic models}
This section presents the mathematical theory used in this thesis. This includes a formulation of the deterministic SIR model, an explanation of seasonality, and discussions of vector-borne disease modelling, the basic reproduction number, a model specific for \textit{Wolbachia} and dengue dynamics, and the latin hypercube sampling (LHS) and  partial rank correlation coefficient (PRCC) methods for multivariate analysis.

\subsection{The Deterministic SIR model}\label{SIRReview}
A deterministic formulation of the SIR model is presented in this section. A deterministic model is appropriate when studying disease transmission dynamics in large populations, as is the case throughout this thesis. The SIR model has been used as a cornerstone for the development of compartment-based models. The general principle of compartment-based models is that the population is divided into different sub-populations, according to their disease status. In the SIR model, the population is divided into three sub-populations, namely \textit{Susceptible} (S), \textit{Infectious} (I) and \textit{Removed} (R).

\indent
Let us consider the simple SIR model without demography and assume homogenous and well-mixed populations. 
Humans are born  \textit{susceptible} to infection and move to the \textit{infectious} class at a rate $\beta$ after contact with infectious individuals. After a period of time $1/\gamma$, they recover. This SIR model is governed by the following system of differential equations

\begin{align}
\frac{dS}{dt}&=-\beta I S,\label{S}\\
\frac{dI}{dt}&=\beta I S-\gamma I,\label{I}\\
\frac{dR}{dt}&=\gamma I\label{R},
\end{align}
where $\beta$ and $\gamma$ are positive constants.  Here the parameter $\beta$ represents a combination of the rate at which contact between susceptible and infectious people is made and the probability that contact results in the successful transmission of the disease, while the parameter $\gamma$ is the recovery rate.

We can determine the threshold above which an epidemic occurs by rearranging the infected population equation (Equation~\ref{I}), that is
$$\frac{dI}{dt}=I\left(\beta S-\gamma\right).\\$$
 It is easy to see that the critical threshold is  $S=\gamma/\beta$, where the populations of $S$ and $I$ are in proportions. When an infectious case is introduced to the population, an epidemic will proceed if the susceptible population is higher than this critical threshold, otherwise an outbreak cannot occur. The inverse of this critical threshold is called the basic reproduction number (further explained in subsection~\ref{R0Review}), which is defined  to be the number of new cases generated by a typical infectious individual in an entirely susceptible population.

The classical SIR model can be extended by adding other classes, depending on the purpose of the model and the characteristics of the disease being studied. For example, if a disease has a latent period, one may add a latent compartment to the system of equations.  Another possible extension would be to include demographic factors such as birth and death rates.  As additional details are incorporated into the model, it becomes more complex, and typically the calculation of the basic reproduction number is conducted by constructing the next generation matrix and finding the spectral radius of that matrix as described in Section~\ref{R0Review}.  Although some analytical expressions such as the steady state or the critical threshold can be derived, deriving analytical solutions in terms of S and I simultaneously is not straightforward~\cite{Keeling2008}. This is particularly true when models become increasingly complex as additional important features are included. One way to find the solutions of a deterministic model is by the use of computational tools. Throughout this thesis, we use \textsc{Matlab}~\cite{Matlab} to generate the solutions of the models.

\subsection{Vector-borne Disease Modelling}\label{VectorBorneModel}
Some diseases such as malaria, chikungunya, and dengue are transmitted to humans via a vector  and consequently their transmission dynamics are affected by the population dynamics of the vector.  At the population level, the transmission dynamics of vector-borne diseases can be investigated using mathematical models~\cite{Yakob2013, Yakob2010, McLennan2014, Beers2015}.  Deterministic models provide one technique for modelling the transmission dynamics of vector-borne diseases, particularly in large populations.   The simple SIR model presented in Section~\ref{SIRReview} can be extended into such a model by including the vector population and introducing parameters which adjust for the  biological features of the vectors and the disease. A review on vector-borne disease mathematical models can be found in Smith \textit{et al.}~\cite{Smith2014}.

When more than one vector is involved in the transmission of a disease, the biological features of each vector such as its  transmission rate, and  birth and death rates  influence disease transmission. In this thesis, we construct deterministic mathematical models including two types of mosquitoes: mosquitoes not carrying \textit{Wolbachia} and mosquitoes carrying \textit{Wolbachia} (referred throughout this thesis as ``non-\textit{Wolbachia}'' and ``\textit{Wolbachia}-carrying'' mosquitoes, respectively), together with the human population with dengue, and use these models to investigate the spread of dengue. We focus on the situation where an outbreak occurs and measure the reduction in the number of people infected due to the introduction of \textit{Wolbachia}-carrying mosquitoes.

\subsection{Seasonality}\label{SeasonalityReview}

The transmission of infectious diseases varies seasonally depending on various factors. Climatic fluctuation contributes to the seasonal dynamics of many infectious diseases because it affects pathogen transmission rates~\cite{Koelle2005}. For vector-borne diseases, climatic factors are linked to the fluctuation of vector populations and, hence, to vector-borne disease transmission. 

\indent
In most mathematical models, the seasonal effect is captured by
the use of sinusoidal and/or square wave functions.  For example, seasonal variation in the transmission rate can be represented using a sinusoidal function such as

\begin{equation}
\beta(t)=\beta_0\left(1+\delta \cos 2\pi t\right)\,,
\label{BetaSeasonally}
\end{equation}
where $\beta(t)$ is the transmission rate which varies seasonally, $\beta_0$ is the average transmission rate, and $\delta$ is the amplitude of seasonal forcing of the transmission rate or ``strength of seasonality'' and  lies between $0 $ and $1$.

Seasonal variation can also be modelled using a square wave function such as  

\begin{equation}
 \beta = \left\{
  \begin{array}{l l}
    \beta_0\left(1+\delta\right), & \quad \textrm{for high seasons},\\
    \beta_0\left(1-\delta\right), & \quad \textrm{for low seasons},
  \end{array} \right.
	\end{equation}
where each year is divided into two different seasons: the high season and the low season. The parameter  $\delta$ represents the degree of seasonality which varies between 0 and 1 (see \cite{Earn2000, Keeling2001, Nguyen2008, Olinky2008, Stone2007}).

\subsection{Basic Reproduction Number}\label{R0Review}

The basic reproduction number is an important quantity in epidemiological modelling which is commonly estimated when data is available~\cite{Glass2011, Chowel2007}. The basic reproduction number, generally denoted by $\mathcal{R}_0$, is defined to be the number of secondary cases generated by a typical infectious individual in an entirely susceptible population. 
  The expression for the basic reproduction number can be obtained by constructing the next generation matrix (NGM) for the model~\cite{Diekmann1990, Diekmann2010, Driessche2002, Heffernan2005} and finding its dominant eigenvalue. Another method to derive the basic reproduction number for a non-seasonal model is to use the ``survival function''. However, this method becomes increasingly difficult when there are more infectious states and hence the use of the next generation approach is more appropriate~\cite{Heffernan2005}.
  
Diekmann \textit{et al.}~\cite{Diekmann2010} give  a clear explanation of how to construct the NGM, which they denote by $K$. The decomposition of the NGM relates to epidemiologically new infections only. For example, changes in the population in a model with latent and infectious compartments consist of new infections into the latent compartment and members of the population moving from the latent to the infectious compartment. The change from the latent to the infectious compartment is not a new infection. A matrix whose decomposition relates to the expected offspring of individuals in any infected state is called a next generation matrix with large domain and denoted by $K_L$.   Diekmann \textit{et al.} showed that the largest eigenvalue of the next generation matrix with large domain is equal to that of the next generation matrix, $\rho\left(K_L\right)=\rho\left(K\right)$, and hence, to the basic reproduction number, $\mathcal{R}_0$. 

 The steps for finding the basic reproduction number
 given in Diekmann \textit{et al}.~\cite{Diekmann2010} are the following.
\begin{enumerate}
\item Let $n$ be the number of infected states, and $\pmb{x}=\left(x_1, x_2, ...,x_n\right)$ be the possible infected states of individuals.  We write a linearised infected subsystem around the disease-free steady state in the form
\begin{equation}
\pmb{\frac{dx}{dt}}=\left(\pmb{T} + \pmb{\Sigma}\right) \pmb{x}\,,
\end{equation}
where   $T$ is the transmission matrix, which encodes the production of new infection, and $\pmb{\Sigma}$ is the transition matrix, which encodes the transition of infected individuals between compartments including those that transfer out of compartments for reasons such as the death of the individual.
\item  The next generation matrix with large domain can be found using the formula $K_L=-T\pmb{\Sigma}^{-1}$. The basic reproduction number is calculated by finding the spectral radius of the matrix $K_L$.
\end{enumerate}

\subsection{Mathematical Models for \textit{Wolbachia} Population Dynamics}\label{WolbachiaModelReview}
A number of mathematical models have been developed to provide insight into the population dynamics of \textit{Wolbachia}-carrying mosquitoes ~\cite{Hancock2011, Hancock2012, Chan2013, Ndii2012,Zhang2015,Zheng2014,Rodrigues2014}. Hancock \textit{et al.}~\cite{Hancock2011} developed a mathematical model to explore the host population dynamics and \textit{Wolbachia} infection frequency with both single and multiple introductions (``seeding'') of \textit{Wolbachia}-carrying mosquitoes. The same authors also developed a metapopulation model to explore spatial dynamics and found that spatial variations in density-dependent competition have an effect on the spread of \textit{Wolbachia} infection~\cite{Hancock2012}. Chan and Kim~\cite{Chan2013} developed a spatial mathematical model incorporating both slow and fast dispersal situations for the spread of mosquitoes.  They found that temperature affects the speed of \textit{Wolbachia} invasion and that the death rate of \textit{Wolbachia}-carrying mosquitoes influences their persistence.  Zhang \textit{et al.}~\cite{Zhang2015} developed a birth-pulse model of \textit{Wolbachia} spread considering  cyptoplasmic incompatibility and different density dependent death rates. They found that it is likely that mosquitoes carrying \textit{Wolbachia} will invade the population. Zheng \textit{et al.}~\cite{Zheng2014} developed a mathematical model to analyse the release of \textit{Wolbachia}-carrying mosquitoes into the population, and found that the release of minimal numbers of mosquitoes with \textit{Wolbachia} can lead to persistence of \textit{Wolbachia}-carrying mosquitoes.

Ndii \textit{et al.}~\cite{Ndii2012} developed a mathematical model of mosquito population dynamics in the presence of \textit{Wolbachia} which incorporated the effects of CI in the mating function and competition for resources in the aquatic stage. They found that  \textit{Wolbachia}-carrying mosquitoes and non-\textit{Wolbachia} mosquitoes can coexist, and showed that the steady state where \textit{Wolbachia}-carrying mosquitoes alone persist only exists when maternal transmission is perfect, a finding which was confirmed  by  Chan and Kim~\cite{Chan2013}. Rodriguez \textit{et al.}~\cite{Rodrigues2014} used a mathematical model to analyse the changes in the proportion of \textit{Wolbachia}-carrying mosquitoes during the host's life cycle on the dynamics of the spread of \textit{Wolbachia}, and found that the spread of \textit{Wolbachia} may be affected by these changes.

In this thesis, we extend the model of Ndii \textit{et al.}~\cite{Ndii2012} to include a human population in which dengue is present. The extended model is used to investigate the performance of \textit{Wolbachia} in reducing dengue. The model for mosquito population dynamics in the presence of \textit{Wolbachia}-carrying mosquitoes as given by Ndii \textit{et al.} \cite{Ndii2012} is governed by the following differential equations

\begin{align}
\frac{dA_{N}}{dt} &= \rho_{N}\frac{F_{N}M_{N}}{P}\left(1-\frac{\left(A_{N}+A_{W}\right)}{K}\right)-\mu_{NA}A_{N}-\tau_{N}A_{N},\\
\frac{dM_{N}}{dt} &= \epsilon_{N}\tau_{N}A_{N}-\mu_{N}M_{N} + \epsilon_{NW}\left(1-\alpha\right)\tau_{W}A_{W},\\
\frac{dF_{N}}{dt} &= \left(1-\epsilon_{N}\right)\tau_{N}A_{N}-\mu_{N}F_{N}+ \left(1-\epsilon_{NW}\right)\left(1-\alpha\right)\gamma_{W}A_{W},\\
\frac{dA_{W}}{dt} &= \rho_{W}\frac{F_{W}\left(M_{W}+M_{N}\right)}{P}\left(1-\frac{\left(A_{N}+A_{W}\right)}{K}\right)-\mu_{WA}A_{W}-\tau_{W}A_{W},\\
\frac{dM_{W}}{dt} &= \epsilon_{W}\alpha\tau_{W}A_{W}-\mu_{W}M_{W},\\
\frac{dF_{W}}{dt} &= \left(1-\epsilon_{W}\right)\alpha\tau_{W}A_{W}-\mu_{W}F_{W},
\label{ModelNdiietal}
\end{align}
where $P=F_N +M_N+F_W +M_W$. Here $A$, $M$ and $F$ denote aquatic, male and female mosquitoes, respectively, and the subscripts $N$ and $W$ differentiate between non-\textit{Wolbachia} and \textit{Wolbachia}-carrying mosquitoes. The parameters  $\epsilon_N$ and $\epsilon_{W}$ denote the respective proportions of non-\textit{Wolbachia} and \textit{Wolbachia}-carrying mosquitoes that are male. 

The effect of CI is represented by the following equations. Non-\textit{Wolbachia} aquatic mosquitoes are produced after non-\textit{Wolbachia} male and female mosquitoes mate.
The population's growth is limited by the carrying capacity $K$ through the term

\begin{equation}
\rho_N \frac{F_N M_N}{P}\left(1-\frac{A_N+A_W}{K}\right).
\end{equation}
The fact that \textit{Wolbachia}-carrying females reproduce after mating with non-\textit{Wolbachia} or \textit{Wolbachia}-carrying males is represented by the term
\begin{equation}
\rho_W \frac{F_W \left(M_N +M_W\right)}{P}\left(1-\frac{A_N+A_W}{K}\right).
\end{equation}

Furthermore, the effects of imperfect maternal transmission are also included in the model. That is, not all \textit{Wolbachia} aquatic mosquitoes  mature to be \textit{Wolbachia}-carrying adult mosquitoes. A proportion of them mature to be non-\textit{Wolbachia} adults.  This proportion is represented by the parameter $\alpha$. A proportion $\epsilon_{NW}$ of the \textit{Wolbachia} aquatic mosquitoes  that mature to become non-\textit{Wolbachia} mosquitoes become male and the rest become female mosquitoes.

\subsection{Mathematical Models for Dengue in the Absence of \textit{Wolbachia}}\label{denguemodelReview}
A number of dengue models have been developed and analysed. A review of various deterministic dengue models can be found in Andraud \textit{et al.} \cite{Andraud2012}. Dengue models were split into two categories: vector-host and host-to-host transmission.

To capture the effect of seasonality, a sinusoidal function is generally used~\cite{McLennan2014,Andraud2013}. For example, McLennan-Smith and Mercer \cite{McLennan2014} developed a mathematical model for dengue which included seasonality in the transmission rate. The model was found to have increasingly complex behaviour as the strength of seasonality was increased.  Andraud \textit{et al.}~\cite{Andraud2013}  used a periodically-forced model to estimate parameter values using data from Singapore.  They found a good fit between the output from the model and the  data from Singapore, and showed that seasonality is an important factor driving dengue dynamics.

\noindent

Yang and Ferreira~\cite{Yang2008} developed a dengue mathematical model with variation in some parameter values over time. The parameters were set to different values to represent favourable and unfavourable conditions for the development of mosquitoes.  Their model divided aquatic and adult mosquito populations into six compartments:  eggs, larvae, pupae, susceptible, exposed and infectious adults. They found that, although dengue persists in favourable conditions, there is a gap of several years between successive outbreaks of dengue.

Several studies have investigated the effects of introducing individuals with dengue into the population~\cite{Williams2015,Bannister2013}. Bannister-Tyrrell \textit{et al.}~\cite{Bannister2013} investigated the variation of dengue activity in Australia using a process based modelling approach in which dengue was introduced bimonthly, monthly and weekly. They found that increasing the frequency of introductions of dengue results in larger outbreaks. Williams \textit{et al.}~\cite{Williams2015} used an existing dengue model, which involved an entomological component (CIMSiM)  and a disease component (DENSiM), to assess the effects of the importation rate on dengue outbreaks. They conducted weekly introductions with the number of imported dengue cases ranging from 1--25 individuals. They found that higher rates of importation each week result in increased monthly dengue incidence. However, these studies did not investigate the importation of different serotypes of dengue, a factor which is known to affect dengue transmission dynamics and have effects on secondary infections.

A number of mathematical models considering multiple serotypes of dengue have been developed~\cite{Ferguson1999,Cummings2005,Bianco2009,Adams2006,Romero2013,Kooi2014,Aquiar2011,Hu2013,Woodall2014}. An important issue that arises when investigating disease dynamics in the presence of multiple serotypes of dengue is the effect of antibody dependent enhancement (ADE) on dengue dynamics.  Adams and Boots~\cite{Adams2006} used a two-serotype dengue model to study dengue transmission dynamics, and found that the enhancement of transmission does not lead to the exclusion of either serotype  when both serotypes have the same basic reproduction number. Ferguson \textit{et al.}~\cite{Ferguson1999} found that enhancement may permit the coexistence of all serotypes. Woodall  and Adams~\cite{Woodall2014} developed a model in which they assumed that only small fraction of primarily infected humans became susceptible to enhanced secondary infection. They found that antibody-dependent enhancement may not be driving the oscillatory dynamics of dengue.  Aguiar \textit{et al.}~\cite{Aquiar2011} developed a seasonal multiple serotype dengue model and performed a comparison study between non-seasonal, low seasonal and high seasonal models with a low importation rate of infected individuals.  They found complex dynamics, and a match between empirical DHF monitoring data and model simulations. Romero \textit{et al.} \cite{Romero2013} found that the presence of ADE heterogenity can facilitate the persistence of dengue serotypes. Kooi \textit{et al.}~\cite{Kooi2014} used a mathematical model to analyse the characteristics of two dengue serotypes for which the force of infection differs. They concluded that models that include the assumption of identical epidemiological characteristics are  useful when the difference between dengue serotype  characteristics is not great, otherwise an exploration of different (asymmetric) epidemiological characteristics  is required. However, most of these studies did not consider different disease introduction strategies (or \textit{Wolbachia}).


Most  multiple serotype dengue models do not take  vector population dynamics into account. Hu \textit{et al.}~\cite{Hu2013} investigated the effect of including the mosquito population and the host incubation period on the dynamics of dengue when multiple serotypes are circulating. In this host-only model, they found a large degree of frequency instability which made future predictions difficult.  However, this behaviour disappears when the vector is included in the model. Their findings emphasize the importance of including vectors of dengue in mathematical models.

Another area of deficiency in the mathematical literature lies in the scarcity of studies investigating \textit{Wolbachia} intervention using mathematical models which include both non-\textit{Wolbachia} and \textit{Wolbachia}-carrying mosquitoes. Given that the presence of multiple serotypes of dengue has been observed to influence dengue dynamics, a study of the effectiveness of the \textit{Wolbachia} intervention in reducing dengue in the presence of more than one dengue serotype is also worthwhile.

\subsection{Mathematical Models for Dengue in the Presence of \textit{Wolbachia}}\label{modeldenguepresenceReview}  

Few mathematical models have been developed that couple a model of the \textit{Wolbachia}-carrying mosquito population with a human dengue model \cite{Hughes2013, Hancock2011a}. Hughes \textit{et al.}~\cite{Hughes2013} developed a mathematical model considering the situation where only either non-\textit{Wolbachia} or  \textit{Wolbachia}-carrying mosquitoes persist in the population. They found that \textit{Wolbachia} can be a powerful control for dengue when $\mathcal{R}_0$ is not too large ($\mathcal{R}_0<6.2$).  Hancock \textit{et al.}~\cite{Hancock2011a} explored a scenario of \textit{Wolbachia} release  and its effects on vector-borne diseases, but did not specifically consider dengue. They found that the male-biased release of \textit{Wolbachia}-carrying mosquitoes can reduce disease transmission. Ferguson \textit{et al.}~\cite{Ferguson2015} formulated and estimated a basic reproduction number ($\mathcal{R}_0$) for dengue, and found that \textit{Wolbachia} can reduce the basic reproduction number by 66--75 \%.   A reduction of the basic reproduction number by 66--75 \% should be sufficient to reduce dengue incidence in areas with low to moderate transmission settings, that is, with typical $\mathcal{R}_0=3-4$~\cite{Ferguson2015}. However, if $\mathcal{R}_0>4$, a reduction of 66-75\% still results in $\mathcal{R}_0>1$ and hence, dengue outbreaks can still occur. Although the effective reproduction number is still higher than unity, the presence of \textit{Wolbachia} can still reduce the number of dengue cases, though the proportional reduction in dengue incidence is not high.  

The existing studies have not investigated the effect of disease introduction scenarios on \textit{Wolbachia} performance. This is important as the introduction of dengue into the population is irregular, and it is uncertain which dengue serotypes will enter the population at any given time. If the characteristics of dengue serotypes differ~\cite{Vaughn2000, Althouse2014, Barraquer2014, Tricou2011, Nishiura2007, Ferguson1999a}, undertanding the effects of these factors on \textit{Wolbachia} performance in reducing dengue incidence is important.  To our knowledge, no population level models have been developed to study the performance of \textit{Wolbachia} in reducing dengue incidence when more than one dengue serotype circulates. In this thesis, we develop  models in which two dengue serotypes circulate and study the performance of \textit{Wolbachia} in reducing dengue incidence under these conditions.

\widowpenalty=1000
\subsection{Latin Hypercube Sampling and Partial Rank Correlation Coefficient}\label{SensitivityReview}

The degree of certainty in models for infectious disease dynamics is not always known. Although some parameter values can be estimated from data, there remains uncertainty in estimates. Other parameter values are difficult to estimate from data. For example, data sets generally do not provide information on the number of asymptomatic cases, or the exact time a pathogen infects people. Therefore, it is important to conduct sensitivity analyses of the model parameters in order to determine which parameters have the greatest effects on the model outcomes of interest, and to assess the likely impact of changes in parameters within reasonable bounds.

If the inputs such as parameters or initial conditions are known with little uncertainty, we can estimate the degree of uncertainty in the output by finding the partial derivative of the output function with respect to the input parameters~\cite{Marino2008, Chitnis2008}. This is known as a local sensitivity analysis. In epidemiological models, most parameters are very uncertain, and so a global sensitivity analysis is more appropriate.  

Latin Hypercube Sampling (LHS), in conjuction with Partial Rank Correlation Coefficient (PRCC) multivariate analysis~\cite{Marino2008, Blower1994}, is often used for global sensitivity analysis. LHS is a stratified Monte Carlo sampling method, where the random parameter distributions are divided into $N$ equal probability intervals~\cite{Marino2008, Wu2013, Blower1994,McKay1979}.  Here $N$ is the sample size. Each interval of each parameter is sampled only once without replacement, and the entire range of each parameter is explored \cite{Marino2008, Wu2013, Blower1994,McKay1979}. Inputs and outputs are then ranked before  PRCC indices are determined.

Suppose we have $K$  parameters and $N$ samples.  Then the LHS matrix, $X$,  and the output, $Y$, are
\[
\stackrel{\mbox{LHS matrix}}{X=
\begin{bmatrix}
p_{11} & p_{12} &\cdots  & p_{1K} \\
p_{21} & p_{22} &\cdots & p_{2K}\\
\vdots & \vdots& \ddots & \vdots \\
p_{N1}& p_{N2} & \cdots & p_{NK}
\end{bmatrix}\,,
}
\qquad\quad
\stackrel{\mbox{Output}}{Y =
\begin{bmatrix}
O_{1} \\
O_{2}\\
\vdots \\
O_{N}
\end{bmatrix}
}.
\]
Here the output is calculated from the solution of the mathematical model using the input parameters  given in the matrix. We can rank the samples of each parameter (the entries in each row of the matrix $X$), and the entries of the output $Y$,  in order of increasing size to give matrices
\[
\stackrel{\mbox{LHS matrix}}{X_R=
\begin{bmatrix}
x_{11} & x_{12} &\cdots  & x_{1K}\\
x_{21} & x_{22} &\cdots & x_{2K}\\
\vdots & \vdots& \ddots & \vdots \\
x_{N1}& x_{N2} & \cdots & x_{NK}
\end{bmatrix}\,,
}
\qquad\quad
\stackrel{\mbox{Output}}{Y_R =
\begin{bmatrix}
y_1 \\
y_2\\
\vdots \\
y_N
\end{bmatrix}\,,
}
\]
with entries $X_R =[x_{ij}]$ and $Y_R =[y_i]$, where $x_{ij} \in \{1,\dots,K\}$ is the rank of parameter $p_{ij}$ in our parameter sample ordering and $y_i \in \{1,\dots,N\}$ is the rank of output $O_i$ in the ordering of our output values.

The partial rank correlation coefficient between $X_R$ and $Y_R$ $\left(\text{PRCC}\left(X_R, Y_R\right)\right)$ is found  using the formula~\cite{Conover1981, Marino2008}
\begin{equation}
\gamma_{xy}=\frac{\text{Cov}(x_j, y)}{\sqrt{\left(\text{Var}(x_j)\text{Var}(y)\right)}}=\frac{\sum_{i=1}^{N}\left(x_{ij}-\mu_x\right)\left(y_{i}-\mu_y\right)}{\sqrt{\sum_{i=1}^{N}\left(x_{ij}-\mu_x\right)^2 \sum_{i=1}^{N}\left(y_{i}-\mu_y\right)^2}}\,,
\end{equation}
where $\mu_x$ and $\mu_y$ are the sample means, and $j=1,2,..,K$.

The partial rank correlation coefficient is then inspected to assess the sensitivity of the parameters. If the coefficient is close to 1 or -1, the parameters are more sensitive, with negative values indicating an inverse relationship between the inputs and outputs. For example, an increase in the inputs leads to a decrease in outputs, and conversely. As the PRCC measures the nonlinear but monotonic relationships between inputs and outputs, then the output should be increasing monotonic function. In our analysis, the output is the cumulative number of infectious humans.

Significance tests can be performed to assess whether PRCC ($\gamma$) is significantly different from zero. This is tested using the  statistic~\cite{Conover1981, Blower1994}

\begin{equation}
T=\gamma \sqrt{\frac{N-2}{1-\gamma^2}}.   
\end{equation}
If the PRCC value is higher than the T-test value, it is statistically significant.

\section{Summary}
This chapter is summarised in two sections, covering biological and mathematical reviews.
\subsection{Biological Aspects}

\begin{enumerate}[wide=0pt,labelwidth = 1.3333em, labelsep = 0.3333em, leftmargin = \dimexpr\labelwidth + \labelsep\relax ]
\item The increasing number of dengue cases in many countries is an indication of the threat it poses.
\item There are four dengue serotypes. Individuals infected by one of the seroptypes gain life long immunity to that serotype, but are more likely to develop a more severe form of dengue called Dengue Haemorhagic Fever (DHF) or Dengue Shock Syndrome (DSS) when  they are later infected by other serotypes. A particular risk associated with secondary infections is higher viral loads, which can lead to higher transmissibility and death.
\item \textit{Aedes albopictus} and \textit{Aedes aegypti} are mosquitoes that transmit dengue. The latter is considered to be the main vector.  The death rate of mosquitoes is environmentally-dependent.
\item \textit{Wolbachia} gives female mosquitoes a reproductive advantage, known as cytoplasmic incompatibility (CI). \textit{Wolbachia} reduces the lifespan of mosquitoes by up to 50\% for the \textit{WMelPop} strain and 10\% for the \textit{WMel} strain.  It also reduces  the reproductive rate and viral load in the mosquitoes.
\item The higher death rates for mosquitoes carrying \textit{Wolbachia} may result in \textit{Wolbachia}-carrying mosquitoes being unable to persist  in the field.  
\item Little research has been conducted to investigate potential reductions in dengue when the \textit{Wolbachia} intervention is implemented in the field.
\item The times of importation of dengue into the population are irregular, and the dengue serotypes that enter the population are uncertain.
\item Little research has considered the effects of different disease importation strategies on \textit{Wolbachia} performance in reducing dengue.
\end{enumerate}

\subsection{Mathematical Aspects}
\begin{enumerate}[wide=0pt,labelwidth = 1.3333em, labelsep = 0.3333em, leftmargin = \dimexpr\labelwidth + \labelsep\relax ]
\item A deterministic model can be utilised to understand disease transmission dynamics, in particular, for large populations.
\item The basic reproduction number $\mathcal{R}_0$ is the average number of new cases generated by one infected individual in an entirely susceptible population. $\mathcal{R}_0$ is used to determine whether  epidemics can occur.
\item  Mathematical models for single and multiple serotypes of dengue have been developed, but only a few include a \textit{Wolbachia}-carrying mosquito population.
\item Seasonal effects can be captured using sinusoidal or square wave functions.
\item A global sensitivity analysis is highly relevant as most parameters of epidemiological models are very uncertain. Latin Hypercube Sampling, in conjunction with the Partial Rank Correlation Coefficient calculation,  can be used to understand the sensitivity of the outcome variables to the input parameters.
\item Only a few mathematical models have been developed to study the effect of \textit{Wolbachia} on dengue transmission dynamics.
\item Very few mathematical models have been developed to study the performance of \textit{Wolbachia} in reducing dengue when more than one dengue serotype circulates in the population.
\end{enumerate}

\chapter*{Part 1: A Single Dengue Serotype}
\addcontentsline{toc}{chapter}{Part 1: A Single Dengue Serotype}

In part 1 of this thesis, we investigate the performance of \textit{Wolbachia} in reducing the incidence of dengue in populations where a  single dengue serotype is circulating. The focus is on single and multiple introductions of dengue cases into the population. This part comprises Chapters~\ref{chap:singlemathmodel}--\ref{chap:multipleintro}, and the results are summarised below.
\section*{Summary}
\begin{itemize}[wide=0pt,labelwidth = 1.3333em, labelsep = 0.3333em, leftmargin = \dimexpr\labelwidth + \labelsep\relax ]
\item Single serotype dengue transmission models in the absence and presence of \textit{Wolbachia}-carrying mosquitoes are developed in Sections~\ref{sec:modelabsence} (Equations~\eqref{eq1}--\eqref{eq8}) and \ref{sec:modelpresence} (Equations~\eqref{10}--\eqref{21}), respectively.
\item We estimated parameter values for the model against data from Cairns, Australia, presented in Table~\ref{table2}, Section~\ref{WlbAbsence}.
\item Sensitivity analyses, conducted in Sections~\ref{WlbAbsence} and~\ref{DengueWolbahiaModel},  reveals that the transmission probability, the biting rate and the average adult mosquito death rate are the most influential parameters (see Figures~\ref{fig5} and~\ref{fig6}).
\item Using the \textit{WMel} strain allows \textit{Wolbachia}-carrying mosquitoes to persist according to our steady state analysis of the model (Section~\ref{WolbachiaParams}, Figure~\ref{fig3}).
\item We show in Section~\ref{BasicReprosingleIntro} that  \textit{Wolbachia} reduces the number of days at which the reproduction number is higher than unity ($\mathcal{R}_0>1$) by up to 78 days (Section~\ref{BasicReprosingleIntro}, Figure~\ref{basicrepro}).
\item  \textit{Wolbachia} introduction reduces the lengths of the potential seasons in which epidemics occur by up to six weeks as shown in Section~\ref{DiffTimeIntro}, Figure~\ref{Potential4Outbreak}.  
\item \textit{Wolbachia} is effective in reducing dengue incidence by up to 80\% in areas or regions with low and medium seasonality and moderate transmission rates as shown in Sections~\ref{SeasonStrength} and ~\ref{TransRate}, Figures~\ref{Eta_results} and ~\ref{Beta_Results}, respectively.
\end{itemize} 

\chapter[Mathematical Modelling of a Single Dengue Serotype]{Mathematical Modelling of a Single Dengue Serotype\footnote{The models are part of the manuscripts that have been published  as ``Modelling the transmission dynamics of dengue in the presence of \textit{Wolbachia}'', Meksianis Z. Ndii, R. I. Hickson, David  Allingham, G.N. Mercer, in \textit{Mathematical Biosciences} (2015) and submitted as ``The effect of \textit{Wolbachia} on dengue outbreaks when dengue is repeatedly introduced'', Meksianis Z. Ndii, David Allingham, R.I. Hickson, Kathryn Glass.}}\label{chap:singlemathmodel} 


\ifpdf
    \graphicspath{{3/figures/PNG/}{3/figures/PDF/}{3/figures/}}
\else
    \graphicspath{{3/figures/EPS/}{3/figures/}}
\fi
  In this chapter, seasonal dengue models for a single serotype in the absence and presence of \textit{Wolbachia}-carrying mosquitoes are developed. The models are an extension of the model developed by Ndii \textit{et al.}~\cite{Ndii2012} for mosquito populations, which was reviewed in Section~\ref{WolbachiaModelReview}. The aim of this chapter is to present a full derivation and important features of the models. In Chapters 4 and 5 we shall use these models to investigate the effects of \textit{Wolbachia} on dengue transmission dynamics under single and multiple introductions of dengue. 

\section{Model in the Absence of \textit{Wolbachia}}\label{sec:modelabsence}

In this section we formulate the baseline model, that is, the model in the absence of \textit{Wolbachia}-carrying mosquitoes. Our model includes populations of humans and non-\textit{Wolbachia} mosquitoes. 



\begin{figure}[h!]
\begin{center}
\includegraphics[width=0.9\linewidth]{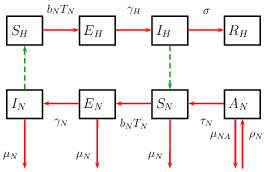}
\end{center}
\caption[Flowchart of the single serotype dengue model in the absence of \textit{Wolbachia}-carrying mosquitoes]{The flowchart of the model in the absence of \textit{Wolbachia}-carrying mosquitoes, Equations \eqref{eq1}--\eqref{eq8}. Solid lines are population progression lines and dashed lines are disease transmission lines. The subscript $H$ is for the human population and $N$ is for the non-\textit{Wolbachia} mosquito population. The compartments are `S' for susceptible, `E' for exposed to dengue, but not yet infectious, `I' for infectious, `R' for recovered, and `A' for the aquatic phase of the mosquito life cycle. The transition rates between compartments are shown next to the progression lines and these are described in the text.}
\label{fig1}
\end{figure}

The human and mosquito populations in this model are assumed to be homogeneous and well-mixed, and we assume that only one dengue serotype is circulating. The model comprises human and vector (mosquito) populations, with births and seasonally forced deaths for the adult mosquito population. As the ratio between male and female mosquitoes is approximately 1.02:1~\cite{Arrivillaga2004}, the numbers of male and female mosquitoes are assumed to be equal. As only female mosquitoes feed on blood from humans as part of their reproductive cycle, only female mosquitoes can transmit dengue.  Consequently, this model only considers female mosquitoes. Vertical transmission of dengue in mosquitoes can occur (see, for example, ~\cite{Gunter2007, Lambrechts2010}), however this tends to be at low levels~\cite{Lambrechts2010}, and there is no evidence this level changes with the introduction of \textit{Wolbachia}, and so we ignore vertical transmission.
A schematic representation of the model is given in Figure \ref{fig1}.  Here  the subscript $H$ denotes the human population and the subscript $N$ denotes the non-\textit{Wolbachia} mosquito population.

The human population is divided into four subpopulations, namely Susceptible ($S_H$), Exposed ($E_H$), Infectious ($I_H$) and Recovered ($R_H$). Furthermore, a constant human population size is assumed, and hence human birth and death rates are assumed to be equal, that is, $B =\mu_H$ and $N_H=S_H+E_H+I_H+R_H$.

The mosquito population is divided into subpopulations of Aquatic ($A_N$) comprised of eggs, larvae and pupae, Susceptible ($S_N$), Exposed ($E_N$) and Infectious ($I_N$) mosquitoes. The total adult female mosquito population is $F_N=S_N+E_N+I_N$. The subscript $N$ is used to denote non-\textit{Wolbachia} mosquitoes.  We use this subscript here for consistency and to differentiate from the \textit{Wolbachia}-carrying mosquitoes included in the models in later sections. We group eggs, larvae and pupae into one compartment as they are not involved in the transmission of dengue. Without loss of generality, they can be represented by a single death rate and maturation rate for the purpose of modelling. No recovered class is required for mosquitoes as they remain infected for the rest of their lives. 

\clearpage
The model is then governed by the following system of differential equations,
\begin{align}
\frac{d S_H}{dt}&=B N_H - \frac{b_N T_{N} I_N}{N_H}S_H-\mu_H S_H,\label{eq1}\\[5mm]
\frac{d E_H}{dt}&=\frac{b_N T_{N} I_N}{N_H}S_H-\gamma_H E_H -\mu_H E_H,\label{eq2}\\[5mm]
\frac{d I_H}{dt}&=\gamma_H E_H-\sigma I_H -\mu_H I_H,\label{eq3}\\[5mm]
\frac{d R_H}{dt}&=\sigma I_H -\mu_H R_H,\label{eq4}\\[5mm]
\frac{d A_N}{dt}&=\rho_N \frac{F_N}{2} \left(1-\frac{A_N}{K}\right)-\left(\tau_N+\mu_{NA}\right) A_N,\label{eq5}\\[5mm]
\frac{d S_N}{dt}&=\tau_N \frac{A_N}{2}-\left(\frac{b_N T_{N}I_H}{N_H}+\mu_N(t)\right) S_N,\label{eq6}\\[5mm]
\frac{d E_N}{dt}&=\left(\frac{b_N T_{N}I_H}{N_H}\right) S_N-\left(\gamma_N+\mu_N(t)\right) E_N,\label{eq7}\\[5mm]
\frac{d I_N}{dt}&=\gamma_N E_N-\mu_N(t) I_N. \label{eq8}
\end{align}
The probability of transmission from a non-\textit{Wolbachia} mosquito to a human is assumed to be the same as the reverse, hence we let $T_{HN}=T_{NH}=T_N$. The variation in the adult mosquito death rate is strongly influenced by environmental factors such as temperature, humidity, and rainfall \cite{Yang2009} and in our model is sinusoidally forced,
\begin{equation}
\mu_N(t)=\mu_{N0}\left(1-\eta \cos \left(\frac{2\pi(t+\omega)}{365}\right)\right) \,. 
\label{deathrateadult} 
\end{equation}
Here $\eta$ is the amplitude of seasonal forcing in the adult death rate, which will be called the \textit{strength of seasonality} throughout this thesis.  Further, $\mu_{N0}$ is the average adult death rate, $t$ is time and $\omega$ is the phase shift, which is used to align the cosine function with the seasonal factors in Far North Queensland. This forcing of the adult death rate results in appropriate seasonal fluctuations in the adult mosquito population, in turn leading to a seasonal aquatic population through the mating function. Hence, there is no need to force the reproductive rate, and we keep it constant.

A human becomes exposed (but not yet infectious) after being bitten by an infectious mosquito at rate $b_N T_{N} I_N /N_H$ (Equations \eqref{eq1} and \eqref{eq2}), where $b_N$ is the successful biting rate and $T_{N}$ is the transmission probability from non-\textit{Wolbachia} mosquitoes to humans and the reverse. Exposed humans then become infectious at rate $\gamma_H$ and recover from dengue at rate $\sigma$.

The aquatic population increases as the male and female mosquitoes mate and breed, but  the population is limited by carrying capacity $K$ through a logistic term
\begin{equation*}
 \rho_N \frac{F_N M_N}{M_N+F_N} \left(1-\frac{A_N}{K}\right) \,.
\end{equation*}
Since there are equal numbers of male and female mosquitoes, $M_N= F_N$, this becomes $\rho_N F_N (1-A_N/K)/2$ (Equation \eqref{eq5}). Members of the aquatic population die at rate $\mu_{NA}$ and mature into susceptible female mosquitoes at rate $\tau_N$, where only half of the maturing aquatics are female.
Susceptible mosquitoes  progress to the exposed class after biting infectious humans at rate $b_N T_{N} I_H/N_H$. They then become infectious at rate $\gamma_N$ (Equation \eqref{eq8}), where $1/\gamma_N$ is the extrinsic incubation period.

The populations of both humans and mosquitoes are non-dimensionalised by letting $\hat{S}_H= S_H/N_H$, $\hat{E}_H= E_H/N_H$, $\hat{I}_H= I_H/N_H$, $\hat{R}_H= R_H/N_H$ and $\hat{A}_N= A_N/K$, $\hat{S}_N= S_N/K$, $\hat{E}_N= E_N/K$, and $\hat{I}_N= I_N/K$. Since the carrying capacity, $K$, is related to the number of available breeding sites, which depends on the number of humans, $K\propto N_H$.  Hence  $K=L N_H$, where $L$ is the ratio of the carrying capacity to the total human population.  The model, after removing the hats for the sake of simplicity, is then reduced to

\begin{align}
\frac{d S_H}{dt}&=B-b_N T_{N} L I_NS_H-\mu_H S_H,\label{nondimeq1}\\[5mm]
\frac{d E_H}{dt}&=b_N T_{N} L I_N S_H-\gamma_H E_H -\mu_H E_H,\label{nondimeq2}\\[5mm]
\frac{d I_H}{dt}&=\gamma_H E_H-\sigma I_H -\mu_H I_H,\label{nondimeq3}\\[5mm]
\frac{d R_H}{dt}&=\sigma I_H -\mu_H R_H,\label{nondimeq4}\\[5mm]
\frac{d A_N}{dt}&=\rho_N \frac{F_N}{2} \left(1-A_N\right)-\left(\tau_N+\mu_{NA}\right) A_N,\label{nondimeq5}\\[5mm]
\frac{d S_N}{dt}&=\tau_N \frac{A_N}{2}-\left(b_N T_{N}I_H+\mu_N(t)\right) S_N,\label{nondimeq6}\\[5mm]
\frac{d E_N}{dt}&=\left(b_N T_{N} I_H\right) S_N-\left(\gamma_N+\mu_N(t)\right) E_N,\label{nondimeq7}\\[5mm]
\frac{d I_N}{dt}&=\gamma_N E_N-\mu_N(t) I_N, \label{nondimeq8}
\end{align}
where $S_H+E_H+I_H+R_H=1$, but the mosquito population does not necessarily sum to one due to the relationship with the carrying capacity and the sinusoidal forcing term.

\section{Model in the Presence of \textit{Wolbachia}}\label{sec:modelpresence}

In this section the formulation of the model in the presence of \textit{Wolbachia} is presented. This extends the model in the absence of \textit{Wolbachia}-carrying mosquitoes given in Equations~\eqref{eq1}--\eqref{eq8} by including a \textit{Wolbachia}-carrying mosquito population.

\begin{figure}[h!]
 \begin{center}
 \includegraphics[width=0.8\linewidth]{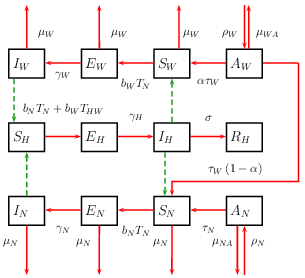}
 \end{center}
 \caption[Flowchart of the single serotype dengue model in the presence of \textit{Wolbachia}-carrying mosquitoes]{The flowchart of the model in the presence of \textit{Wolbachia}-carrying mosquitoes, Equations~\eqref{10}--\eqref{21}. Solid lines are population progression lines and dashed lines are disease transmission lines. The subscript $H$ is for the human population, $N$ is for the non-\textit{Wolbachia} mosquito population, and $W$ is for the \textit{Wolbachia}-carrying mosquito population. The compartments are `S' for susceptible, `E' for exposed to dengue but not yet infectious, `I' for infectious, `R' for recovered, and `A' for the aquatic phase of the mosquito life cycle. The transition rates between compartments are shown by the progression lines and described in the text.}
 \label{fig2}
 \end{figure}

The population of \textit{Wolbachia}-carrying mosquitoes is divided into subpopulations of Aquatic $(A_W)$ comprised of eggs, larvae and pupae, Susceptible ($S_W$),  Exposed ($E_W$) and Infectious ($I_W$) mosquitoes, where $S_W+E_W+I_W=F_W$. The subscript $W$ is used to denote \textit{Wolbachia} and to differentiate between non-\textit{Wolbachia} and \textit{Wolbachia}-carrying mosquitoes. The model comprises twelve compartments in total, four  for each of the three  modelled populations: two mosquito populations and one human population.
 A schematic representation of the model is given in Figure \ref{fig2}.

\clearpage
The model is governed by the following system of differential equations
\begin{align}
\frac{dS_H}{dt}&=B N_H -\frac{bT_{N} I_N}{N_H} S_H- \frac{b_W T_{HW}I_W}{N_H} S_H-\mu_H S_H,\label{10}\\[5mm]
\frac{dE_H}{dt}&=\frac{bT_{N} I_N}{N_H} S_H+ \frac{b_W T_{HW}I_W}{N_H} S_H -\gamma_H E_H -\mu_H E_H,\label{11}\\[5mm]
\frac{dI_H}{dt}&=\gamma_HE_H-\sigma I_H -\mu_H I_H,\label{12}\\[5mm]
\frac{dR_H}{dt}&=\sigma I_H -\mu_H R_H,\\[5mm]
\frac{dA_N}{dt}&=\rho_N\frac{F_N^2}{2(F_N+F_W)}\left(1-\frac{(A_N+A_W)}{K}\right)-\left(\tau_N+\mu_{NA}\right)A_N,\label{14}\\[5mm]
\frac{dS_N}{dt}&=\tau_N\frac{A_N}{2}+\left(1-\alpha\right)\tau_W\frac{A_W}{2}-\left(\frac{b_NT_{N}I_H}{N_H}+\mu_N(t)\right)S_N,\label{15}\\[5mm]
\frac{dE_N}{dt}&=\frac{b_NT_{N}I_H}{N_H} S_N-\left(\gamma_N+\mu_N(t)\right)E_N,\label{16}\\[5mm]
\frac{dI_N}{dt}&=\gamma_NE_N-\mu_N(t)I_N,\label{17}\\[5mm]
\frac{dA_W}{dt}&=\rho_W\frac{F_W}{2}\left(1-\frac{(A_N+A_W)}{K}\right)-\left(\tau_W+\mu_{WA}\right)A_W,\label{18}\\[5mm]
\frac{dS_W}{dt}&=\tau_W\alpha\frac{A_W}{2}-\left(\frac{b_WT_{N}I_H}{N_H}+\mu_W(t)\right)S_W,\label{19}\\[5mm]
\frac{dE_W}{dt}&=\frac{b_WT_{N}I_H}{N_H}S_W-\left(\gamma_W+\mu_W(t)\right)E_W,\label{20}\\[5mm]
\frac{dI_W}{dt}&=\gamma_WE_W-\mu_W(t)I_W . \label{21}
\end{align}
In this model, the exposure rates for humans   are  different  from those in the model in the absence of \textit{Wolbachia}, as seen in Equations \eqref{eq1}. In this model, a susceptible human becomes exposed after being bitten by either non-\textit{Wolbachia} or \textit{Wolbachia}-carrying infectious mosquitoes at rate $b_NT_NI_N/N_H$ or $b_WT_{HW}I_W/N_H$, respectively (see Equations \eqref{10} and \eqref{11}).   Here $b_W$ is the  biting rate for \textit{Wolbachia}-carrying mosquitoes and $T_{HW}$ is the transmission probability from \textit{Wolbachia}-carrying mosquitoes to humans. Note that the transmission probability from humans to \textit{Wolbachia}-carrying mosquitoes is assumed to be equal to that of humans to non-\textit{Wolbachia} mosquitoes, so $T_{WH}=T_N$. By contrast, there are differences in the transmission probabilities  of dengue from mosquitoes to humans for \textit{Wolbachia} and non-\textit{Wolbachia} mosquitoes (see Section \ref{WolbachiaParams} for an explanation).

The effects of cytoplasmic incompatibility and imperfect maternal transmission on the mosquito populations are included in this model. The effect of CI is incorporated by differences in the mating  functions. That is, the non-\textit{Wolbachia} females can only reproduce when mating with \textit{Wolbachia} mosquito males, giving
\begin{equation}
\frac{\rho_N F_N M_N}{P}\,,
\end{equation}
where $P=F_N+M_N+F_W+M_W$. As the ratio of male to female mosquitoes is assumed to be 1:1, this is reduced to  $\rho_NF_N^2/(2(F_N+F_W))$ (see Equation \eqref{14}). The aquatic  \textit{Wolbachia} mosquitoes are produced when \textit{Wolbachia}-carrying female mosquitoes mate with either non-\textit{Wolbachia} or \textit{Wolbachia} males, giving the term

\begin{equation}
\frac{\rho_WF_W\left(M_N+M_W\right)}{P}\,,
\end{equation}
where $P=F_N+M_N+F_W+M_W$, which simplifies to $\rho_WF_W/2$ (Equation \eqref{18}). The growth of aquatic mosquitoes is limited by carrying capacity $K$, so that each mating function is multiplied by
\begin{equation}
\frac{A_N + A_W}{K}.
\end{equation}
 \textit{Wolbachia}-carrying aquatic mosquitoes mature to be \textit{Wolbachia}-carrying adult mosquitoes at rate $\tau_W$.  To capture the imperfect maternal transmission of \textit{Wolbachia}~\cite{Walker2011,Hoffmann1990}, we assume that a proportion $\alpha$ of them become \textit{Wolbachia}-carrying adults and a proportion ($1-\alpha$) become non-\textit{Wolbachia} adults (see Equations \eqref{15} and \eqref{19}). In a similar manner to that for  non-\textit{Wolbachia} mosquitoes, the death rate of \textit{Wolbachia}-carrying adult mosquitoes varies seasonally ($\mu_W=f \mu_N$. More explanation of about this relationship is given in Section~\ref{WolbachiaParams}).

The non-dimensionalised model form is a useful generalisation, with the independence of variable scale allowing the solution to be applicable in other settings, such as for a different population size. Therefore, as in the model in the absence of \textit{Wolbachia}-carrying mosquitoes, the populations of both human and mosquitoes are non-dimensionalised by dividing the human subpopulations by $N_H$ and the mosquito subpopulations by $K$. The model is then reduced to
\begin{align}
\frac{dS_H}{dt}&=B-b_NT_{N}LI_N S_H- b_WT_{HW}LI_W S_H-\mu_H S_H,\label{nondim10}\\[5mm]
\frac{dE_H}{dt}&=b_NT_{N}LI_NS_H+b_WT_{HW}LI_WS_H-\gamma_H E_H -\mu_H E_H,\label{nondim11}\\[5mm]
\frac{dI_H}{dt}&=\gamma_HE_H-\sigma I_H -\mu_H I_H,\label{nondim12}\\[5mm]
\frac{dR_H}{dt}&=\sigma I_H -\mu_H R_H,\\[5mm]
\frac{dA_N}{dt}&=\rho_N\frac{F_N^2}{2(F_N+F_W)}(1-(A_N+A_W))-\left(\tau_N+\mu_{NA}\right)A_N,\label{nondim14}\\[5mm]
\frac{dS_N}{dt}&=\tau_N\frac{A_N}{2}+\left(1-\alpha\right)\tau_W\frac{A_W}{2}-\left(b_NT_{N}I_H+\mu_N(t)\right)S_N,\label{nondim15}\\[5mm]
\frac{dE_N}{dt}&=b_NT_{N}I_H S_N-\left(\gamma_N+\mu_N(t)\right)E_N,\label{nondim16}\\[5mm]
\frac{dI_N}{dt}&=\gamma_NE_N-\mu_N(t)I_N,\label{nondim17}\\[5mm]
\frac{dA_W}{dt}&=\rho_W\frac{F_W}{2}\left(1-(A_N+A_W)\right)-\left(\tau_W+\mu_{WA}\right)A_W,\label{nondim18}\\[5mm]
\frac{dS_W}{dt}&=\tau_W\alpha\frac{A_W}{2}-\left(b_WT_{N}I_H+\mu_W(t)\right)S_W,\label{nondim19}\\[5mm]
\frac{dE_W}{dt}&=b_WT_{N}I_HS_W-\left(\gamma_W+\mu_W(t)\right)E_W,\label{nondim20}\\[5mm]
\frac{dI_W}{dt}&=\gamma_WE_W-\mu_W(t)I_W . \label{nondim21}
\end{align}
The parameter descriptions and values can be found in Table~\ref{table2}, Page~\pageref{table2}.

\section{Basic and Effective Reproduction Number}\label{sec:R0}

The basic reproduction number $(\mathcal{R}_0)$ is one of the most important quantities in infectious disease modelling, acting as an epidemic threshold. If $\mathcal{R}_0 >1$, an outbreak may occur, whereas it cannot if $\mathcal{R}_0 <1$. We determine $\mathcal{R}_0$ for our model in order to investigate the effect of introducing \textit{Wolbachia}-carrying mosquitoes on the epidemic threshold.
\indent

Following Diekmann, Heesterbeek and Roberts \cite{Diekmann2010}, to construct a next generation matrix, we need to first identify the subsystem of ODEs that describes the production of new cases and the changes between infectious classes or groups, which is called the infected subsystem. From the model~\eqref{10}--\eqref{21}, the infected subsystem is

\begin{align}
\frac{dE_H}{dt}&=\frac{b_NT_{N}I_N}{N_H}S_H+ \frac{b_WT_{HW}I_W}{N_H}S_H-\gamma_H E_H -\mu_H E_H,\label{infecsubs1}\\[5mm]
\frac{dI_H}{dt}&=\gamma_HE_H-\sigma I_H -\mu_H I_H,\label{infecsubs2}\\[5mm]
\frac{dE_N}{dt}&=\frac{b_NT_{N}I_H}{N_H} S_N-\left(\gamma_N+\mu_N(t)\right)E_N,\label{infecsubs3}\\[5mm]
\frac{dI_N}{dt}&=\gamma_NE_N-\mu_N(t)I_N,\label{infecsubs4}\\[5mm]
\frac{dE_W}{dt}&=\frac{b_WT_{N}I_H}{N_H}S_W-\left(\gamma_W+\mu_W(t)\right)E_W,\label{infecsubs5}\\[5mm]
\frac{dI_W}{dt}&=\gamma_WE_W-\mu_W(t)I_W . \label{infecsubs6}
\end{align}
At the infection-free steady state, $E_H=I_H=E_N=I_N=E_W=I_W=0$, and $S_H=N_H$. For small $\left(E_H, I_H, E_N, I_N, E_W, I_W\right)$, the linearised infected subsystem is approximated by Equations~\eqref{infecsubs1}--\eqref{infecsubs6}, with $S_H=N_H$.

Let $\mathbf{x}=\left(E_H, I_H, E_N, I_N, E_W, I_W\right)'$.   We want to write the linearised infected subsystem in the form
\begin{equation*}
\mathbf{\frac{dx}{dt}} =\left(\mathbf{T} + \pmb{\Sigma}\right)\mathbf{x},
\end{equation*}
where  $\mathbf{T}$ is the transmission matrix, whose entries correspond to transmission events, and $\pmb{\Sigma}$ is the transition matrix, whose entries correspond to movement between the infected compartments including deaths. 
The transmission matrix $\mathbf{T}$ is
$$\mathbf{T} =
 \begin{pmatrix}
  0&0&0&b_{N}T_{N}&0&b_{W}T_{HW}\\
  0&0&0&0 &0&0 \\
  0&\frac{b_{N}T_{N}}{N_H}S_N(t)&0&0&0&0\\
  0&0&0&0&0&0\\
  0&\frac{b_{W}T_{N}}{N_H}S_W(t)&0&0&0&0\\
  0&0&0&0&0&0
 \end{pmatrix}\,,$$
where $S_N(t)$ and $S_W(t)$ are the population of mosquitoes with and without \textit{Wolbachia}.  As mosquito populations vary seasonally, both of these are functions of time. The transition matrix $\pmb{\Sigma}$ is
 
 $$\pmb{\Sigma} =
  \begin{pmatrix}
   -\left(\gamma_H+\mu_H\right)&0&0&0&0&0 \\
   \gamma_H&-\left(\sigma+\mu_H\right)&0&0&0&0 \\
   0&0&-\left(\gamma_N+\mu_N(t)\right)&0&0&0 \\
   0&0&\gamma_N&-\mu_N(t)&0&0\\
   0&0&0&0&-\left(\gamma_W+\mu_W(t)\right)&0\\
   0&0&0&0&\gamma_W&-\mu_W(t)
  \end{pmatrix}.$$
  
We next find the inverse of the transition matrix $\pmb{\Sigma}^{-1}$:
$$\pmb{\Sigma}^{-1} =
  \begin{pmatrix}
   -\frac{1}{\left(\gamma_H+\mu_H\right)}& 0& 0& 0& 0& 0 \\
   -\frac{\gamma_H}{\left(\gamma_H+\mu_H\right)\left(\sigma+\mu_H\right)}&-\frac{1}{\left(\sigma+\mu_H\right)}& 0&0&0&0 \\
   0&0&-\frac{1}{\left(\gamma_N+\mu_N(t)\right)}&0&0&0 \\
   0&0&-\frac{\gamma_N}{\mu_N(t)\left(\gamma_N+\mu_N(t)\right)}&-\frac{1}{\mu_N(t)}&0&0\\
   0&0&0&0&-\frac{1}{\left(\gamma_W+\mu_W(t)\right)}&0\\
   0& 0& 0& 0&-\frac{\gamma_W}{\mu_W(t)\left(\gamma_W+\mu_W(t)\right)}&-\frac{1}{\mu_W(t)}
  \end{pmatrix}. $$
  

We multiply the transmission matrix and the inverse of transition matrix to obtain
$$-\mathbf{T}\pmb{\Sigma}^{-1} =
 \begin{pmatrix}
  0&0&\frac{b_{N}T_{N}\gamma_N}{\mu_N\left(\gamma_N+\mu_N\right)}&\frac{b_{N}T_{N}}{\mu_N}& \frac{b_{W}T_{HW}\gamma_W}{\mu_W\left(\gamma_W+\mu_W\right)}& \frac{b_{W}T_{HW}}{\mu_W}\\
  0&0&0&0 &0&0 \\
  0&\frac{b_{N}T_{N} \gamma_H S_N(t)}{\left(\sigma+\mu_H\right)\left(\gamma_H + \mu_H\right)N_H}&\frac{b_{N}T_{N}S_N(t)}{\left(\sigma+\mu_H\right)N_H}&0&0&0\\
  0&0&0&0&0&0\\
  0&\frac{b_{W}T_{N} \gamma_H S_W(t)}{\left(\sigma+\mu_H\right)\left(\gamma_H + \mu_H\right)N_H}&\frac{b_{W}T_{N}S_W(t)}{\left(\sigma+\mu_H\right)N_H}&0&0&0\\
  0&0&0&0&0&0
 \end{pmatrix}.$$
 
The spectral radius of the matrix $-\mathbf{T}\pmb{\Sigma}^{-1}$ is the basic reproduction number, that is,

\begin{equation}
\resizebox{0.9\hsize}{!}{$\mathcal{R}_P=\sqrt{\frac{b_N^2  T_N^2 \gamma_N \gamma_H S_N(t)}{\mu_N(t)\left(\gamma_N + \mu_N(t)\right)\left(\sigma +\mu_H\right)\left(\gamma_H + \mu_H\right)N_H} +\frac{b_W^2  T_{HW} \gamma_W T_N \gamma_H S_W(t)}{\mu_W(t)\left(\gamma_W + \mu_W(t) \right)\left(\sigma + \mu_H\right)
\left(\gamma_H + \mu_H\right)  N_H}}$.}
\label{eq:R0}
\end{equation}

Equation~\eqref{eq:R0} gives the basic reproduction number in the presence of \textit{Wolbachia}. In the absence of \textit{Wolbachia}, after setting all \textit{Wolbachia}-related parameters to zero, Equation~\ref{eq:R0} is reduced to
\begin{equation}
\mathcal{R}_A=\sqrt{\frac{b_N^2  T_N^2 \gamma_N \gamma_H S_N(t)}{\mu_N(t)\left(\gamma_N + \mu_N(t)\right)\left(\sigma +\mu_H\right)\left(\gamma_H + \mu_H\right)N_H}}.
\label{eq:R0Abs}
\end{equation}

The disease-free equilibrium results in all humans becoming susceptible ($S_H= N_H$).  However, the equilibrium mosquito population includes both susceptible adults ($S_N$ and $S_W$) and mosquitoes in the aquatic phase ($A_N$, $A_W$).  Thus we explicitly include $S_N(t)$ and $S_W(t)$ in the formula for $\mathcal{R}_P$ and $\mathcal{R}_A$.

The basic reproduction number found here can be considered to be an instantaneous basic reproduction number which depends on fluctuations in  the mosquito population, $S_N(t)$ and $S_W(t)$.

Besides determining the basic reproduction number, it is also important to determine time-dependent variations in the potential for transmission of infectious disease. This is usually conducted by evaluating the effective reproduction number. The effective reproduction number is the average number of secondary cases generated by a primary case~\cite{Nishiura2009, Yang2014}. The effective reproduction number is different to the basic reproduction number in that it takes into account the depletion of susceptible individuals and any public health intervention~\cite{Glass2011}. To make it easier for readers to differentiate between basic and effective reproduction numbers in this thesis we will refer to the reproduction number as the basic reproduction number if the population is entirely susceptible, even where there is a \textit{Wolbachia} intervention.

The effective reproduction number can be obtained by multiplying the basic reproduction number by the proportion of the population that is susceptible populations~\cite{Nishiura2009}. The first term of Equation~\eqref{eq:R0} is the average number of secondary infectious humans  produced by one primary human case in a completely susceptible population  of humans and non-\textit{Wolbachia} mosquitoes. The second term of Equation~~\eqref{eq:R0} is the same as the first term, but where mosquitoes carry \textit{Wolbachia}. The effective reproduction number can be obtained as the sum of the first term multiplied by $(S_H(t)/N_H)(S_N(t)/F_N(t))$ and the second term multiplied by $(S_H(t)/N_H)(S_W(t)/F_W(t))$. That is, 

\begin{equation}
\mathcal{R}_{P\mbox{eff}}=\sqrt{\mathcal{R}_{1\mbox{eff}}+\mathcal{R}_{2\mbox{eff}}}\,,
\label{ReffPres}
\end{equation}
where
\begin{equation}
\mathcal{R}_{1\mbox{eff}}=\frac{b_N^2  T_N^2 \gamma_N \gamma_H S_N(t)}{\mu_N(t)\left(\gamma_N + \mu_N(t)\right)\left(\sigma +\mu_H\right)\left(\gamma_H + \mu_H\right)N_H} \frac{S_H(t)}{N_H} \frac{S_N(t)}{F_N(t)}\,,
\end{equation}
and

\begin{equation}
\mathcal{R}_{2\mbox{eff}}=\frac{b_W^2  T_{HW} \gamma_W T_N \gamma_H S_W(t)}{\mu_W(t)\left(\gamma_W + \mu_W(t) \right)\left(\sigma + \mu_H\right)
\left(\gamma_H + \mu_H\right)N_H} \frac{S_H(t)}{N_H} \frac{S_W(t)}{F_W(t)}.
\end{equation}

%
%

In the absence of \textit{Wolbachia}-carrying mosquitoes, the effective reproduction number is

\begin{equation}
\mathcal{R}_{A\mbox{eff}}=\sqrt{\mathcal{R}_{1\mbox{eff}}}\,.
\label{ReffAbs}
\end{equation}

\section{Discussion}
Mathematical models developed in this chapter are used to investigate the effectiveness of a \textit{Wolbachia} control strategy in reducing the number of human dengue cases under scenarios of single and multiple introductions of human dengue cases. In these
models, we assume a well-mixed and homogeneous population and ignore heterogeneity in the population, although our models could be extended to include this. Furthermore, although we assume the transmission probabilities from humans to mosquitoes and mosquitoes to human are the same, the force of infections of humans and mosquitoes are different due to differences in the proportions of infectious humans and mosquitoes (Equations \eqref{eq1} and \eqref{eq6} , respectively). Vertical transmission of dengue in mosquitoes can occur (see, for example, ~\cite{Gunter2007, Lambrechts2010}), however this tends to be at low levels~\cite{Lambrechts2010}. Furthermore, Adams and Boots~\cite{Boots2010} found that vertical transmission is not an important factor for determining dengue persistence.
As the purpose of this thesis is to determine the indicative reduction in the number of dengue cases due to a \textit{Wolbachia} intervention, the assumptions used in this thesis are sufficient. Further explorations can be conducted and suggestions are detailed in Section~\ref{FutureDirection}.

\section{Summary}
\begin{itemize}[wide=0pt,labelwidth = 1.3333em, labelsep = 0.3333em, leftmargin = \dimexpr\labelwidth + \labelsep\relax ]
\item Single serotype dengue models in the absence and presence of \textit{Wolbachia}-carrying mosquitoes are developed (Sections~\ref{sec:modelabsence} and \ref{sec:modelpresence}).
\item We provide dimensionalised and non-dimensionalised models (Sections~\ref{sec:modelabsence} and \ref{sec:modelpresence}).
\item The death rate of mosquitoes is sinusoidally forced (Section~\ref{sec:modelabsence}, Equation~\eqref{deathrateadult}).
\item The basic and effective reproduction number is derived, which can be considered to be an instantaneous basic reproduction number (Section~\ref{sec:R0}, Equations~\eqref{eq:R0}--\eqref{ReffAbs}).
\item The basic and effective reproduction number depends on fluctuations in the mosquito population (Section~\ref{sec:R0}, Equations~\eqref{eq:R0}--\eqref{ReffAbs}).
\end{itemize}

\chapter[Dengue Transmission Dynamics for a Single Introduction Event]{Dengue Transmission Dynamics for a \\Single Introduction Event \footnote{This work is part of the manuscript that has been published as Meksianis Z. Ndii, R. I. Hickson, David Allingham, G. N. Mercer. Modelling the transmission dynamics of dengue in the presence of \textit{Wolbachia}. Mathematical Biosciences. 262:157--166. 2015.}}\label{chap:singleintro} 


\ifpdf
    \graphicspath{{4/figures/PNG/}{4/figures/PDF/}{4/figures/}}
\else
    \graphicspath{{4/figures/EPS/}{4/figures/}}
\fi

 The aim of this chapter is to study the effects of \textit{Wolbachia} on dengue transmission dynamics in the case where there is a single introduction of dengue cases into the human population.

The mathematical models developed in Chapter~\ref{chap:singlemathmodel} are used to investigate the effects of \textit{Wolbachia} on dengue transmission dynamics for a single strain. We consider a single outbreak with a duration of approximately one year, and hence omit human births and deaths from our models  by setting $B=\mu_H=0$. We work with our non-dimensionalised models in the absence (Equations \eqref{nondimeq1}--\eqref{nondimeq8}) and presence (Equations \eqref{nondim10}--\eqref{nondim21}) of \textit{Wolbachia}-carrying mosquitoes.

\section{Data and Parameter Estimation}\label{ParameterEstimation}
Cairns is the largest city in the region of Australia where \textit{Aedes aegypti} are present and local dengue transmission occurs, and is also where the \textit{Wolbachia} field trials, which began in 2011, are being conducted~\cite{Hofman2011}.  In summer 2008/2009, there was a DEN3 outbreak in Cairns~\cite{Ritchie2013} and we use data from this outbreak to estimate the parameter values of transmission probability ($T_{N}$), strength of seasonality ($\eta$) and seasonal phase ($\omega$) for the baseline model, for example, in the absence of \textit{Wolbachia}, represented by Equations \eqref{nondimeq1}--\eqref{nondimeq8}. The other parameter values were obtained from the literature and are given in Table \ref{table2}. The Cairns data covers the period from 2nd November 2008 to 31st May 2009 and was extracted from Figure 2 of the paper by Ritchie \textit{et al.}~\cite{Ritchie2013}. As our model is formulated as a proportion of the population, each data point is divided by 150,000, which was the approximate population of Cairns in 2008~\cite{CairnsPop2009}.

We minimise the sum of the squared error between the model and data, which is given by
\begin{equation}
RSS= \sum\limits_{i=1}^n \left(y_i-f_i(x)\right)^2,
\label{RSS}
\end{equation}
where $y_i$ is the total proportion of human dengue cases up to the $i^{th}$ week from the observed data, and $f_i(x)$ is the total proportion of human dengue cases up to the $i^{th}$ week from the model simulations. The {\tt lsqnonlin} built-in function in \textsc{Matlab}~\cite{Matlab} is then used to estimate the $T_N$, $\eta$ and $\omega$ parameter values.

\section{\textit{Wolbachia} Parameters}\label{WolbachiaParams}
In this section, model parameters relating to \textit{Wolbachia}-carrying mosquitoes are presented and discussed. Most of these are given in terms of non-\textit{Wolbachia} mosquitoes, following the conventions in the literature. The relationships between \textit{Wolbachia} and non-\textit{Wolbachia} parameters, 

\begin{align}
 \rho_W &= c \rho_N\,,\label{relationrhoW}\\
  T_{HW} & = d T_N\,,\label{relationTHW}\\
 \mu_W & = f \mu_N\,,\label{relationmuW}\\
  b_W & = g b_N\,.\label{relationbW}
\end{align}


The reproductive rate of \textit{Wolbachia}-carrying mosquitoes is generally lower than that for non-\textit{Wolbachia}-carrying mosquitoes ($c<1$).  There is a significant difference in reproductive rates for the \textit{WMelPop} and a marginal difference for the \textit{WMel} strain.  This is because \textit{WMelPop} decreases the viability of eggs~\cite{McMeniman2010}, whereas \textit{WMel} does not have a significant effect on them~\cite{Walker2011}.

The death rate of \textit{Wolbachia}-carrying mosquitoes is higher than that of non-\textit{Wolbachia} mosquitoes ($f>1$) because \textit{Wolbachia} reduces the mosquito lifespan~\cite{Yeap2011, Walker2011}. \textit{WMel} and \textit{WMelPop} reduce the lifespan of mosquitoes by up to 10\% and 50\%, respectively~\cite{Yeap2011, Walker2011}.

\textit{Wolbachia} also inhibits viral replication and dissemination in the mosquitoes~\cite{Bian2010, Moreira2009, Hofman2011, Walker2011, Frentiu2014}. This results in a lower dengue viral load in the \textit{Wolbachia}-carrying mosquitoes, and to reflect this, we set the transmission probability from infectious \textit{Wolbachia}-carrying mosquitoes to humans to less than that for \textit{Wolbachia}-free mosquitoes ($d<1$). Additionally, \textit{Wolbachia} causes a condition known as bendy proboscis \cite{Turley2009} which inhibits feeding and lowers the successful biting rate ($g<1$). This lower biting rate also captures the effect that, due to the inhibition of viral replication, some \textit{Wolbachia}-carrying mosquitoes are effectively not infected with dengue, and so the overall transmission rate from humans is lower (that is, $b_W T_N<b_N T_N$).

\begin{figure}[ht]
  \begin{center}
  \includegraphics[width=0.95\linewidth]{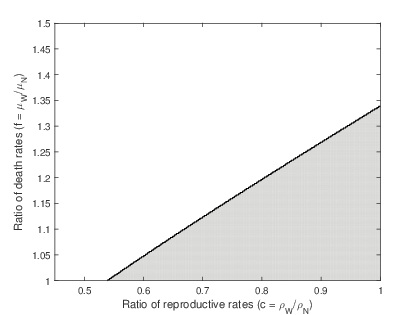}
  \end{center}
  \caption[A plot of the persistence of the \textit{Wolbachia}-carrying mosquito population]{A plot of the persistence, shown by the shaded region, of the \textit{Wolbachia}-carrying mosquito population, over a range of the \textit{Wolbachia} reproductive and death rates, expressed as ratios of the non-\textit{Wolbachia} rates. In the unshaded region only non-\textit{Wolbachia}-carrying mosquitoes persist.}
  \label{fig3}
\end{figure}

For the parameter values used in this paper, there are only two realistic stable states, which are that only non-\textit{Wolbachia}-carrying mosquitoes persist, and that both populations persist (Figure \ref{fig3}). An additional stable state does exist in which only the \textit{Wolbachia}-carrying mosquitoes persist, but this requires the perfect maternal transmission of \textit{Wolbachia}, and may not be realistic~\cite{Walker2011}. This state was also found by Ndii \textit{et al.}~\cite{Ndii2012} for an autonomous system. When both populations persist in this model, the proportion of \textit{Wolbachia}-carrying mosquitoes is around 86\%, which compares well with the 90\% observed in Hoffmann \textit{et al.}~\cite{Hofman2011, Hoffmann2014}.

\section{Measurement of the Effect of \textit{Wolbachia}  on Dengue Outbreaks}

A measure is needed to assess the impact of the \textit{Wolbachia} intervention on dengue transmission.  We do this by comparing the total numbers of human dengue cases in the absence and presence (for example, persistence) of \textit{Wolbachia}-carrying mosquitoes. The relative effect is expressed as a percentage, given by
\begin{equation}
\kappa=100\times\left(\frac{H_A-H_P}{H_A}\right)\,\% \,,
\label{RE}
\end{equation}
where $H$ is the final proportion of dengue cases, with subscripts to denote the absence (A) and presence (P) of \textit{Wolbachia}-carrying mosquitoes.

\section{Sensitivity Analysis} \label{methods:sensitivity}
A global sensitivity analysis is performed using the standard combination of Latin Hypercube Sampling (LHS) and Partial Rank Correlation Coefficient (PRCC) multivariate analysis~\cite{Marino2008, McKay1979, Wu2013, Blower1994}. LHS is a stratified Monte Carlo sampling method, in which the random parameter distributions are divided into $N$ equal probability intervals and samples are taken from each~\cite{Marino2008,McKay1979, Wu2013, Blower1994}.  Here $N$ is the sample size for each parameter.  Each interval of each parameter is sampled exactly once without replacement, so that the entire range of each parameter is explored~\cite{Marino2008, Wu2013, Blower1994}. Parameters are sampled from a triangular probability distribution because we expect that the values close to the peak of the triangular distribution pattern are those which are more likely to occur. The minimum, maximum and expected values are given in Table~\ref{table2}.

PRCC is an efficient method for measuring the nonlinear, but monotonic relationship between inputs and the model outcome of interest~\cite{Marino2008, Wu2013, Blower1994, Conover1981}. In this paper, the inputs are the parameters as well as the initial number of exposed humans, while the model outcome is the cumulative proportion of infectious individuals, which is the solution of the differential equation
\begin{equation*}
\frac{dC_{IH}}{dt}=\gamma_H E_H .
\end{equation*}
The ranges of the input parameters are available in the literature and only samples of the parameter values that result in the persistence of mosquitoes are included in the calculation. The PRCC is computed for the full length of 31 weeks, the period of time for which  data is available (see Richie \textit{et. al}\cite{Ritchie2013}).
The most significant parameters are those for which a small change in value leads to a significant change in the output, that is in the cumulative number of infectious humans.

\section{Results} \label{Results}

In this section the model simulation and sensitivity analysis results are presented for both models. The governing systems of differential equations are integrated using MATLAB's inbuilt routine ``ode45'', with the parameter values given in Table~\ref{table2}. We run the model until the mosquito population reaches the periodic stable state, before the infected humans are introduced into the population on the $2^\text{nd}$ of November. That is, the transient dynamics of the introduction of the \textit{Wolbachia} mosquitoes into the system are not considered. In Far North Queensland, dengue is not endemic, hence dengue outbreaks occur as dengue cases are introduced into the population. For both models, the initial conditions for the human population are $E_{H0}=2/(1.5\times10^5)$, $I_{H0}=0=R_{H0}$ and $S_{H0}=1-E_{H0}-I_{H0}-R_{H0}$.

\begin{table}
\centering
   \begin{tabular}{l p{5cm} l l l l l}
    \hline  \hline
    Symbol & Description & Min& Expected & Max & Unit & Source \\
    \hline
    $\alpha$& Maternal transmission& 0.85&0.9&1&N/A&\cite{Walker2011, Hoffmann1990, Ndii2012}\\
    $b_N$ &Biting rate&0&0.63&1& day$^{\textrm{-1}}$&\cite{Scott2000}\\
    $c=\rho_W/\rho_N$& Ratio of reproductive rate W \textit{c.f.} non-W& 0.7&0.95&1&N/A&\cite{Walker2011}\\      
    $d=T_{HW}/T_{N}$& Ratio of transmission probability W \textit{c.f.} non-W &0&0.5&1 &N/A&\cite{Bian2010}\\
    $E_{H0}$& Initial exposed humans & 1 & 2 & 5 &N/A&\\
    $\eta$& Strength of seasonality&0&0.6228&1&N/A&Fitted\\
   $f=\mu_W/\mu_N$& Ratio of death rate W \textit{c.f.} non-W& 1&1.1&1.25&N/A &\cite{Yeap2011, Walker2011}\\
   $g=b_W/b_N$ &Ratio of biting rates W \textit{c.f.} non-W&0&0.95&1& N/A&\cite{Turley2009}\\
    $\gamma_H$&Progression rate from exposed to infectious human&1/7&1/5.5& 1/4& day$^{\textrm{-1}}$&\cite{Gubler1998}\\
    $\gamma_{N}$& Progression from exposed to infectious non-W& 1/12&1/10&1/8& day$^{\textrm{-1}}$&\cite{Chowel2007}\\
    $\gamma_{W}$& Progression rate from exposed to infectious& 1/12&1/10&1/8 & day$^{\textrm{-1}}$&\cite{Chowel2007}\\
     $L=K/N_H$& Ratio of carrying capacity to total human population&& 3&&N/A&\cite{Chowel2007}\\
    $\mu_{N0}$& Average adult mosquito death rate (non-W)&1/30&1/14&1/10&day$^{\textrm{-1}}$&\cite{Yang2009}\\
    $\mu_{NA}$& Death rate of aquatic non-W&1/20&1/14&1/7&day$^{\textrm{-1}}$&\cite{Yang2009}\\
    $\mu_{WA}$& Aquatic death rate&1/20&1/14&1/7& day$^{\textrm{-1}}$&\cite{Yang2009}\\
    $\omega$&Phase&0&20.61&365&day&Fitted\\
   $\rho_{N}$& Reproductive rate of non-W& 1&1.25&2.5& day$^{\textrm{-1}}$&\cite{Ndii2012}\\ 
   $\sigma$&Recovery rate&1/14&1/5&1/3& day$^{\textrm{-1}}$&\cite{Gubler1998}\\ 
   $T_{N}$& Transmission probability& 0 &  0.2614 & 1 & N/A&Fitted\\
   $\tau_{N}$& Maturation rate of non-W& 1/17&1/10&1/6& day$^{\textrm{-1}}$&\cite{Yang2009}\\
   $\tau_{W}$& Maturation rate of W& 1/12&1/10&1/8& day$^{\textrm{-1}}$&\cite{Yang2009} \\  
  \hline
 \end{tabular}
 \caption[Parameter descriptions, values and sources for dengue models in the absence and presence of \textit{Wolbachia}-carrying mosquitoes with a single dengue serotype]{Parameter descriptions, values and sources for both models. The \textit{Wolbachia}-related parameters are for the \textit{WMel} strain. Here ``Non-W'' refers to non-\textit{Wolbachia} mosquitoes and ``W'' refers to \textit{Wolbachia}-carrying mosquitoes. Further explanation of the parameter values is given in the text.}
 \label{table2}
 \end{table}

\subsection{Parameter Estimation and Sensitivity of the Model in the Absence of \textit{Wolbachia}}\label{WlbAbsence}
In this section we consider the model in the absence of \textit{Wolbachia}, that is, Equations~\eqref{nondimeq1}--\eqref{nondimeq8}, as described in Chapter~\ref{chap:singlemathmodel}, which serves as a baseline model for comparison with the \textit{Wolbachia} intervention. Most of the parameters are obtained from the literature, as per the references in Table \ref{table2}. The remaining three parameters, $T_{N}$, $\eta$ and $\omega$, are estimated via optimisation  using \textsc{Matlab}'s {\tt lsqnonlin} function. We constrain the optimisation by physical limits, that is, between 0 and 1 for $T_N$ since it is a probability, 0 to 365 for the seasonal phase $\omega$, and 0 to 1 for $\eta$, so that the death rate $\mu_N$ remains positive at all times. The ranges of parameter values are chosen such a way that it is biologically realistic and result in the persistence of \textit{Wolbachia}-carrying mosquitoes.

\begin{figure}[h!]
\begin{center}
\includegraphics[width=0.95\linewidth]{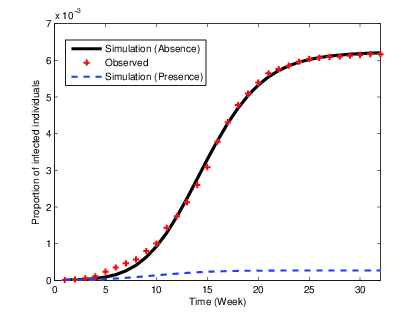}
\end{center}
\caption[Plots of observed data and the output of the fitted models]{Plots of observed data (red crosses)  and the output of the fitted models in the absence of \textit{Wolbachia}-carrying mosquitoes (black solid lines; Equations~\eqref{nondimeq1}--\eqref{nondimeq8}), and the model in the presence of \textit{Wolbachia} (Equations \eqref{nondim10}--\eqref{nondim21}). The data covers the time period $2^\text{nd}$ November 2008 to $31^\text{st}$ May 2009, extracted from Figure 2 of Ritchie \textit{et al.}~\cite{Ritchie2013}. The parameter values are given in Table \ref{table2}. The initial human subpopulations are $E_{H0}=2/(1.5\times10^5)$, $I_{H0}=R_{H0}=0$ and $S_{H0}=1-E_{H0}-I_{H0}-R_{H0}$, for both models. The initial mosquito subpopulations for the model in the absence of \textit{Wolbachia} are  $A_{N0}=0.8210$, $S_{N0}=1.2634$ and $E_{M0}=I_{M0}=0$, and for the model in the presence of \textit{Wolbachia} they are $A_{N0}=0.0138$, $S_{N0}=0.1326$, $A_{W0}=0.7535$, $S_{N0}=0.9400$ and $E_{M0}=I_{M0}=E_{W0}=I_{W0}=0$. The initial proportion of exposed individuals are introduced after the mosquito population reaches the stable periodic state.}
\label{fig4}
\end{figure}

We fit the parameters as follows. The model is run to its periodic stable state and then two exposed humans are introduced. The sum of squares error is calculated using the cumulative 31-week proportional weekly data of Ritchie \textit{et al.}~\cite{Ritchie2013}. The parameters are then estimated using \textsc{Matlab}'s {\tt lsqnonlin} function, with a final sum of squares error equal to $3\times10^{-7}$.   The resulting parameter values $T_{N}$, $\eta$ and $\omega$ are given in Table \ref{table2}. To assess the goodness-of-fit, we calculated the $\chi^2$ statistic for the model residuals using \textsc{Matlab}'s {\tt chi2gof} function.  This returned a p-value of 0.3713, and so we failed to reject the hypothesis that the residuals are from a normal distribution. There is some systematic bias in the residuals around the start of the outbreak, evident in Figure~\ref{fig4}, where the observed outbreak rises slightly faster than the model.  While model embellishments could be added to account for this, here we consider that the standard SEIR model of infectious disease is sufficient for our intended purpose: comparing the outbreak dynamics in the absence and presence of \textit{Wolbachia}.

\begin{figure}[h!]
 \begin{center}
 \includegraphics[width=0.95\linewidth]{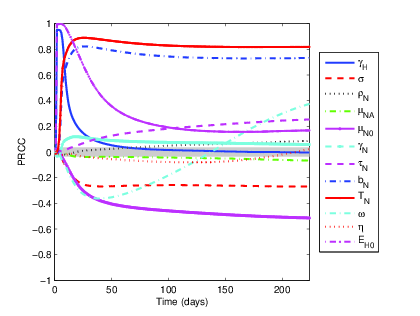}
 \end{center}
 \caption[Plot of the PRCC over time of the model in the absence of \textit{Wolbachia}]{Plot of the PRCC over time of the model in the absence of \textit{Wolbachia}. The PRCC is calculated with respect to the cumulative number of infectious individuals.  The grey area indicates the region where the PRCC is not significantly different from zero (significance level 0.01), using 5,200 samples.}
 \label{fig5}
 \end{figure}

The sensitivity analysis was performed by executing 5,200 runs (to reduce variance) to assess the model's sensitivity to the parameters. The parameter ranges used are given in Table~\ref{table2}. The range of the number of initially-exposed humans is taken to be between one and five individuals. This is realistic as only a small number of initial cases is needed to trigger an outbreak~\cite{Ritchie2013}.

The changes in parameter sensitivity exhibited over time are common for SEIR models due to the changes in disease dynamics over time, as seen in Figures \ref{fig5} and \ref{fig6}. Figure \ref{fig5} shows that, for most of the time period, the most influential parameters are the transmission probability ($T_N$), the biting rate ($b_N$) and the average adult death rate ($\mu_{N0}$), where the latter has a negative correlation. The phase, $\omega$, influences the outbreak size by shifting the peaks and troughs of the mosquito population in time; outbreaks tend to take off around the peaks.  Because of the sinusoidal nature of the seasonality, the correlation between $\omega$ and the outbreak size changes sign over time. The parameter $\gamma_H$ determines the progression rate of humans from the exposed to the infectious class. In the early days, when  exposed individuals are introduced, this parameter drives an increase in the number of infectious humans. If this parameter has a high value, the initial introduced cases will quickly move into the infectious class.

As the epidemic takes off, the cumulative number of infectious individuals is determined more by the biting rate $b_N$ and the transmission probability $T_N$, and so $\gamma_H$ declines in importance.  Moreover, the dynamics of the mosquito population have a larger influence on the disease dynamics once the epidemic has taken off.

When there are many susceptible mosquitoes in the population, there will be many infectious mosquitoes and hence many infected humans. An increase in the number of susceptible mosquitoes is regulated by the parameter $\tau_N$ and so this parameter has an important impact in the later period after the epidemic has taken off. As expected, the cumulative number of infectious individuals is most sensitive to the number of initially-exposed humans ($E_{H0}$) at early times since they are immediately added to this output. However, for the range considered, $E_{H0}$ does not change the proportion of dengue reduction due to the introduction of \textit{Wolbachia}-carrying mosquitoes.

\subsection{Sensitivity Analysis of the Model in the Presence of \textit{Wolbachia}}\label{DengueWolbahiaModel}

In this section the effects of \textit{Wolbachia}-carrying mosquitoes on the dengue dynamics are investigated. Variations in parameter values are explored to quantify the persistence of \textit{Wolbachia}-carrying mosquitoes and its effects on the number of human dengue cases.  The initial conditions for the two mosquito populations are $A_{N0}=0.0138$, $S_{N0}=0.1326$, $E_{N0}=I_{N0}=0$, $A_{W0}=0.7535$, $S_{W0}=0.9400$, and $E_{W0}=0=I_{W0}$, which are obtained by running \textsc{Matlab} simulations of the model in the absence of dengue to the periodic stable state.  Since the \textit{WMel} strain of \textit{Wolbachia} is used in the Cairns field experiments \cite{Hofman2011}, we use its expected parameter values in our model.


 \begin{figure}[ht]
  \begin{center}
  \includegraphics[width=0.95\linewidth]{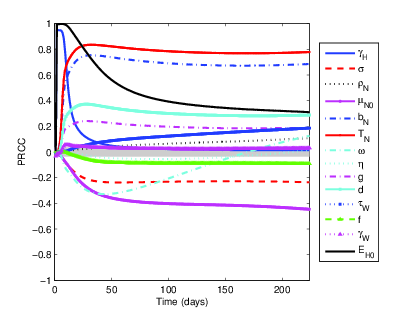}
  \end{center}
  \caption[Plot of PRCC over time for the model in the presence of \textit{Wolbachia}]{Plot of PRCC over time for the model in the presence of \textit{Wolbachia}. For clarity we show only those parameters that have a PRCC outside the range (-0.05,0.05) over the whole time window.  The grey area indicates the region in which the PRCC is not significantly different from zero (significance level 0.01), using 5,190 samples that result in the persistence of \textit{Wolbachia}-carrying mosquitoes.}
  \label{fig6}
  \end{figure}

The results of the sensitivity analysis are similar to those for the model in the absence of \textit{Wolbachia}, and are shown in Figure \ref{fig6}. The transmission probability ($T_N$), the biting rate ($b_N$) and the average adult mosquito death rate ($\mu_{N0}$) are the most influential parameters on the model outcome. Furthermore, at early times, the cumulative number of infectious individuals is sensitive to $E_{H0}$ and $\gamma_H$. An exploration of the ratios of transmission probability, $d=T_{HW}/T_N$, biting rate, $g=b_W/b_N$, and death rate, $f=\mu_W/\mu_N$, is also carried out as they relate to the most influential parameters. We also explore the ratio of  the reproductive rates for \textit{Wolbachia} and non-\textit{Wolbachia} mosquitoes, $c=\rho_W/\rho_N$, to obtain information regarding its effect on dengue disease dynamics and on the parameter range for which  \textit{Wolbachia}-carrying mosquitoes persist.

\subsection{Parameter Exploration} \label{parameterExploreation}

 \begin{figure}[ht]
  \begin{center}
  \includegraphics[width=0.95\linewidth]{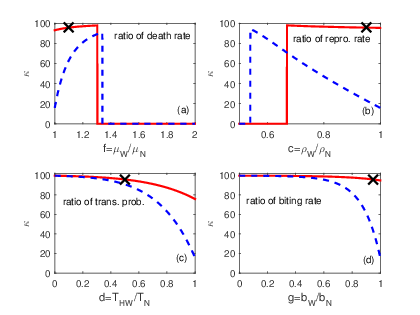}
  \end{center}
  \caption[Plot of proportional reduction in dengue due to \textit{Wolbachia} against ratios of death rates, reproductive rates, transmission probabilities and biting rates]{Plot of the relative effect, $\kappa$ from Equation~\eqref{RE}, on human dengue cases in the absence and presence of \textit{Wolbachia}-carrying mosquitoes against the ratios of (a) adult death rates ($f=\mu_W/\mu_N$), (b) reproductive rates ($c=\rho_W/\rho_N$), (c) transmission probabilities ($d=T_{HW}/T_N$) and (d) biting rates ($g=b_W/b_N$).  In each case the other ratios are set to the expected values for \textit{WMel} from Table~\ref{table2} (solid lines), or 1 (dashed lines).  Crosses mark the case where all ratios are set to their expected \textit{WMel} values.}
  \label{fig7}
  \end{figure}
  
 Although we are using values for $c$, $d$, $f$ and $g$ from the literature, the provided values are generally qualitative descriptions, or come from laboratory trials which may not be representative of what happens in the field, and hence the correct parameter values may differ. Therefore, in this Section, we vary these parameters, one at a time, to determine their effects on human dengue incidence as measured by the relative effect given in Equation~\eqref{RE}. Thus, we quantify the effect of introducing \textit{Wolbachia}-carrying mosquitoes on dengue transmission for a range of realistic parameter values.

We consider two scenarios, and their resulting effects on dengue cases are shown in Figure~\ref{fig7}. The first uses the expected values for \textit{WMel} for the non-varied \textit{Wolbachia} parameters (indicated by the solid line in the Figure). The second scenario uses the non-\textit{Wolbachia}  mosquito values, allowing us to explore the effect of cytoplasmic incompatibility (CI) on the dengue transmission dynamics, assuming that \textit{Wolbachia} has no other effect on  mosquito physiology. This second scenario is represented by the dashed line in the figure.

The vertical lines in Figure~\ref{fig7}(a) and~(b), for $f$ and $c$, respectively, denote the boundary between persistence ($\kappa>0$) and non-persistence ($\kappa=0$) for the \textit{Wolbachia}-carrying mosquitoes.  Both scenarios show that as the death rate of \textit{Wolbachia}-carrying mosquitoes increases, the relative effect on dengue incidence also increases up to the point where the \textit{Wolbachia}-carrying mosquitoes no longer persist, around $f=1.30$ for the solid line and $f=1.34$ for the dashed line. The default values, with $f=1.1$, are identified on the figure by a cross, showing that \textit{WMel} should reduce human dengue incidence by approximately 96\%.
Interestingly, in the second scenario, when $c=d=f=g=1$, we see that even if all of the \textit{Wolbachia} parameters are equal to their non-\textit{Wolbachia} counterparts,  human dengue incidence is still reduced by approximately 16\%. This occurs due to the combination of CI and competition, with the total female mosquito population ($F_N+F_W$) being reduced. 

The CI means  \textit{Wolbachia}-free and \textit{Wolbachia}-carrying mosquitoes have different mating functions, and, in particular, that no non-\textit{Wolbachia} offspring are produced from the combination of a non-\textit{Wolbachia} female and a \textit{Wolbachia}-carrying male (there is effectively a $0\times F_N F_W$ term in Equation~\ref{14}). This difference in the mating functions, combined with the competition in the aquatic phase, results in a reduced total mosquito population, and hence less dengue. This provides the new insight that CI, when considered in isolation, alters the dengue transmission dynamics.

The solid line with expected \textit{WMel} values in Figure \ref{fig7}(b) shows that when the reproductive rate of \textit{Wolbachia}-carrying mosquitoes is too low compared to non-\textit{Wolbachia} mosquitoes, the \textit{Wolbachia}-carrying mosquitoes no longer persist, and hence have no effect on human dengue incidence. This occurs at approximately $c=0.66$ for the expected parameter values (solid line), and $c=0.52$ for $d=f=g=1$ (dashed line). When \textit{Wolbachia}-carrying mosquitoes do persist, their effect on human dengue incidence decreases with increasing $c$ since the total number of mosquitoes increases, despite CI. However, this effect is negligible for the expected parameter values (solid line),
with an over 90\% reduction in dengue for all values where \textit{Wolbachia}-carrying mosquitoes persist.

The results given in Figure \ref{fig7}(c)  show that even if $d \approx 1$ (transmission is not directly affected by \textit{Wolbachia}), a reduction in human dengue incidence of approximately 76\% is still obtained due to the other effects of \textit{Wolbachia} on mosquito physiology.
In addition, as shown in Figure \ref{fig7}(d), a 90\% reduction in human dengue incidence is obtained even though the biting rate of \textit{Wolbachia}-carrying mosquitoes is close to that of non-\textit{Wolbachia} mosquitoes ($g \approx 1$).

\subsection{The Basic Reproduction Number}\label{BasicReprosingleIntro}

\begin{figure}[ht]
  \begin{center}
  \includegraphics[width=0.9\linewidth]{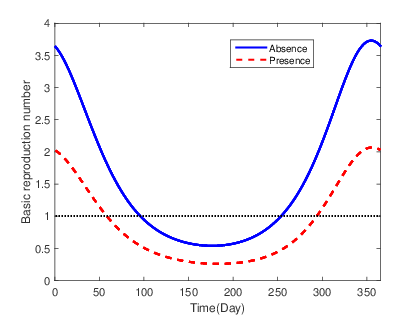}
  \end{center}
  \caption[The basic reproduction numbers for both the absence and presence of \textit{Wolbachia}-carrying mosquitoes]{The basic reproduction numbers for both the absence (solid lines) and presence (dashed lines) models, with  sinusoidally forced mosquito adult death rates. The dotted horizontal line shows the important threshold value, $\mathcal{R}_0=1$.}
  \label{basicrepro}
\end{figure}

Using the formula for the basic reproduction number in Chapter~\ref{chap:singlemathmodel}, we investigate the basic reproduction number. Figure \ref{basicrepro} shows the effects of seasonal forcing on the basic reproduction number. This can be considered to be an instantaneous basic reproduction number. The results show that the basic reproduction number is greater than one, meaning an outbreak may occur, for approximately 208 out of 365 days for the  model where \textit{Wolbachia} is absent, and 130 days for the  model where \textit{Wolbachia} is present. That is,  \textit{Wolbachia} intervention reduces the number of days when an outbreak can occur by 78 days a year in this deterministic model. If infected humans are introduced in favourable conditions, where the basic reproduction number is higher than unity, then outbreaks can still occur~\cite{Grassly2006, Coutinho2006}. This means that introducing \textit{Wolbachia}-carrying mosquitoes reduces the chance that a dengue outbreak will happen by reducing the number of days when the conditions are favourable. The basic reproduction number is a function of the parameter values, and is affected in the following way by key ones. If the transmission probability increases, the basic reproduction number also increases. When the ratio of death rate of \textit{Wolbachia}-carrying to non-\textit{Wolbachia} mosquitoes (f) increases, the basic reproduction number decreases, but then increases when f > 1:3. This is realistic because if f > 1:3, \textit{Wolbachia}-carrying mosquitoes die out and non-\textit{Wolbachia} mosquitoes dominate the population, which then results in a higher $\mathcal{R}_0$. If the vertical transmission is not ignored, the basic reproduction number would be different. That is, some proportion of mosquitoes mature to be exposed mosquitoes. However, the modelling method would be the same.

\section{Discussion and Conclusions}\label{Discussion}
We have developed a mathematical model for dengue transmission dynamics in the presence of a \textit{Wolbachia}-carrying mosquito population in order to quantify the effects of  a \textit{Wolbachia} intervention on human dengue incidence. Our model incorporates seasonal forcing through the adult mosquito death rate, the effect of cytoplasmic incompatibility, and competition for resources in the aquatic stage. This model considers the circulation of only a single dengue serotype with the mosquito population in a periodic stable state, appropriate to the study of a single dengue outbreak. It is assumed that half of the mosquito population mature into female adults, that a single introduction event occurs, and that there is no initial dengue immunity in the human population. These effects will have an impact on dengue transmission dynamics, with initial dengue immunity reducing the relative effect of the \textit{Wolbachia} introduction due to smaller outbreak sizes. However, the purpose of this model was to obtain an indication of the effect of the introduction of \textit{Wolbachia}-carrying mosquitoes on dengue transmission dynamics.  Extensions to the model which consider the effect of immunity in the human population are considered in Chapters~\ref{chap:multipleintro}, \ref{chap:symmetry}, and \ref{chap:asymmetry}.

A model which does not incorporate seasonality can only fit the data for Far North Queensland in one of two ways. First, nearly all humans need to have been previously infected and now recovered ($R_{H0}\approx 1$), which is not realistic.  Second, all mosquitoes can die out, as they would at some time in a one year period in a model which incorporated seasonal effects. However, unlike in a model which includes seasonal effects, the mosquito population cannot re-emerge, so this finding is again not realistic. Thus seasonality must be incorporated into our model if we wish it to fit the data for Far North Queensland.

The key parameter ranges were explored in order to determine the level of dengue incidence reduction due to \textit{Wolbachia} and the persistence of \textit{Wolbachia}-carrying mosquitoes.  This was an important part of our work as definite values for many parameters are not known. A global sensitivity analysis was used to determine the most influential parameters.

We found that inoculating mosquitoes using the \textit{Wolbachia} strain \textit{WMel}, which reduces the mosquito lifespan by at most 10\%, allows \textit{Wolbachia}-carrying mosquitoes to persist. This result is compatible with the recommendations of  Walker \textit{et al.}~\cite{Walker2011} and the results from field experiments~\cite{Hofman2011,Hoffmann2014}. Furthermore, we obtain new insight from Figure~\ref{fig7} into the importance of cytoplasmic incompatibility as a biological factor. CI not only gives a reproductive advantage to \textit{Wolbachia}-carrying female mosquitoes, but also influences the transmission dynamics of dengue as it reduces the mosquito population. The release of more \textit{Wolbachia}-carrying males than females is being considered by Hancock \textit{et al.}~\cite{Hancock2011a}, and their analysis may be important in considering the effect of cytoplasmic incompatibility on dengue transmission dynamics, and the possible feminisation of the \textit{Wolbachia}-carrying mosquito population. However, our results suggest that, with equal numbers of males and females, \textit{Wolbachia}-carrying mosquitoes can persist, and hence a reduction in dengue human incidence can be obtained through the use of this intervention.

The ratios of the transmission probability, $d=T_{HW}/T_N$, and of the biting rate, $g=b_W/b_N$, determine the level of reduction in human dengue incidence due to the introduction of \textit{Wolbachia}. By contrast, the ratios of the death rate, $f=\mu_W/\mu_N$, and of the reproductive rate, $c=\rho_W/\rho_N$, determine the persistence of \textit{Wolbachia}-carrying mosquitoes. This finding is corroborated by  our sensitivity analysis. Additionally, the biting and reproductive rates are linked, since female mosquitoes feed on human blood during their reproductive cycle. However, we treat the biting and reproductive rates as independent when exploring the parameter space and note that the relative behaviour of these parameters is as shown in Figure \ref{fig7}.
A significant reduction in human dengue incidence due to the introduction of \textit{Wolbachia} is also corroborated by the results found for the reproduction number. Figure \ref{basicrepro} shows that the number of days where this quantity is higher than unity is reduced by around 78 days when \textit{Wolbachia}-carrying mosquitoes are introduced, indicating that the chance of a dengue outbreak occurring is reduced. This indicates that, in our seasonal model, the time of introduction of dengue is important. When individuals with dengue enter the population at the appropriate time of the year, an outbreak can occur. This is further explored in Chapter~\ref{chap:multipleintro}. 

In conclusion, a one-strain dengue mathematical model incorporating \textit{Wolbachia}-carrying mosquitoes has been developed and analysed for a single outbreak. We quantify the effect that \textit{Wolbachia} will have on human dengue cases, once the mosquito populations have reached their periodic stable states. \textit{Wolbachia} comes with a fitness cost  but also a reproductive advantage (CI) to \textit{Aedes aegypti} mosquitoes, and we explore the parameter values for which \textit{Wolbachia} will persist in the mosquito population. Our results show that for realistic \textit{WMel} parameter values, \textit{Wolbachia}-carrying mosquitoes can greatly reduce the transmission of dengue. We also show that in the absence of any other effect on the mosquito physiology by \textit{Wolbachia}, cytoplasmic incompatibility alone affects the transmission of dengue, reducing the human incidence.

The results discovered this chapter provide an indication that \textit{Wolbachia} is effective in reducing dengue transmission. This chapter focuses on a single introduction of dengue cases. Generally, individuals with dengue enter the population multiple times, and the elapsed time between introductions is irregular. Therefore, understanding the effect of \textit{Wolbachia} on dengue transmission dynamics when individuals carrying  dengue enter the population multiple times is important. In the next chapter, multiple introductions of dengue cases into the population are investigated to assess their effects on the performance of \textit{Wolbachia} in reducing dengue transmission.

\section{Summary}
\begin{itemize}[wide=0pt,labelwidth = 1.3333em, labelsep = 0.3333em, leftmargin = \dimexpr\labelwidth + \labelsep\relax ]
\item We estimated the parameter values against data for Cairns, Australia in Section~\ref{WlbAbsence}.
\item \textit{Wolbachia} is expected to reduce dengue transmission by over 90\% (Section~\ref{parameterExploreation}, see Figure~\ref{fig7}).
\item Using the \textit{WMel} strains that reduces the mosquito lifespan by approximately 10\% allows the mosquitoes carrying \textit{Wolbachia} to persist in the wild (Sections~\ref{WolbachiaParams} and \ref{parameterExploreation}, Figures~\ref{fig3} and \ref{fig7}(a).)
\item A global sensitivity analysis reveals that the transmission probability, the average death rate and the biting rate are the important parameters (Sections~\ref{WlbAbsence} and \ref{DengueWolbahiaModel}, see Figures~\ref{fig5} and ~\ref{fig6}).
\item The presence of \textit{Wolbachia} reduces the number of days at which basic reproduction number is higher than one by around 78 days in a year (Section~\ref{BasicReprosingleIntro}, Figure~\ref{basicrepro}).
\item In the absence of any other effect on the mosquito physiology by \textit{Wolbachia}, cytoplasmic incompatibility (CI) alone reduces the dengue incidence (Section~\ref{parameterExploreation}, Figure~\ref{fig7}).
\end{itemize}

\chapter[Dengue Transmission Dynamics for Multiple Introductions]{Dengue Transmission Dynamics for Multiple Introductions\footnote{This work is part of the manucript that has been submitted for publication as Meksianis Z. Ndii, David Allingham, R.I. Hickson, Kathryn Glass. The effect of \textit{Wolbachia} on dengue outbreaks when dengue is repeatedly introduced. }}\label{chap:multipleintro}

\ifpdf
    \graphicspath{{5/figures/PNG/}{5/figures/PDF/}{5/figures/}}
\else
    \graphicspath{{5/figures/EPS/}{5/figures/}}
\fi

In the previous chapter, we studied the effect of \textit{Wolbachia} on dengue transmission dynamics for a single introduction of dengue cases. In this chapter, we study the effect of \textit{Wolbachia} on dengue transmission dynamics for multiple introductions of dengue. The aim of this chapter is to investigate the performance of \textit{Wolbachia} in reducing dengue incidence when dengue cases are introduced into the population multiple times.

We work with our dimensionalised models from Chapter~\ref{chap:singlemathmodel} in the absence (Equations \eqref{eq1}--\eqref{eq8}) and presence (Equations \eqref{10}--\eqref{21}) of \textit{Wolbachia}-carrying mosquitoes.

\section{Introduction}

The transmission dynamics of dengue are very seasonal due to fluctuations in the mosquito population as a result of variations in  climatic conditions such as temperature and rainfall. In regions with strongly seasonal climatic conditions,  epidemics generally occur at certain times of the year and are mostly triggered by the importation of dengue cases. This means that the time period in which an infected individual enters the population is an important factor in determining whether an outbreak will occur.

Only a few research projects have considered the effects of multiple introduction times  on dengue outbreaks (see Section~\ref{denguemodelReview}). Some studies have investigated the effects of the frequency of  dengue introductions on dengue transmission dynamics~\cite{Bannister2013, Williams2015}.  Bannister-Tyrrell \textit{et al.}~\cite{Bannister2013} investigated the variation in dengue activity in Australia using a process-based modelling approach in which dengue cases were introduced  once a year, bimonthly, monthly and weekly.  Williams \textit{et al.}~\cite{Williams2015} used an existing dengue model, which involved an entomological component (CIMSiM)  and a disease component (DENSiM), to assess the effects of the importation rate on dengue outbreaks.  They assumed that between 1 and 25 dengue cases were introduced into the population at weekly intervals. A few mathematical models have been developed to study the effects of \textit{Wolbachia} on dengue transmission dynamics.  In this chapter, we investigate the combined influence of these two factors on dengue transmission dynamics by exploring the performance of \textit{Wolbachia} in reducing dengue incidence when dengue cases are repeatedly introduced into the population.

Mathematical models describing dengue transmission dynamics in the absence and presence of \textit{Wolbachia}-carrying mosquitoes are given in Chapter~\ref{chap:singlemathmodel}. The model incorporates seasonality through a function controlling the adult mosquito death rate and is parameterised for Cairns, Australia. However, this model can be generalised to simulate dengue transmission dynamics in other places where dengue outbreaks are triggered by the importation of dengue cases.  We measure the effectiveness of the  \textit{Wolbachia} intervention by comparing the incidence of dengue in the absence and presence of \textit{Wolbachia}, and use this to compute the proportional reduction in dengue incidence. The formula for this reduction is given in Equation~\eqref{RE} on Page~\pageref{RE}.

\subsection{Simulation Procedure}\label{simulationprocedure}

We run simulations for our model using the built-in
MATLAB functions {\tt ode45} and {\tt odextend}.  The initial conditions used for the simulations are $S_{H0} = 150,000$~\cite{CairnsPop2009}, $A_{N0} = 3 \times S_{H0}$,  $S_{N0}=3 \times S_{H0}$~\cite{Chowel2007}, $S_{W0}= 2 \times S_{N0}$~\cite{Walker2011}, and the remaining populations are set to zero.  In experiments conducted to assess the spread of \textit{Wolbachia} and its persistence, \textit{Wolbachia} and non-\textit{Wolbachia} mosquitoes were released in the ratio 2:1~\cite{Walker2011}, and so we set $S_{W0}= 2 \times S_{N0}$. In the absence of \textit{Wolbachia}-carrying mosquitoes, $S_{W0}=0$.  In this analysis, the phase, $\omega$, is shifted 60 days ($\omega = 80.61$ days) as our start time is January.

We use the following procedure for our simulations.  First, the model is run without dengue using the initial conditions described above until the mosquito population reaches the periodic stable state, and then individuals with dengue are introduced. Second, we continue the simulation using ``odextend'' for another week, after which dengue is reintroduced. The reintroduction procedure is repeated until the final introduction for the year. Step two is repeated for the same periods of each following year until the human infected population has remained constant for at least 75 consecutive years. As we assume a constant human population, when individuals with dengue are introduced, the same number of individuals are removed from the susceptible human class, $S_{H}$. We are particularly interested in the occurrence of outbreaks as the aim of this chapter is to determine the reduction in dengue outbreak size due to \textit{Wolbachia}. To ensure that the disease does not persist when the  infected human population is less than one individual, the infected population is set to zero if it falls below a threshold of 0.5, which is a deterministic proxy for stochastic fade-out.

\subsection{Introduction Scenarios}

\begin{figure}[ht]
\begin{center}
\includegraphics[width=0.9\textwidth]{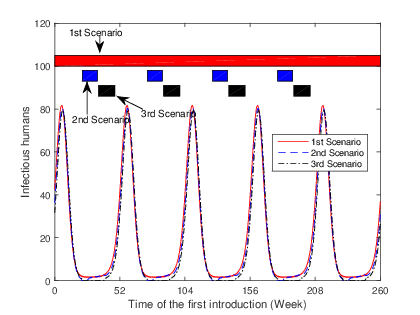}
\end{center}
\caption[Plot of the time series of dengue outbreaks with weekly introductions]{Plots of the time series of dengue outbreaks with weekly introductions. In each of the 1st, 2nd and 3rd scenarios, dengue cases are introduced weekly year-round, in June to August, and in September to November, respectively, as represented by  rectangles at the top of the figure.}
\label{fig:TimeSeriesScenario}
\end{figure}

We consider weekly year-round introductions and weekly three-month introduction periods.  Plots of the time series for infected humans when we introduce dengue every week in a year (1$^{st}$ scenario) and every week in a three month period (2$^{nd}$ and 3$^{rd}$ scenarios) are given in Figure~\ref{fig:TimeSeriesScenario}. In the second scenario, dengue cases are introduced weekly from June to August, while in the third scenario, dengue cases are introduced weekly from September to November. The introduction periods are represented by rectangles at the top of Figure~\ref{fig:TimeSeriesScenario}. In the second and third scenarios, the disease dies out over winter,  but the overall dynamics for each of the three scenarios are very similar  due to the level of introduction.

The true number of imported cases is very uncertain and differs across geographic regions~\cite{Schwartz2008}. In Queensland, for instance, the number of imported dengue cases was around 202 in 2010 and 118 in 2011~\cite{Knope2013}. The number of imported cases in Taiwan has increased over time from one individual in 1998 to around 15 individuals in 2007~\cite{Shang2010}. In our model, two dengue cases are introduced each week, with a total of 104 imported cases for year-round introduction (scenario 1) and 26 imported cases for each of the three month periods of introduction (scenarios 2 and 3).

\section{Results}
In this section, we assess the potential for an outbreak, the effect of the strength of seasonality and the transmission rate, and investigate the effective reproduction number.
\subsection{Potential for an Outbreak}\label{DiffTimeIntro}

\begin{figure}[ht]
\begin{center}
\includegraphics[width=1\textwidth]{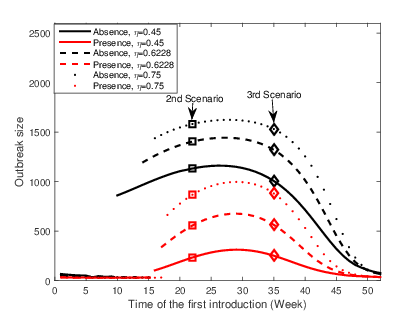}
\end{center}
\caption[Plot of the first week of introduction of introduced dengue cases versus outbreak size]{Plot of the first week of introduction of introduced dengue cases versus outbreak size. The square and diamond are the first week of the introduction for scenarios 2 and 3 respectively. Parameter values are given in Table~\ref{table2}. `Absence' means that \textit{Wolbachia}-carrying mosquitoes have not been introduced. `Presence' means that \textit{Wolbachia}-carrying mosquitoes have been introduced. The parameter $\eta$ represents the seasonality strength.}
\label{Potential4Outbreak}
\end{figure}

Here we explore the impact of the season in which dengue is imported. Figure \ref{Potential4Outbreak}  shows the outbreak size with and without \textit{Wolbachia} for different values of $\eta$, the parameter representing the strength of seasonality. In all simulations, dengue is introduced over a three month period (Scenarios 2 and 3), with the week of the first introduction given on the x-axis.  Figure \ref{Potential4Outbreak} shows that the lengths of the potential outbreak seasons have been reduced in the presence of \textit{Wolbachia}-carrying mosquitoes for all three values of $\eta$. The maximum  potential outbreak size occurs when dengue is first introduced between June and September. Furthermore,  the outbreak season contracts as the strength of seasonality ($\eta$) increases.

In investigating the effects of the transmission rate and the strength of seasonality under the assumption that dengue is introduced over a period of three months, we use the June to August (scenario 2) and September to November  (scenario 3) time periods of introduction as  outbreaks can potentially occur when dengue is introduced in these periods. We use three-month periods of introduction as these are equivalent to the approximate
length of a climatic season such as Summer or Winter.

\subsection{The Effect of the Strength of Seasonality}\label{SeasonStrength}
\begin{figure}[ht]
\begin{center}
\includegraphics[width=1\textwidth]{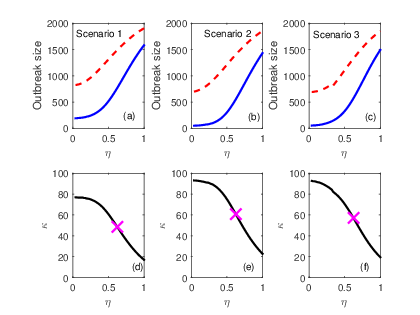}
\end{center}
\caption[The effect of the strength of seasonality on the proportional reduction in dengue]{The effect of the strength of seasonality on the proportional reduction in dengue. Top: Plot of seasonality strength ($\eta$) vs. outbreak size in the absence (dashed line) and presence (solid line) of \textit{Wolbachia}. Bottom: Plot of strength of seasonality ($\eta$) vs. proportional reduction in dengue ($\kappa$). Plots (a), (b) and (c) show outbreak size for weekly introduction year-round (scenario 1), in June to August (scenario 2), and in September to November (scenario 3), respectively. Plots (d), (e), and (f) are the proportional reduction in dengue for the same scenarios as in plots (a), (b), and (c). The cross indicates the value of seasonality estimated by fitting the model to data from Cairns, Australia ($\eta=0.6228$). Other parameter values are given in Table~\ref{table2}.}
\label{Eta_results}
\end{figure}

The strength of seasonality is one of the key factors to explore since it affects vector dynamics. As the strength of seasonality ($\eta$) increases, the outbreak size increases both in the absence and presence of \textit{Wolbachia}-carrying mosquitoes (Figure~\ref{Eta_results}). When we look at the overall effects of \textit{Wolbachia} on dengue, we see that the proportional reduction in dengue decreases as the strength of seasonality increases. Furthermore, the time of introduction also affects the impact of \textit{Wolbachia} on dengue incidence.  The reduction in dengue incidence when cases are introduced weekly throughout the year (scenario 1) is lower than that for either of the scenarios modelling three months of introduction, and hence \textit{Wolbachia} is less effective in this scenario. For low values of the strength of seasonality, the proportional reduction in dengue is approximately 80\% for year-round introduction (scenario 1) and over 80\% for the three month periods of introduction (scenarios 2 and 3). In general, the \textit{Wolbachia} intervention is most effective when the amplitude of seasonality is low.

\subsection{The Effect of the Transmission Rate}\label{TransRate}
The transmission probability $\left(T_N\right)$ and the biting rate $\left(b_N\right)$ are the most influential parameters in the model~\cite{Ndii2015} and so we explore the effects of these parameters on dengue dynamics in the presence of \textit{Wolbachia}.  As these two parameters have similar effects on the disease dynamics, we combine them into one parameter,  $\beta_N=b_N T_N$ which we refer to as the transmission rate. In Chapter~\ref{chap:singleintro}, the fitted value for $\beta_N$, appropriate for Cairns in northern Australia, was  calculated to be 0.1648 day$^{-1}$.

\begin{figure}[ht]
\begin{center}
\includegraphics[width=1\textwidth]{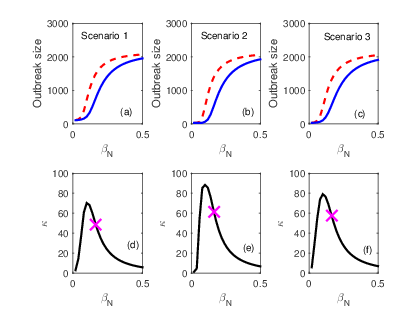}
\end{center}
\caption[Plot of transmission rate ($\beta_N$) vs. outbreak size in the absence and presence of \textit{Wolbachia} and $\beta_N$ vs. the proportional reduction in dengue]{Top: Plot of transmission rate ($\beta_N$) vs. outbreak size in the absence (dashed line) and presence (solid line) of \textit{Wolbachia}. Bottom: Plot of transmission rate ($\beta_N$) vs. proportional reduction in dengue ($\kappa$). Plots (a), (b) and (c) are the outbreak size for weekly introduction year-round (scenario 1), in June to August (scenario 2), and in September to November (scenario 3), respectively. Plots (d), (e), and (f) are the proportional reduction in dengue for the same scenarios as in plots (a), (b), and (c). The cross indicates the value of the transmission rate estimated by fitting the model to data from Cairns, Australia, ($\beta_N=0.1648$). Other parameter values are given in Table~\ref{table2}.}
\label{Beta_Results}
\end{figure}

Figure \ref{Beta_Results} shows that as $\beta_N$ increases, the proportional reduction in dengue, $\kappa$, peaks and then decreases.  Figure \ref{Beta_Results} also reveals that the maximum proportion of dengue reduction is  slightly less than 80\% for the year-round introduction (scenario 1), and around 80\% for weekly introduction during June to August (scenario 2) and September to November (scenario 3). For Cairns, the proportional reduction in dengue is around 50\% for year-round introduction and 60\% for three-month introductions.

\subsection{The Effective Reproduction Number}\label{Effrepro}

\begin{figure}[h!]
\begin{center}
\includegraphics[width=1\textwidth]{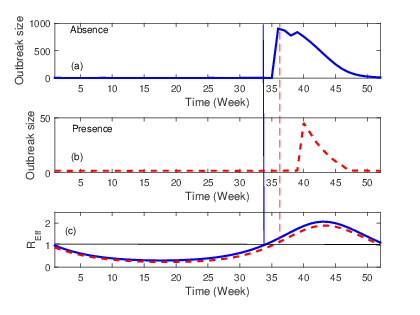}
\end{center}
\caption[Plot of the week of introduction of dengue cases vs. outbreak size in the absence and presence of \textit{Wolbachia}-carrying mosquitoes, and the effective reproduction number.]{ Plot of the week of introduction of dengue cases vs. outbreak size in the absence and presence of \textit{Wolbachia}-carrying mosquitoes (plots (a) and (b), respectively), and the effective reproduction number (plot (c)). The vertical lines indicate the week at which $\mathcal{R}_{\mbox{Eff}}=1$ for model in the absence (solid blue line) and presence (dashed red line) of \textit{Wolbachia}-carrying mosquitoes. In all plots, the solid red lines and the dashed blue lines are for absence and presence of \textit{Wolbachia}-carrying mosquitoes, respectively. The strength of seasonality is 0.6228. Other parameter values are given in Table~\ref{table2}.}
\label{REffModerate}
\end{figure}

Using the formula for the effective reproduction number in Chapter~\ref{chap:singlemathmodel}, Page~\pageref{ReffPres} for the model in dimensionalised form, we investigate the transmission potential of dengue in the long-term. The equilibrium susceptible populations are obtained by simulating the model with weekly year-round introduction until the human infected population remains the same for 75 years. The equilibrium susceptible populations are then used to calculate the effective reproduction number. We then simulate our model where two dengue cases are introduced once a year in a specific week (we implement a thresholding of 0.5 to keep consistency), and compare this with the results of the effective reproduction number.

A comparison of the effective reproduction number ($\mathcal{R}_\text{Eff}$) and with the outbreak size shows a similar threshold (Figure~\ref{REffModerate}). If we introduce dengue cases in the week at which $\mathcal{R}_\text{Eff}>1$, an outbreak is likely to occur (Figure~\ref{REffModerate}). There is a slight difference between the week in which $\mathcal{R}_\text{Eff}>1$ and the week in which outbreak takes off. The outbreak takes off around two weeks later in the absence of \textit{Wolbachia} and four weeks later in the presence of \textit{Wolbachia}-carrying mosquitoes.

\section{Discussion and conclusions}

We investigated the effects of \textit{Wolbachia} on dengue transmission dynamics using a mathematical model which includes the importation of humans infected with dengue.

We found that, in the presence of \textit{Wolbachia}-carrying mosquitoes, the time-window in which epidemics can occur is reduced by between two and six weeks (see Figure~\ref{Potential4Outbreak}), depending on the strength of seasonality. This is driven by a decline in the mosquito population size, likely caused by a reduction in the mosquito lifespan and, thus, in the mosquito's reproductive rate. An increase in the strength of seasonality further reduces the time-window in which an outbreak can occur due to the presence of a limited number of mosquitoes over winter. In this case, the decline in  the mosquito population numbers is due to a short lifespan driven by seasonality. The model provides the insight that the presence of \textit{Wolbachia}-carrying mosquitoes reduces the length of the potential seasons in which an outbreak can occur.    

We also compared the time-dependent reproduction number and the simulation, and found the time at which an outbreak occurs qualitatively matches. That is, an outbreak occurs when the reproduction number is greater than one. The models agree within 2--4 weeks and this slight discrepancy is likely due to a combination of disease importation assumptions and thresholding rules that lead to outbreak fadeout for some reproductions numbers close to one. As the purpose of this thesis is to obtain general understanding of the performance of \textit{Wolbachia} in reducing dengue incidence, the general consistency in these approaches is sufficient and reassuring.

In comparing the outbreak sizes, as the strength of seasonality is increased, we found that outbreak sizes increase and the benefits of \textit{Wolbachia} decrease. This increase in outbreak size is caused by a larger mosquito population since the mosquito lifespan is longer when seasonality is at its highest. This larger mosquito population increases the chance that a human in the population is infected. In contrast, as the transmission rate increases, there is an intermediate level of transmission (around 0.08 to 0.12) at which \textit{Wolbachia} is most effective.  At these transmission rates, outbreaks will occur in the absence of \textit{Wolbachia}, but only either very small or no outbreaks  will occur in the presence of \textit{Wolbachia}-carrying mosquitoes. When the strength of seasonality  is around 0.02-0.30 and the transmission rates are around 0.08 to 0.12 day$^{-1}$, dengue incidence can be reduced by more than 80\%, depending on the time of year in which dengue cases are introduced. In reality, infected individuals enter populations year-round, according to a number of different factors. However, our results across scenarios 1--3 show that the effect of \textit{Wolbachia} is largely independent of the time of introduction. Overall, our results suggest that dengue incidence can be reduced by up to 80\% in populations where dengue is repeatedly introduced.

In this chapter, multiple introductions of a single dengue strain were explored. Although these findings provide insights into the performance of \textit{Wolbachia} in reducing dengue incidence, generally individuals with different dengue serotypes enter the population at random times.   Moreover, since four dengue serotypes are circulating in most places~\cite{Limkittikul2014, Mia2013}, the effects of \textit{Wolbachia} on dengue transmission dynamics with multiple serotypes are of interest.   This  may influence \textit{Wolbachia} performance in reducing dengue. However, generally only one serotype dominates a given outbreak~\cite{Stoddard2014,Limkittikul2014,Mia2013}. In the following three chapters, the performance of \textit{Wolbachia} in reducing dengue when more than one dengue serotype circulates in the population is investigated. 

\section{Summary}
\begin{itemize}[wide=0pt,labelwidth = 1.3333em, labelsep = 0.3333em, leftmargin = \dimexpr\labelwidth + \labelsep\relax ]
\item We conducted multiple introductions of a single dengue serotype to investigate the performance of \textit{Wolbachia} in reducing dengue incidence.
\item When dengue is not endemic, timing of imported cases determines the
likelihood of an outbreak (Sections~\ref{DiffTimeIntro} and \ref{Effrepro}, see Figures~\ref{Potential4Outbreak} and ~\ref{REffModerate}).
\item We found that the time window in which outbreak occurs has been reduced by up to six weeks in the presence of \textit{Wolbachia}-carrying mosquitoes (Section~\ref{DiffTimeIntro}, Figure~\ref{Potential4Outbreak}).
\item A reduction of 80\% in dengue incidence can be obtained when the strength of seasonality  and the transmission rates are around 0.02--0.3 and 0.08--0.12, respectively (Sections~\ref{SeasonStrength} and \ref{TransRate}, see Figures~\ref{Eta_results} and ~\ref{Beta_Results}).
\end{itemize}
 %
\chapter*{Part 2: Two Dengue Serotypes}
\addcontentsline{toc}{chapter}{Part 2: Two Dengue Serotypes}

In this part of the thesis, the performance of the \textit{Wolbachia} intervention when two dengue serotypes  circulate is studied.  In addition,  the effects of symmetric and asymmetric epidemiological characteristics on \textit{Wolbachia} performance in reducing dengue incidence are explored. This part comprises Chapters~\ref{chap:modelmultiplestrains}--\ref{chap:asymmetry}, and the results are summarised below.
\section*{Summary}
\begin{itemize}[wide=0pt,labelwidth = 1.3333em, labelsep = 0.3333em, leftmargin = \dimexpr\labelwidth + \labelsep\relax ]
\item Two-serotype dengue models in the absence and presence of \textit{Wolbachia}-carrying mosquitoes are developed (Chapter~\ref{chap:modelmultiplestrains}, Equations~\eqref{multiabsSH}--\eqref{multiabsIN} and ~\eqref{MultiW1}--\eqref{MultiWend}). 
\item The procedure for simulation is the same as that in Chapter~\ref{chap:multipleintro}, except here the dengue serotype being introduced varies according to the scenario of dengue introductions (Section~\ref{ScenarioIntroMulti}).
\item Different disease introduction scenarios are explored and the results show that varying the disease introduction scenario does not affect  \textit{Wolbachia} performance in reducing dengue incidence, with up to 80\% reduction in human dengue cases (Chapter~\ref{chap:symmetry}, Sections~\ref{1stScenario} and ~\ref{2ndScenarioIntro}).
\item Although the antibody-dependent enhancement parameter affects the performance of \textit{Wolbachia} in reducing dengue incidence, the transmission probability remains the key parameter regulating the dengue dynamics, and, hence, \textit{Wolbachia} performance in reducing dengue incidence (Chapters~\ref{chap:symmetry} and \ref{chap:asymmetry}).
\item When one serotype is more transmissible, that serotype will dominate the primary infections, and the secondary infections will be dominated by the less transmissible serotype.  The performance of \textit{Wolbachia} in reducing dengue infection caused by the more transmissible serotype subsequently declines (Section~\ref{AssTN}).
\item The proportional reduction in secondary infections due to the \textit{Wolbachia} intervention is higher than that for primary infections (Chapters~\ref{chap:symmetry} and \ref{chap:asymmetry}).
\end{itemize} 

\chapter[Mathematical Modelling of Two Dengue Serotypes]{Mathematical Modelling of Two Dengue Serotypes dengue\footnote{The mathematical models presented here are part of the manuscript that has been submitted for publication as M.Z. Ndii, D. Allingham, R. I. Hickson, K. Glass. \textit{The effect of \textit{Wolbachia} on dengue dynamics in the presence of multiple serotypes of dengue: symmetric and asymmetric epidemiological characteristics.}  }}\label{chap:modelmultiplestrains} 


\ifpdf
    \graphicspath{{6/figures/PNG/}{6/figures/PDF/}{6/figures/}}
\else
    \graphicspath{{6/figures/EPS/}{6/figures/}}
\fi

In order to investigate the performance of \textit{Wolbachia} in reducing dengue incidence when more than one dengue serotype circulates, we develop two-serotype models for dengue in the absence and presence of \textit{Wolbachia}-carrying mosquitoes. In this chapter, we present a full derivation of these models and explain their important features. In Chapters~\ref{chap:symmetry} and~\ref{chap:asymmetry} we use our models to study the effects of symmetric and asymmetric epidemiological characteristics of the serotypes on dengue transmission dynamics.

\section{Model Formulation}

In these models, human and mosquito populations are assumed to be well-mixed and homogenous, and we include two serotypes of dengue. The ratio between male and female mosquitoes is  1.02:1~\cite{Arrivillaga2004}.  For simplicity, we assume that the numbers of male and female mosquitoes are equal, that is,  $F_N=M_N$ and $F_W=M_W$. Subscripts $H$, $N$, $W$ denote humans, non-\textit{Wolbachia}  and \textit{Wolbachia}-carrying mosquitoes, respectively. Superscripts $i$ and $j$ denote serotypes $1$ or $2$. This chapter is organised as follows. First, we present the formulation of the model in the absence of \textit{Wolbachia}, and then the model in the presence of \textit{Wolbachia}. Finally, we remind the reader of the relationships between and the values of several important parameters for the models.

\subsection{Model in the Absence of \textit{Wolbachia}}\label{sec:multiabsencemodel}
\begin{figure}
 \begin{center}
\includegraphics[width=0.9\linewidth]{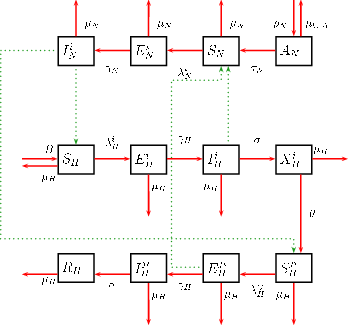} 
 \end{center}
 \caption[Schematic representation of model with two dengue serotypes in the absence of \textit{Wolbachia}-carrying mosquitoes.]{Schematic representation of model with two serotypes in the absence of \textit{Wolbachia}-carrying mosquitoes. Solid lines are population progression lines and dashed lines are disease transmission lines. The subscript $H$ is for the human population, $N$ is for the non-\textit{Wolbachia}-carrying mosquito population and $i$ is for the serotype. The compartments are `S' for susceptible, `E' for exposed to dengue but not yet infectious, `I' for infectious, `R' for recovered from all serotypes, `X' for temporary immunity to all serotypes and `A' for the aquatic phase of the mosquito life cycle. The transition rates between compartments are shown next to the progression lines and are described in the text.}
 \label{fig:MultiFlowChartAbesence}
 \end{figure}

A two-serotype dengue model in the absence of \textit{Wolbachia}-carrying mosquitoes is formulated. This model is an extension of the single serotype dengue model in the absence of \textit{Wolbachia}-carrying mosquito population given in Equations~\eqref{eq1}--\eqref{eq8} on page~\pageref{eq1}. The model comprises human and non-\textit{Wolbachia} mosquito populations, and the schematic representation of the model is given in Figure~\ref{fig:MultiFlowChartAbesence}. Superscripts  $i$ and $j$ denote dengue serotypes, where $i=1,2$ and $j = 3-i$.

The human population is divided into 14 different classes:
\begin{itemize}
\item  Susceptible to all serotypes, $S_H$.
\item  Exposed and infectious with serotype $i$, $E_{H}^i$ and $I_{H}^i,$ respectively.
\item  Temporarily immune to all serotypes, but previously infected by serotype $i$, $X_H^i$.
\item Susceptible to serotype $j$ and previously infected by serotype $i$, $S^{ji}_{H}$.
\item Exposed and infectious to serotype $j$ and previously infected by serotype $i$, $E^{ji}_{H}$ and  $I^{ji}_{H},$  respectively.
\item Recovered from all serotypes, $R_H$.
\end{itemize}
The total human population is then given by the equation
\begin{equation}
N_H=S_H+ R_H +\sum_{i=1}^{2}\left(E_{H}^i + I_{H}^i + X_H^i\right) + \mathop{\sum_{i=1}^2}_{j = 3-i} \left(S_H^{ji} + E_H^{ji} + I_H^{ji} \right) .
\label{eq:NH}
\end{equation}

The non-\textit{Wolbachia} mosquito population is divided into six compartments:
\begin{itemize}
\item Mosquitoes in the aquatic stage, $A_N$.
\item Susceptible, $S_N$.
\item Exposed to either serotype, $E_{N}^i$.
\item Infectious with either serotype, $I_{N}^i$. 
\end{itemize}
The mosquitoes remain infectious throughout their lifetime, hence no recovered class for mosquitoes is required. We also assume that the mosquitoes are never co-infected with both serotypes. The total population of  non-\textit{Wolbachia} mosquitoes, $F_N$, is given by the equation
\begin{equation}
F_N = S_N+\sum_{i=1}^2 \left(E_{N}^i+ I_{N}^i\right).
\label{eq:FN}
\end{equation}

The model is governed by the following system of differential equations
\begin{align}
\frac{dS_H}{dt}&=B N_H - \sum_{i=1}^{2} \lambda^i_{H} S_H  -\mu_H S_H,\label{multiabsSH}\\
\frac{dE_{H}^i}{dt}&= \lambda^i_{H} S_H - \gamma_H E_{H}^i -\mu_H E_{H}^i,\label{multiabsEH}\\
\frac{dI_{H}^i}{dt}&=\gamma_H E_{H}^i- \sigma I_{H}^i  -\mu_H I_{H}^i,\label{multiabsIH}\\
\frac{dX_H^i}{dt}&= \sigma I_{H}^i -\theta X_H^i -\mu_H X_H^i,\label{multiabsXH}\\
\frac{dS_{H}^{ji}}{dt}&=\theta X_H^i -  \lambda_H^j  S_{H}^{ji} -\mu_H S_{H}^{ji},\label{multiabsSH2}\\
\frac{E_{H}^{ji}}{dt}&= \lambda^j_H  S_{H}^{ji} -\gamma_H E_{H}^{ji} -\mu_H E_{H}^{ji},\label{multiabsEH2}\\
\frac{dI_{H}^{ji}}{dt}&= \gamma_H E_{H}^{ji}- \sigma I_{H}^{ji} -\mu_H I_{H}^{ji},\label{multiabsIH2}\\
\frac{dR_{H}}{dt}&=\mathop{\sum_{j=1}^{2}}_{j\neq i}\sigma  I_{H}^{ji}  -\mu_H R_H,\label{multiabsRH}\\
\frac{dA_N}{dt}&=\frac{\rho_N F_N}{2}\left(1-\frac{A_N}{K}\right)-\left(\tau_N + \mu_{NA}\right),\label{multiabsAN}\\
\frac{dS_N}{dt}&=\frac{\tau_N A_N}{2} -\sum_{i=1}^{2}\lambda_{N}^i S_N -\mu_N(t) S_N,\label{multiabsSN}\\
\frac{dE_{N}^i}{dt}&=\lambda_{N}^i S_N -\gamma_N E_{Ni}-\mu_N(t) E_{N}^i,\label{multiabsEN}\\
\frac{dI_{N}^i}{dt}&=\gamma_{N} E_{N}^i -\mu_N(t) I_{N}^i.\label{multiabsIN}
\end{align}
Humans are born susceptible at rate $B$ and then exposed to dengue after being bitten by non-\textit{Wolbachia} infectious mosquitoes at rate  $\lambda^i_H$, which is governed by the equation
\begin{equation}
\lambda^i_H=\frac{b_NT_{N}^i I_{N}^i} {N_H}. 
\label{eq:FoIHuman}
\end{equation}
Here $b_N$ is the biting rate of non-\textit{Wolbachia} mosquitoes, $T_{N}^i$ is the transmission probability  from non-\textit{Wolbachia} mosquitoes with serotype $i$ to humans, and $N_H$ is the total human population, given in Equation~\ref{eq:NH}. The transmission probability between humans and non-\textit{Wolbachia} mosquitoes is assumed to be the same, and is denoted by $T_N^i$.

Exposed individuals progress to the infectious class at rate $\gamma_H$, and then to the temporary  immunity class at rate $\sigma$. After spending $1/\theta$ time units in the temporary immunity class, they become susceptible to the other serotype $j$, where $j\neq i$.  They may then become exposed to dengue after being bitten by non-\textit{Wolbachia} mosquitoes carrying dengue serotype $j$ at rate
\begin{equation}
\lambda^j_H=\frac{b_NT_{N}^j I_{N}^j} {N_H}.
\label{FoIHumanSecond}
\end{equation}
Following re-exposure, they become infectious at rate $\gamma_H$, and  recover from the secondary infection at rate $\sigma$. We assume that dengue is not fatal, so that humans die only at the  natural death rate, $\mu_H$. As explained in Section~\ref{dengue}, fatality due to the severe forms of dengue is only 1\% in developed nations and we focus on the difference in dengue cases in the presence versus absence of \textit{Wolbachia}-carrying mosquitoes, so use this simplification. 

Non-\textit{Wolbachia} aquatic mosquitoes are produced when non-\textit{Wolbachia} female $F_N$ and male $M_N$ mosquitoes  mate. The growth of the  aquatic mosquito population is limited by the carrying capacity $K$, through the inclusion of the term:
\begin{equation}
\frac{\rho_N F_N M_N}{\left(F_N + M_N\right)}\left(1 - \frac{A_N }{K}\right).
\label{eq:ANReproduced}
\end{equation}
As the ratio between males and females is around 1.02:1~\cite{Arrivillaga2004}, we assume that there are equal populations of male and female mosquitoes, $F_N = M_N$. Therefore, Equation~\ref{eq:ANReproduced} is reduced to
\begin{equation}
\frac{\rho_N F_N}{2}\left(1-\frac{A_N}{K}\right).
\end{equation}
Non-\textit{Wolbachia} aquatic mosquitoes become adult susceptible mosquitoes at rate $\tau_N$. The susceptible mosquitoes are infected with dengue after biting infectious humans with either primary or secondary infections at rate
\begin{equation}
\lambda^i_{N}=\frac{b_N T_{N}^i I_{H}^i}{NH}+\frac{\phi^i b T_{N}^i  I_{H}^{ij}}{NH}.  
\label{eq:lambdaN}
\end{equation}
Here $\phi^i$ is the antibody-dependent enhancement of serotype $i$, which is higher than one since infectious individuals with a secondary infection have a higher viral load, and hence, a higher transmission rate. The exposed mosquitoes become infectious at rate $\gamma_H$.
The death rate of adult mosquitoes is sinusoidally forced according to the equation
\begin{equation}
\mu_N=\mu_{N0}\left(1 - \eta\, \cos \left(\frac{2\pi\left(t +\omega\right))}{365} \right)\right), 
\end{equation}
where $\mu_{N0}$ is average death rate, $\eta$ is the amplitude of seasonality and $\omega$ is a phase shift as also done in Section~\ref{sec:modelabsence}. 

\subsection{Model in the Presence of \textit{Wolbachia}}\label{MultiWolbachiaMOdel}
\begin{figure}
 \begin{center}
  \includegraphics[width=0.8\linewidth]{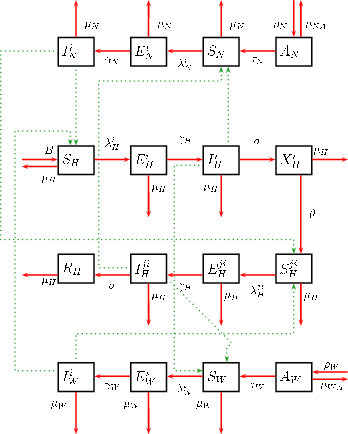}
 \end{center}
 \caption[Schematic representation of the model with two dengue serotypes in the presence of \textit{Wolbachia}-carrying mosquitoes]{Schematic representation of model with two dengue serotypes in the presence of \textit{Wolbachia}-carrying mosquitoes (Equations~\eqref{MultiW1}--\eqref{MultiWend}). Solid lines are population progression lines and dashed lines represent disease transmission. The subscript $H$ is for the human population, $N$ is for the non-\textit{Wolbachia} mosquito population and $W$ is for the \textit{Wolbachia}-carrying mosquito population and $i$ is for the serotype. The compartments are `S' for susceptible, `E' for exposed to dengue but not yet infectious, `I' for infectious, `R' for recovered from all serotypes, `X' for temporary immunity to all serotypes and `A' for the aquatic phase of the mosquito life cycle. The transition rates between compartments are shown next to the progression lines and are described in the text.}
 \label{fig:MultiFlowChartPresence}
 \end{figure}

In this section, a two-serotype dengue model in the presence of \textit{Wolbachia}-carrying mosquitoes is formulated. The model is an extension of the model in the absence of \textit{Wolbachia}-carrying mosquitoes given in Equations~\eqref{multiabsSH}--\eqref{multiabsIN} of the previous section. The flowchart for this model is given in Figure \ref{fig:MultiFlowChartPresence}. The two-serotype dengue model in the presence of \textit{Wolbachia}-carrying mosquitoes is governed by Equations~\eqref{MultiW1}--\eqref{MultiWend}. 

The model comprises human, non-\textit{Wolbachia} and \textit{Wolbachia}-carrying mosquito populations. The human and non-\textit{Wolbachia} mosquito compartments are the same as in the model in the absence of \textit{Wolbachia}-carrying mosquitoes. The population of \textit{Wolbachia}-carrying mosquitoes is grouped into six compartments, which are mosquitoes in the aquatic stage, $A_W$, susceptible, $S_W$, exposed to either serotype $i$, $E_W^i$, and infectious with either serotype $i$, $I_W^i$. The total population of \textit{Wolbachia}-carrying  mosquitoes is
\begin{equation}
F_W = S_W+\sum_{i=1}^2\left( E_{W}^i+ I_{W}^i\right),
\end{equation}
and the total population of non-\textit{Wolbachia} mosquitoes is given in Equation~\eqref{eq:FN}.

Unlike in the model in the absence of \textit{Wolbachia}, susceptible humans are exposed to dengue after being bitten by either infectious non-\textit{Wolbachia} or \textit{Wolbachia}-carrying mosquitoes, that is, at a rate governed by
\begin{equation}
\lambda^i_H=\frac{b_NT_{N}^i I_{N}^i} {N_H} + \frac{b_W T_{HW}^i I_{W}^i} {N_H}.
\label{FoIHumanWlb}
\end{equation}
Here $T_{HW}$ is the transmission probability from \textit{Wolbachia}-carrying mosquitoes to humans. The parameter $b_W$ is the successful biting rate of \textit{Wolbachia}-carrying mosquitoes, where $b_W < b_N$  since \textit{Wolbachia} reduces the biting rate~\cite{Turley2009}. Exposed individuals progress to infectious, temporary immunity and secondary susceptible classes at the same rates as in model in the absence of \textit{Wolbachia}-carrying mosquitoes. Individuals who were previously infected by serotype $i$ are re-infected by dengue serotype $j$ at rate
\begin{equation}
\lambda^j_H=\frac{b_NT_{N}^j I_{N}^j} {N_H} + \frac{b_W T_{HW}^j I_{W}^j} {N_H}, 
\label{FoIHumanWlbSecond}
\end{equation}
where $j\neq i$. Infectious individuals then move to secondary infectious and recovered classes at the same rates as in the model in the absence of \textit{Wolbachia}. All humans die at the natural death rate,~$\mu_H$.

\begin{align}
\frac{dS_H}{dt}&=B N_H - \sum_{i=1}^{2} \lambda^i_{H} S_H  -\mu_H S_H,\label{MultiW1}\\
\frac{dE_{H}^i}{dt}&= \lambda^i_{H} S_H - \gamma_H E_{H}^i -\mu_H E_{H}^i,\\
\frac{dI_{H}^i}{dt}&=\gamma_H E_{H}^i- \sigma I_{H}^i  -\mu_H I_{H}^i,\\
\frac{dX_H^i}{dt}&= \sigma I_{H}^i -\theta_i X_H^i -\mu_H X_H^i\,\\
\frac{dS_{H}^{ji}}{dt}&=\theta_i X_H^i -  \lambda^j_H  S_{H}^{ji} -\mu_H S_{H}^{ji},\\
\frac{dE_{H}^{ji}}{dt}&= \lambda^j_H  S_{H}^{ji} -\gamma_H E_{H}^{ji} -\mu_H E_{H}^{ji},\\
\frac{dI_{H}^{ji}}{dt}&= \gamma_H E_{H}^{ji}- \sigma I_{H}^{ji} -\mu_H I_{H}^{ji},\\
\frac{dR_{H}}{dt}&=\mathop{\sum_{j=1}^{2}}_{j\neq i}\sigma  I_{H}^{ji}  -\mu_H R_H,\\
\frac{dA_N}{dt}&=\frac{\rho_N F_N^2}{2\left(F_N + F_W\right)}\left(1-\frac{A_N+A_W}{K}\right)-\left(\tau_N + \mu_{NA}\right),\\
\frac{dS_N}{dt}&=\frac{\tau_N A_N}{2} +\frac{\left(1-\alpha\right)\tau_W A_W}{2} -\sum_{i=1}^{2}\lambda_{N}^i S_N -\mu_N(t) S_N,\\
\frac{dE_{N}^i}{dt}&=\lambda_{N}^i S_N -\gamma_N E_{N}^i-\mu_N(t) E_{N}^i,\\
\frac{dI_{N}^i}{dt}&=\gamma_{N} E_{N}^i -\mu_N(t) I_{N}^i,\\
\frac{dA_W}{dt}&=\frac{\rho_W F_W}{2}\left(1-\frac{A_N+A_W}{K}\right) -\mu_{WA} A_W -\tau_W A_W,\\
\frac{dS_W}{dt}&= \frac{\alpha \tau_W A_W}{2} - \sum_{i=1}^{2}\lambda_{W}^i S_W -\mu_W S_W,\\
\frac{dE_W^i}{dt}&= \lambda_{W}^i S_W -\gamma_W E_{W}^i-\mu_W E_{W}^i,\\
\frac{dI_W^i}{dt}&=\gamma_W E_{W}^i -\mu_W I_{W}^i.\label{MultiWend}
\end{align}
In the presence of \textit{Wolbachia}, the effects of cytoplasmic incompatibility (CI) on mosquito reproduction are included in the model.  Non-\textit{Wolbachia} mosquitoes are produced when non-\textit{Wolbachia} females and males mate, and the growth of the aquatic non-\textit{Wolbachia} mosquito population is limited by the carrying capacity, $K$, through the expression
\begin{equation}
\frac{\rho_N F_N M_N}{\left(F_N + M_N + F_W + M_W\right)}\left(1 - \frac{A_N + A_W}{K}\right).
\end{equation}
As we assume that there are equal numbers of male and female mosquitoes, $F_N =M_N$ and $F_W =M_W$ , this expression reduces to
\begin{equation}
\frac{\rho_N F_N^2}{2\left(F_N + F_W\right)}\left(1 - \frac{A_N + A_W}{K}\right).
\end{equation}
\textit{Wolbachia}-carrying mosquitoes are reproduced after \textit{Wolbachia}-carrying females mate with either \textit{Wolbachia}-carrying  or non-\textit{Wolbachia} males, according to the expression

\begin{equation}
\frac{\rho_W F_W \left(M_N+ M_W\right)}{\left(F_N + M_N + F_W + M_W\right)}\left(1 - \frac{A_N + A_W}{K}\right),
\label{eq:ReproW}
\end{equation}
which, can be similarly reduced to
\begin{equation}
\frac{\rho_W F_W}{2}\left(1 - \frac{A_N + A_W}{K}\right).
\end{equation}
Non-\textit{Wolbachia} mosquitoes in the aquatic stage mature into susceptible non-\textit{Wolbachia} mosquitoes at rate $\tau_N$. \textit{Wolbachia}-carrying mosquitoes in the aquatic stage mature into susceptible mosquitoes at rate $\tau_W$, with $\alpha$ of them becoming \textit{Wolbachia}-carrying mosquitoes and the remaining $1-\alpha$ becoming non-\textit{Wolbachia} mosquitoes. Mosquitoes in the aquatic stage die at rate $\mu_{NA}$.
Susceptible mosquitoes are exposed to dengue after biting infectious humans with either primary or secondary infections at rates
\begin{equation}
\lambda^i_{W}=\frac{b_W T_{N}^i I_{H}^i}{NH}+\frac{\phi^i b_W T_{N}^i  I_{H}^{ij}}{NH},
\label{eq:lambdaW}
\end{equation}
for \textit{Wolbachia}-carrying mosquitoes and $\lambda^i_{N}$ as given in Equation~\eqref{eq:lambdaN} for non-\textit{Wolbachia} mosquitoes. The transmission probability from humans to \textit{Wolbachia}-carrying mosquitoes is assumed to be the same as that between humans and non-\textit{Wolbachia} mosquitoes, which is represented by $T_N$. The exposed mosquitoes progress to the infectious class at rate $\gamma_N$ for non-\textit{Wolbachia} mosquitoes and $\gamma_W$ for \textit{Wolbachia}-carrying mosquitoes. 


\section{Parameter Descriptions}
Parameter descriptions, values and sources are given in Table~\ref{parametervaluesmulti}. Most of parameters are the same as in Chapter~\ref{chap:singleintro}, but we remind the reader of the relationships between and the values of several important parameters. The transmission probabilities between humans and non-\textit{Wolbachia} mosquitoes, and from humans to \textit{Wolbachia}-carrying mosquitoes are assumed to be equal and are represented by $T_N$. However, the transmission probability from \textit{Wolbachia}-carrying mosquitoes to humans, $T_{HW}$, is lower than $T_N$ since \textit{Wolbachia} reduces the level of dengue virus in the salivary glands of the mosquito. The biting and reproductive rates of \textit{Wolbachia}-carrying mosquitoes are reduced, and hence $b_W<b_N$ and $\rho_W<\rho_N$, respectively. Furthermore, because \textit{Wolbachia} reduces the mosquito lifespan, the death rate of the \textit{Wolbachia}-carrying mosquitoes is higher than that of non-\textit{Wolbachia} mosquitoes ($\mu_W>\mu_N$).

The period of temporary immunity is taken to be half a year or 6 months~\cite{Wearing2006}, hence $\theta = 1/(0.5 \times 365)$ day$^{-1}$. As individuals with secondary infections have higher viral loads, this leads to higher transmission rates. Therefore, the antibody-dependent enhancement, $\phi$, is set to $1.1$~\cite{Hu2013}.


\label{tablemulti}
\begin{table}

\small
\centering
   \begin{tabular}{l p{6cm}  p{3cm} l l}
    \hline  \hline
    Symbol & Description & Value& Unit & Source\\
    \hline
    $\alpha$&Maternal transmission&0.9& N/A&\cite{Walker2011, Hoffmann1990, Ndii2012}\\
     $B$     & Human birth rate & $1/(70 \times 365)$& day$^{\textrm{-1}}$ &\cite{WHO1}\\
     $b_N$ & Biting rate of non-W mosquitoes&0.63&day$^{\textrm{-1}}$&\cite{Scott2000}\\
     $b_W$ & Biting rate of W mosquitoes&0.95 $b_N$&day$^{\textrm{-1}}$&\cite{Turley2009}\\ 
     $\eta$& Strength of seasonality& 0.6228& N/A&Fitted\\
      $\gamma_H$&Progression rate from exposed to infectious & 1/5.5 & day$^{\textrm{-1}}$&\cite{Gubler1998}\\
       $\gamma_N$&Progression from exposed to infectious class of non-W mosquitoes & 1/10 &  day$^{\textrm{-1}}$&\cite{Chowel2007}\\
       $\gamma_W$&Progression from exposed to infectious class of W mosquitoes&1/10& day$^{\textrm{-1}}$&\cite{Chowel2007}\\
     $K$ & Carrying capacity&$3\times N_H$&N/A&\cite{Chowel2007}\\
     $\lambda$& Force of infection&Eqs~\eqref{eq:FoIHuman}, \eqref{eq:lambdaN}, \eqref{FoIHumanWlb},\eqref{eq:lambdaW}&\\
      $\mu_{N0}$&Average adult mosquito death rate (non-W) & 1/14 & day$^{\textrm{-1}}$ &\cite{Yang2009}\\
      
     $\mu_H$ & Human death rate	& 1/$(70\times 365)$& day$^{\textrm{-1}}$ &\cite{WHO1}\\
     $\mu_{NA}$&Death rate of aquatic non-W mosquitoes& 1/14& day$^{\textrm{-1}}$&\cite{Yang2009}\\
      $\mu_W(t)$&Adult aquatic death rate& 1.1$\mu_N(t)$&day$^{\textrm{-1}}$&\cite{Yeap2011, Walker2011}\\
      $\mu_{WA}$&Death rate of W mosquitoes &1/14&day$^{\textrm{-1}}$&\cite{Yang2009}\\
    $\omega$ &Phase& 80.61& day &Fitted\\
     $\phi$& ADE & 1.1& N/A&\cite{Hu2013}\\
      $\rho_N$&Reproductive rate of non-W mosquitoes& 1.25 &day$^{\textrm{-1}}$&\cite{Ndii2012}\\
    $\rho_W$&Reproductive rate of W-mosquitoes& $0.95\rho_N$&day$^{\textrm{-1}}$&\cite{Walker2011}\\
    
     $\sigma$&Recovery rate& 1/5 &day$^{\textrm{-1}}$ & \cite{Gubler1998} \\
     $T_{N}$& Transmission probability from non-W mosquitoes to human  & 0.2614& N/A &Fitted\\
          $T_{HW}$&Transmission probability from W mosquitoes to human&  $0.5 T_N$& N/A&\cite{Bian2010, Ndii2015}\\
     $\theta$& Progression rate from temporary immunity class to susceptible class& $1/(0.5\times 365)$&day$^{\textrm{-1}}$ &\cite{Wearing2006}\\
     
     $\tau_N$&Maturation rate of non-W mosquitoes& 1/10& day$^{\textrm{-1}}$ &\cite{Yang2009} \\   
     $\tau_W$&Maturation rate of W mosquitoes& 1/10 & day$^{\textrm{-1}}$&\cite{Yang2009}\\
  \hline
  \end{tabular}
 \caption[Parameter descriptions, values and sources for dengue models in the absence and presence of  \textit{Wolbachia}-carrying mosquitoes with two dengue serotypes]{Parameter descriptions, values and sources for both models. Further explanation of the parameter values is given in the text. Note that W is used to indicate \textit{Wolbachia}-carrying mosquitoes in the parameter descriptions and N is for non-\textit{Wolbachia}. $N_H = 1.5\times 10^5$. The phase is shifted 60 days as we start from January.}
 \label{parametervaluesmulti}
 \end{table}
\newpage
\section{Summary}
\begin{itemize}[wide=0pt,labelwidth = 1.3333em, labelsep = 0.3333em, leftmargin = \dimexpr\labelwidth + \labelsep\relax ]
\item Two-serotype dengue models in the absence and presence of \textit{Wolbachia}-carrying mosquitoes are developed in Sections~\ref{sec:multiabsencemodel} and ~\ref{MultiWolbachiaMOdel}, Equations~\eqref{multiabsSH}--\eqref{multiabsIN} and \eqref{MultiW1}--\eqref{MultiWend}, respectively.
\item In the absence of \textit{Wolbachia}-carrying mosquitoes, individuals are infected after being bitten by infectious non-\textit{Wolbachia} mosquitoes.
\item In the presence of \textit{Wolbachia}-carrying mosquitoes, individuals are infected after being bitten by infectious non-\textit{Wolbachia} or \textit{Wolbachia}-carrying mosquitoes.
\item Antibody dependent enhancement (ADE) is included, where the ADE is represented by $\phi$. Antibody dependent enhancement increases infectivity of individuals with secondary infections. This can lead to more severe forms of dengue.
\item The temporary immunity class is included in the model.  Individuals in this class have recovered from primary infections and have a temporary immunity to all serotypes of dengue.  They later become susceptible to the  serotypes with which they have not previously been infected. 
\end{itemize}

\chapter[Two Serotypes with Symmetric Epidemiological Characteristics]{Two serotypes with symmetric epidemiological characteristics \footnote{The results of scenario one  form part of the manuscript that has been submitted for publication as M.Z. Ndii, D. Allingham, R.I. Hickson, K. Glass. \textit{The effect of \textit{Wolbachia} on dengue dynamics in the presence of two serotypes of dengue: symmetric and asymmetric epidemiological characteristics.} }}\label{chap:symmetry} 


\ifpdf
    \graphicspath{{7/figures/PNG/}{7/figures/PDF/}{7/figures/}}
\else
    \graphicspath{{7/figures/EPS/}{7/figures/}}
\fi

In Chapters~\ref{chap:singleintro}--\ref{chap:multipleintro}, we studied the effect of \textit{Wolbachia} on dengue dynamics when a single serotype of dengue was circulating.   We considered scenarios when  dengue cases were introduced into the population once, and multiple times. In this chapter and the next, we use the mathematical models formulated in Chapter~\ref{chap:modelmultiplestrains} to study the effect of \textit{Wolbachia} on dengue dynamics when two serotypes of dengue are circulating, assuming  first symmetric and then asymmetric epidemiological characteristics of the serotypes.  The aim of this chapter is to investigate the effectiveness of \textit{Wolbachia} in reducing dengue transmission for two serotypes with symmetric epidemiological characteristics under different disease introduction scenarios.


\section{Introduction}
Several mathematical models have been developed to investigate dengue dynamics in the absence of  \textit{Wolbachia}-carrying mosquitoes when more than one dengue serotype circulates~\cite{Ferguson1999,Cummings2005,Bianco2009,Adams2006,Romero2013,Kooi2014,Aquiar2011,Hu2013,Woodall2014}. However, these models did not consider differing disease introduction scenarios, a factor which can influence disease transmission dynamics. In addition, to the best of our knowledge, no modelling studies have been conducted to investigate the effects of \textit{Wolbachia} on dengue dynamics when more than one dengue serotype circulates under different disease introduction scenarios.

Understanding the effect of different disease introduction scenarios on  \textit{Wolbachia} performance in reducing dengue incidence is of importance, because the exposure of the population to different dengue serotypes may alter the disease dynamics.

In this chapter, we assume that the two circulating dengue serotypes have identical epidemiological characteristics (the symmetric case).

\section{Scenarios of Dengue Introduction}\label{ScenarioIntroMulti}

Deciding on an exact strategy for the introduction of dengue serotypes into the population is a challenging task because  individuals with a particular dengue serotype enter the population at irregular intervals~\cite{Limkittikul2014}. Furthermore, the dominant dengue serotype in any one location varies over time. For example in Thailand, DEN3 was common during 2002--2004 and 2008--2010, whereas DEN4 was common during 2003--2008~\cite{Limkittikul2014}.

\begin{figure}[ht]
\centering
\includegraphics[width=1\linewidth]{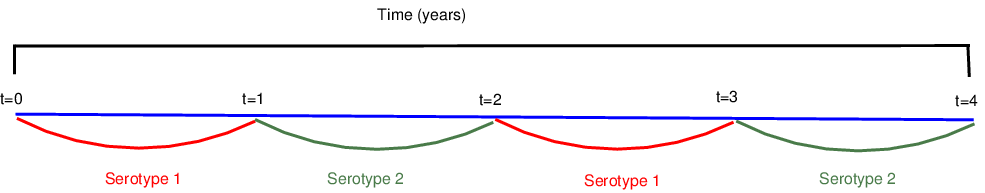}
\caption[Dengue introduction scenario 1: each dengue serotype is introduced in alternate year]{Introduction scenario 1: Individuals with a particular dengue serotype are introduced weekly in alternate years. In this scenario, individuals carrying dengue serotype 1 are introduced weekly for one year, and then individuals carrying dengue serotype 2 are introduced weekly in the subsequent year. Throughout this chapter, the terminology, scenario 1, or, first scenario, refers to this disease
introduction strategy. In this figure, $t$ denotes the time in years.}
\label{Scenario1}
\end{figure}
\begin{figure}[ht]
\centering
\includegraphics[width=1\linewidth]{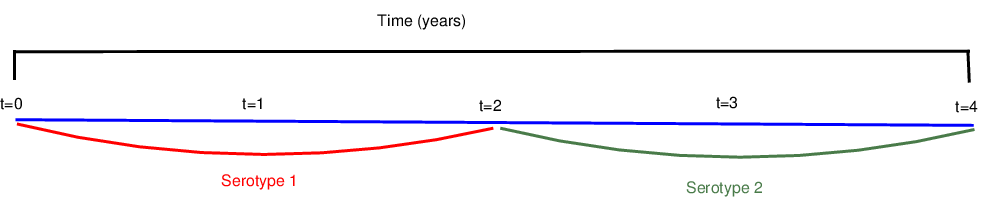}
\caption[Dengue introduction scenario 2: one dengue serotype is introduced weekly for two years and then another dengue serotype is introduced weekly for another two years]{Introduction scenario 2: Individuals with a particular dengue serotype are introduced weekly, where individuals infected carrying dengue serotype 1 are introduced weekly for two years, and then individuals infected carrying dengue serotypes 2 are introduced weekly for two years. Throughout this chapter, the terminology, scenario 2, or, second scenario, refers to this disease introduction strategy. In this figure, $t$ is the time in years.}
\label{Scenario2}
\end{figure}

Here we consider two disease introduction scenarios as illustrated in Figures~\ref{Scenario1} and \ref{Scenario2}. In the first scenario, individuals carrying dengue serotype 1 are introduced weekly over a one year period, and individuals carrying another dengue serotype are introduced weekly over an additional one year period (Figure~\ref{Scenario1}). In the second scenario, individuals carrying dengue serotype 1 are introduced weekly for a duration of two years, and then individuals carrying another dengue serotype are introduced weekly over an additional two year period (Figure~\ref{Scenario2}).  These patterns are then repeated for many years. Throughout this chapter and in Chapter~\ref{chap:asymmetry}, serotypes 1 and 2 refer to two different dengue serotypes, but do not specifically refer to DEN1 and DEN2.

We use simulations to investigate the effect of \textit{Wolbachia} when two dengue serotypes circulate under two disease introduction scenarios. Our simulations are conducted in the same way as that  presented in Chapter~\ref{chap:multipleintro}, Page~\pageref{simulationprocedure}, except for differences in the dengue serotypes,  depending on the disease introduction scenario.  The effects of the antibody-dependent enhancement rate ($\phi$) and the transmission probability ($T_N$) on \textit{Wolbachia} performance in reducing dengue are examined for all scenarios.

\begin{figure}[ht]
  \begin{center}
  \includegraphics[width=0.95\linewidth]{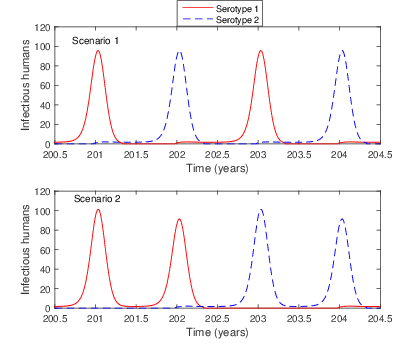}
  \end{center}
  \caption[Time series plot for dengue introduction scenario 1 (top) and 2 (bottom)]{Time series plot for dengue introduction scenario 1 (top) and 2 (bottom). The solid red line represents the outbreak due to serotype 1 and the dashed blue line is the outbreak due to serotype 2.}
  \label{ScenarioIlustration}
\end{figure}

The effectiveness of the \textit{Wolbachia} intervention is measured by comparing the relative differences in the outbreak sizes in the absence and presence of \textit{Wolbachia}-carrying mosquitoes, which is given by
\begin{equation}
\kappa=\left(\frac{H_A-H_P}{H_A}\right)100\% \,.
\end{equation}
Here $H$ is the final number of dengue cases in the human population, with subscripts to denote in the absence (A) and presence (P) of \textit{Wolbachia}. In order to ensure that the two disease introduction scenarios are comparable, a four-year outbreak size is used to calculate the proportional reduction in dengue due to \textit{Wolbachia}.  This gives two outbreaks of dengue caused by each serotype in both disease introduction scenarios as shown in Figure~\ref{ScenarioIlustration}.

\section{Results}

The performance of \textit{Wolbachia} in reducing dengue is explored under two disease introduction scenarios. The antibody-dependent enhancement (ADE) and the transmission probabilities for the two serotypes are set to be equal. That is, $\phi^1=\phi^2 = \phi$ and $T_N^1=T_N^2 = T_N$. The parameter $T_N$ is the transmission probability between humans and non-\textit{Wolbachia}  mosquitoes and from humans to \textit{Wolbachia}-carrying mosquitoes. The transmission probability from \textit{Wolbachia}-carrying mosquitoes to humans is set to $T_{HW}= 0.5 T_N$, because \textit{Wolbachia} reduces the level of dengue virus in mosquitoes~\cite{Bian2010, Moreira2009, Hofman2011, Walker2011, Frentiu2014, Ye2015} and hence reduces transmission.

\subsection{First Dengue Introduction Scenario}\label{1stScenario}

\begin{figure}[h!]
  \begin{center}
  \includegraphics[width=0.95\linewidth]{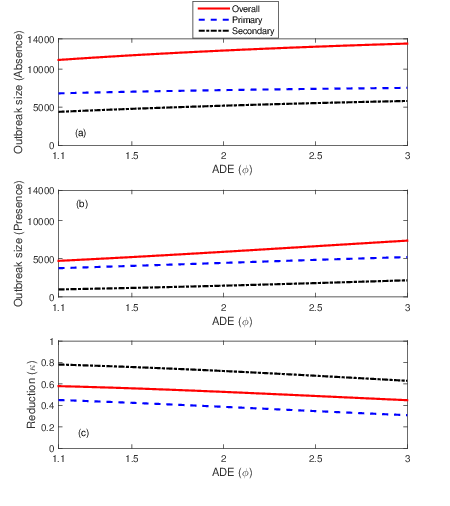}
  \end{center}
  \caption[The effect of changes in the ADE rate for both dengue serotypes under the first scenario of dengue introduction]{The effect of changes in the ADE rate for both dengue serotypes under the first scenario of dengue introduction. All plots show overall (solid red lines), primary (blue dashed line) and secondary (black dash-dot line) infections. Plots (a) and (b) show the outbreak size in the absence and presence of \textit{Wolbachia}-carrying mosquitoes, respectively. The plot (c)  shows the proportional reduction in dengue due to \textit{Wolbachia}.}
  \label{Thesis_Scenario1_Panel}
\end{figure}

When we explore the case where the dengue serotypes have the same ADE, we find that as the ADE increases, the overall outbreak size and the outbreak sizes for the primary and secondary infections in the absence and presence of \textit{Wolbachia}-carrying mosquitoes increase.  In addition, the proportional reductions in the overall numbers of dengue cases, and  the reductions in the numbers of primary and secondary infections due to the \textit{Wolbachia} intervention  decrease (Figure~\ref{Thesis_Scenario1_Panel}). The proportional reduction in the incidence of secondary infections is higher than that of primary infections, with the maximum reduction in secondary infections found to be approximately 78\%, compared to a reduction in primary infections of around 45\%.

\begin{figure}[h!]
  \begin{center}
  \includegraphics[width=0.95\linewidth]{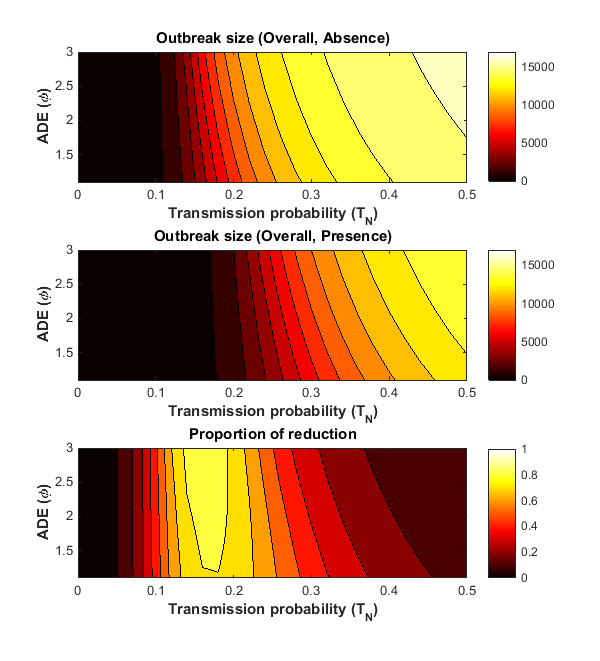}
  \end{center}
  \caption[Contour plots showing simultaneous changes to the ADE rate and the transmission probability for the first scenario of dengue introduction.]{Contour plots showing simultaneous changes to the ADE ($\phi$) and the transmission probability ($T_N$) for the first scenario of dengue introduction. The top and middle plots give overall outbreak sizes in the absence and presence of \textit{Wolbachia}-carrying mosquitoes, and the bottom plot shows the overall reduction in dengue due to \textit{Wolbachia}.}
  \label{Thesis_Scenario1_Contour_Overall}
\end{figure}

\begin{figure}[h!]
  \begin{center}
 \includegraphics[width=0.95\linewidth]{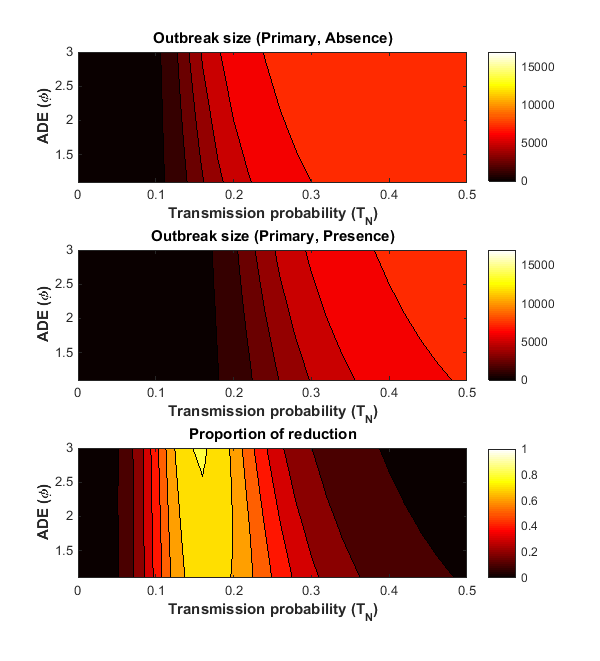}
  \end{center}
  \caption[Contour plots showing simultaneous changes to the ADE rate ($\phi$) and the tranmission probability ($T_N$) for primary infections under the first scenario of dengue introduction]{Contour plots showing simultaneous changes to the ADE rate ($\phi$) and the transmission probability ($T_N$) for primary infections under the first scenario of dengue introduction. The top and middle plots give the outbreak sizes for  primary infections in the absence and presence of \textit{Wolbachia}-carrying mosquitoes, and the bottom plot shows the overall reduction in primary infections due to \textit{Wolbachia}.}
  \label{Thesis_Scenario1_Contour_Primary}
\end{figure}

\begin{figure}[h!]
  \begin{center}
  \includegraphics[width=0.95\linewidth]{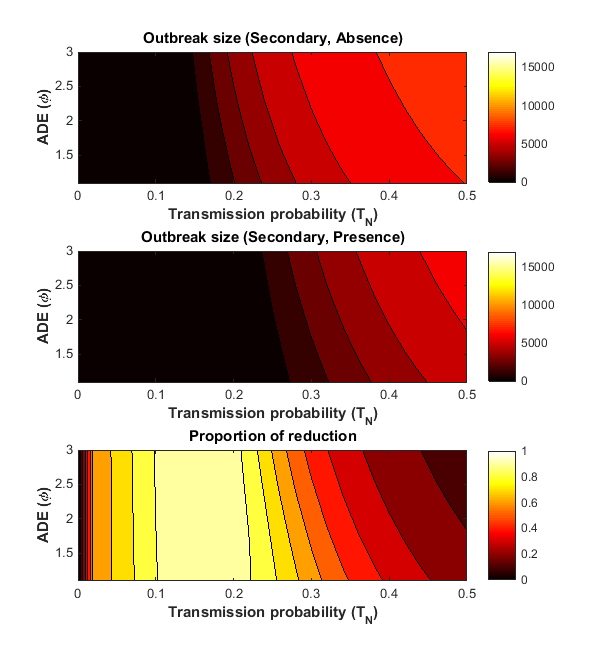}
  \end{center}
  \caption[Contour plots showing simultaneous changes to the ADE ($\phi$) and the transmission probability ($T_N$) under the first scenario of dengue introduction in the secondary infections]{Contour plots showing simultaneous changes to the ADE ($\phi$) and the transmission probability ($T_N$) under the first scenario of dengue introduction. The top and middle plots give the outbreak sizes for secondary infections in the absence and presence of \textit{Wolbachia}-carrying mosquitoes, and the bottom plot shows the reduction in secondary infections due to \textit{Wolbachia}.}
  \label{Thesis_Scenario1_Contour_Secondary}
\end{figure}

When we vary the ADE and the transmission probability simultaneously, we find that the transmission probability has the greater  effect on the performance of \textit{Wolbachia} in reducing dengue incidence  (see Figures~\ref{Thesis_Scenario1_Contour_Overall}--\ref{Thesis_Scenario1_Contour_Secondary}). Furthermore, lower transmission probability values do not result in outbreaks.   In the presence of \textit{Wolbachia}-carrying mosquitoes, the range of transmission probabilities for which epidemics do not occur is larger than that in the absence of \textit{Wolbachia}-carrying mosquitoes (for overall cases, it is around 0--0.1 in the absence of \textit{Wolbachia}, compared to 0--0.18 in the presence of \textit{Wolbachia}-carrying mosquitoes).  This means that the presence of \textit{Wolbachia}-carrying mosquitoes raises the threshold at which epidemics occur. The maximum overall reduction in dengue incidence due to \textit{Wolbachia} is around 70-80\%, which occurs when the transmission probability lies in the range 0.14--0.22.  In this range of transmission probabilities,  epidemics either do not occur or the outbreak size is smaller when  \textit{Wolbachia}-carrying mosquitoes are present (Figures~\ref{Thesis_Scenario1_Contour_Overall}). Moreover, as the transmission probability increases, the reduction in secondary infections due to \textit{Wolbachia} is greater than that of primary infections (Figures~\ref{Thesis_Scenario1_Contour_Primary} and \ref{Thesis_Scenario1_Contour_Secondary}).  This reduction in secondary infections can reach up to 90\%.

\subsection{Second Dengue Introduction Scenario of Dengue}\label{2ndScenarioIntro}

\begin{figure}[h!]
  \begin{center}
  \includegraphics[width=0.95\linewidth]{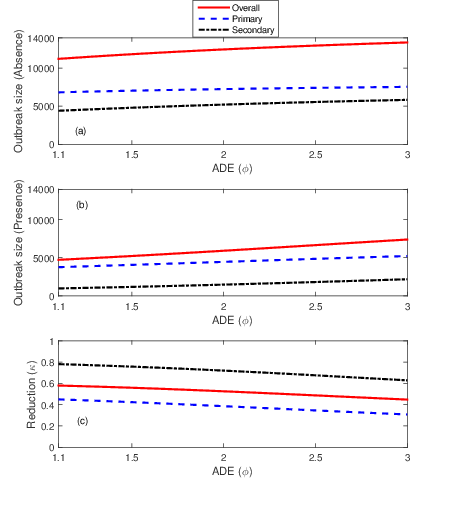}
  \end{center}
  \caption[The effect of changes in ADE for both serotypes in dengue cases under second scenario of dengue introduction]{The effect of changes in the ADE rate for both dengue serotypes under the second scenario of dengue introduction. All plots show overall (solid red lines), primary (blue dashed line) and secondary (black dash-dot line) infections. Plots (a) and (b) show the outbreak size in the absence and presence of \textit{Wolbachia}-carrying mosquitoes, respectively. Plot (c)  shows the proportional reduction in dengue due to \textit{Wolbachia}.}
  \label{Thesis_Scenario2_Panel}
\end{figure}

The performance of \textit{Wolbachia} in reducing dengue in the second scenario of dengue introduction is similar to that in the first scenario. An increase in the ADE rate leads to an increase in the outbreak size of dengue infections in both in the absence and in the presence of \textit{Wolbachia}-carrying mosquitoes (Figure~\ref{Thesis_Scenario2_Panel}).  When the ADE is higher, the level of reduction in dengue incidence caused by \textit{Wolbachia} slightly decreases. The reduction in the number of secondary infections is higher than that of primary infections, with the level of overall reduction and reduction in primary and secondary infections  similar to those found for the first scenario of disease introduction.

\begin{figure}[h!]
  \begin{center}
  \includegraphics[width=0.95\linewidth]{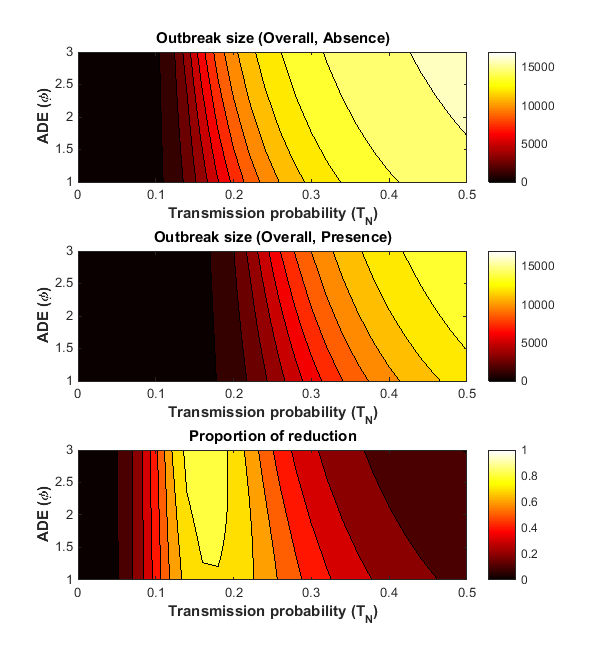}
  \end{center}
  \caption[Contour plots showing simultaneous changes to the ADE and the transmission probability under the second scenario of dengue introduction. ]{Contour plots showing simultaneous changes to the ADE ($\phi$) and the transmission probability ($T_N$) under the second scenario of dengue introduction. The top and middle plots give overall outbreak size in the absence and presence of \textit{Wolbachia}-carrying mosquitoes, and the bottom plot shows the overall reduction in dengue due to \textit{Wolbachia}.}
  \label{Thesis_Scenario2_Contour_Overall}
\end{figure}

\begin{figure}[h!]
  \begin{center}
  \includegraphics[width=0.95\linewidth]{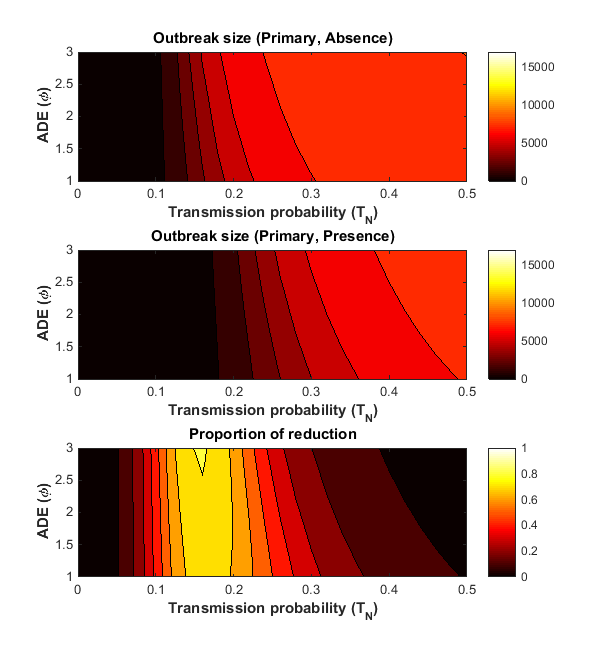}
  \end{center}
  \caption[Contour plots showing simultaneous changes to the ADE rate and the transmission probability for primary infections under the second scenario of dengue introduction]{Contour plots showing simultaneous changes to the ADE ($\phi$) and the transmission probability ($T_N$) under the second scenario of dengue introduction. The top and middle plots give the outbreak sizes for primary infections in the absence and presence of \textit{Wolbachia}-carrying mosquitoes, and the bottom plot shows the reduction in primary infections due to \textit{Wolbachia}.}
  \label{Thesis_Scenario2_Contour_Primary}
\end{figure}

\begin{figure}[h!]
  \begin{center}
  \includegraphics[width=0.95\linewidth]{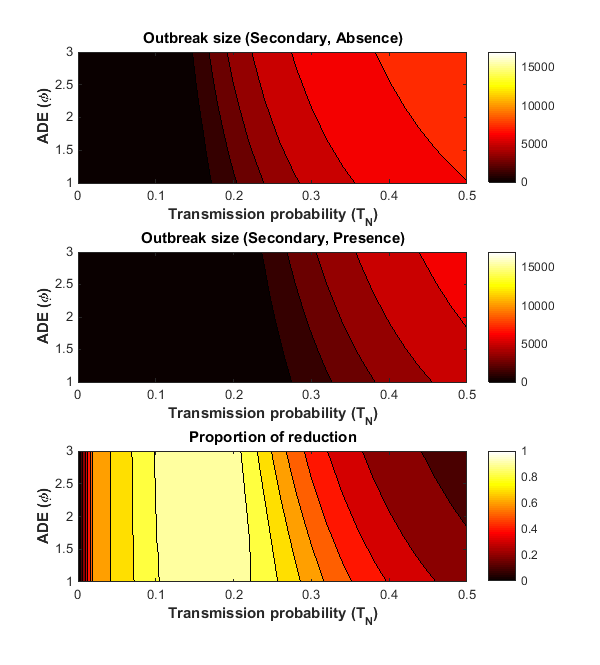}
  \end{center}
  \caption[Contour plots showing simultaneous changes to the ADE and the transmission probability for secondary infections under the second scenario of dengue introduction]{Contour plots showing simultaneous changes to the ADE ($\phi$) and the transmission probability ($T_N$) under the second scenario of dengue introduction. The top and middle plots give  outbreak sizes for  secondary infections in the absence and presence of \textit{Wolbachia}-carrying mosquitoes, and the bottom plot shows the reduction in secondary infections due to \textit{Wolbachia}.}
  \label{Thesis_Scenario2_Contour_Secondary}
\end{figure}

When the ADE and the transmission probability are varied simultaneously, we find that the transmission probability has more impact on the performance of \textit{Wolbachia} in reducing dengue incidence than the ADE. The presence of \textit{Wolbachia} raises the threshold at which epidemics occur (see Figures~\ref{Thesis_Scenario2_Contour_Overall}--\ref{Thesis_Scenario2_Contour_Secondary}). The maximum overall reduction in dengue reaches 80\% for a similar range of  transmission probabilities (around 0.14-0.22) to the one found for the first disease introduction scenario  (Figure~\ref{Thesis_Scenario2_Contour_Overall}).  The effects of ADE become stronger for higher values of the transmission  probability.  The proportional reduction in secondary infections is higher than that for primary infections.   The maximum reduction in primary infections is 80\%, compared to more than 90\% for secondary infections.

\section{Discussion and Conclusions}

In this chapter, we analysed the effect of \textit{Wolbachia} on dengue dynamics under different disease introduction scenarios, when two dengue serotypes with symmetric epidemiological characteristics are circulating. We found that the benefits of \textit{Wolbachia} are similar regardless of the disease introduction scenario.  Since the time of  introduction of a particular dengue serotype into the population is very  uncertain, it is not easy to predict the dominant serotype before an epidemic occurs. Predicting the dominant dengue serotype before an epidemic takes off may be important from the public health perspective as the presence of different serotypes may change the severity of the epidemic~\cite{Nishiura2007, Tricou2011, Vaughn2000}.  However, our findings suggest that uncertainty around dengue serotypes is not a hurdle for the implementation of the \textit{Wolbachia} intervention. Although all dengue serotypes circulate in the population, and the times at which individuals infected with different dengue serotypes enter the population are irregular, \textit{Wolbachia} will still reduce the number of dengue cases. This is particularly true for secondary infections which can result in more severe forms of dengue.

Our results for both disease introduction scenarios  show that the outbreak size increases as the transmission probability increases. Furthermore, we found that the ADE does not have a noticeable impact on the effectiveness of \textit{Wolbachia} unless the transmission probability is high.  The maximum overall reduction in dengue cases is obtained for transmission probabilities in the range 0.14-0.22, and a considerable reduction in secondary infections of 70-90\% is achieved. However, if the transmission probability is high, reductions in dengue incidence of only 20-40\% are likely.  This implies that \textit{Wolbachia} will be most effective in reducing dengue transmission if dengue serotypes are not strongly transmissible, a similar finding to the results of Chapter~\ref{chap:multipleintro} for a single serotype.   As the transmission probability is one of the parameters that regulates the basic reproduction number, our results are consistent with the finding by Hughes and Britton~\cite{Hughes2013} and Ferguson \textit{et al}~\cite{Ferguson2015} that \textit{Wolbachia} will be effective if the basic reproduction number is not too high. Our finding that differences in the ADE levels between serotypes have much less impact on dengue dynamics than differences in the transmission probability indicates that a better understanding of serotype-specific transmission probabilities may be needed to optimize the delivery of the \textit{Wolbachia} intervention.

In order to confirm our results, two further scenarios of dengue introductions were explored: introducing individuals carrying serotype~1 for the first six months and individuals carrying another dengue serotype for a further six months, and introducing individuals carrying dengue serotypes~1 and~2 at the same time. We found similar results to those presented in this chapter.

Research has shown that the epidemiological characteristics of dengue serotypes may be different~\cite{Vaughn2000, Althouse2014, Barraquer2014, Tricou2011}. Although our results in this chapter, assuming symmetric epidemiological characteristics, provide insights into the effectiveness of \textit{Wolbachia}  in reducing human dengue incidence in the presence of more than one dengue serotype, an analysis of asymmetric epidemiological characteristics is of importance to further our understanding of \textit{Wolbachia} performance in reducing human dengue incidence. The effects of \textit{Wolbachia} on dengue transmission dynamics when the epidemiological  characteristics  between serotypes differ are explored in the next chapter.

\section{Summary}
\begin{itemize}[wide=0pt,labelwidth = 1.3333em, labelsep = 0.3333em, leftmargin = \dimexpr\labelwidth + \labelsep\relax ]
\item We investigate the effect of \textit{Wolbachia} in reducing dengue incidence under different disease introduction scenarios when the antibody-dependent enhancement (ADE) and the transmission probability of serotypes are the same.
\item We found that different disease introduction scenarios do not affect the performance of \textit{Wolbachia} in reducing dengue incidence (Sections~\ref{1stScenario} and~\ref{2ndScenarioIntro} ). 
\item The level of ADE does not have a noticeable effect on dengue transmission dynamics (Sections~\ref{1stScenario} and~\ref{2ndScenarioIntro}, see Figures~\ref{Thesis_Scenario1_Panel}--\ref{Thesis_Scenario2_Contour_Secondary}).
\item The proportional reduction in secondary infections (up to 90\%) is higher than that in the primary infections (up to 80\%; Sections~\ref{1stScenario} and~\ref{2ndScenarioIntro}, see Figures~\ref{Thesis_Scenario1_Panel}--\ref{Thesis_Scenario2_Contour_Secondary}).
\end{itemize}


\chapter[Two Serotypes with Asymmetric Epidemiological Characteristics]{Two Serotypes with Asymmetric Epidemiological Characteristics\footnote{Results and discussions in this chapter form part of the manuscript that has been submitted for publication as M.Z. Ndii, D. Allingham, R.I. Hickson, K. Glass. The effect of \textit{Wolbachia} on dengue dynamics in the presence of two serotypes of dengue: symmetric and asymmetric epidemiological characteristics.}}\label{chap:asymmetry} 


\ifpdf
    \graphicspath{{8/figures/PNG/}{8/figures/PDF/}{8/figures/}}
\else
    \graphicspath{{8/figures/EPS/}{8/figures/}}
\fi

In the previous chapter, we studied the effect of \textit{Wolbachia} on dengue transmission dynamics under different disease introduction scenarios when the epidemiological characteristics of the serotypes were assumed to be symmetric.   In this chapter, we study the effect of \textit{Wolbachia} on dengue transmission dynamics when the epidemiological characteristics of the serotypes are asymmetric. That is, we explore the case where one serotype is more transmissible or has a higher   ADE  than the other. 

\section{Introduction}
Research has shown that the characteristics of dengue serotypes may be different~\cite{Vaughn2000, Althouse2014, Barraquer2014, Tricou2011, Nishiura2007, Ferguson1999a}. For example, the severity of disease varies between serotypes~\cite{Vaughn2000}, DEN1 may result in more severe symptoms than DEN4~\cite{Nishiura2007}, and infections with DEN1 may result in higher viraemia levels than infections with DEN2~\cite{Tricou2011}. These differences may affect the performance of \textit{Wolbachia} in reducing dengue transmission.

Modelling studies have investigated the effects of asymmetric epidemiological characteristics on dengue dynamics in the absence of  \textit{Wolbachia}-carrying mosquitoes~\cite{Romero2013, Kooi2014}. However, little has been done to investigate the effects of asymmetric epidemiological characteristics of dengue serotypes on the performance of \textit{Wolbachia} in reducing dengue transmission.  In this chapter, we  use the mathematical models formulated in Chapter~\ref{chap:modelmultiplestrains} to study the effects of \textit{Wolbachia} in reducing dengue incidence when the epidemiological characteristics of the two circulating serotypes differ, Equations~\eqref{multiabsSH}--\eqref{multiabsIN} and Equations~\eqref{MultiW1}--\eqref{MultiWend}.  The characteristics of interest are the  antibody-dependent enhancement parameter, $\phi$,  and the transmission probability, $T_N$.

  The equations from Chapter~\ref{chap:modelmultiplestrains} describing the forces of infection for humans, non-\textit{Wolbachia} mosquitoes and \textit{Wolbachia}-carrying mosquitoes are, respectively, 

\begin{equation}
\lambda^i_H=\frac{b_NT_{N}^i I_{N}^i} {N_H} + \frac{b_W T_{HW}^i I_{W}^i} {N_H},
\label{eq:FoIHumanAssChap}
\end{equation}

\begin{equation}
\lambda^i_{N}=\frac{b_N T_{N}^i I_{H}^i}{N_H}+\frac{\phi^i b_N T_{N}^i  I_{H}^{ij}}{N_H},
\label{eq:FoIMossiesAssChap}
\end{equation}

\begin{equation}
\lambda^i_{W}=\frac{b_W T_{N}^i I_{H}^i}{N_H}+\frac{\phi^i b_W T_{N}^i  I_{H}^{ij}}{N_H}.
\label{eq:FoIMossiesWAssChap}
\end{equation}
Here $T_N^i$ is the transmission probability of serotype $i$ between humans and non-\textit{Wolbachia} mosquitoes, and from humans to \textit{Wolbachia}-carrying mosquitoes. The parameter $\phi^i$ is the  ADE for serotype $i$, and the parameters $b_N$ and $b_W$ are the biting rates of non-\textit{Wolbachia} and \textit{Wolbachia}-carrying mosquitoes, respectively. The parameter $T_{HW}^i$ is the transmission probability from \textit{Wolbachia}-carrying mosquitoes with serotype $i$ to humans, and it is set to $T_{HW}^i = 0.5 T_N^i$. In our investigations, the  ADE and the transmission probability for serotype~2 are set to be higher than those for serotype~1 as follows:

\begin{equation}
\begin{aligned}
\phi^2&=\phi^1+\epsilon_\phi\,,\\
T_N^2&=T_N^1+\epsilon_T.
\end{aligned}
\end{equation}
Here $\phi^1=1.1$~\cite{Hu2013} and $T_N^1=0.2614$~\cite{Ndii2015}. The parameter  $\epsilon_\phi$ varies between 0 and 1.9 ($0\leq\epsilon_\phi\leq 1.9$)  and  the parameter $\epsilon_T$  varies between 0 and 0.2387 ($0\leq \epsilon_T\leq 0.2387$).   Thus $1.1 \leq \phi^2 \leq 3$ and $0.2614\leq T_N^2\leq 0.5$, which are reasonable ranges for ADE and the transmission probability~\cite{Hu2013,Andraud2012}. Therefore, the force of infection for humans, non-\textit{Wolbachia} mosquitoes and \textit{Wolbachia}-carrying mosquitoes of serotype~2 is higher than that of serotype 1: 
$$
\lambda_H^2\geq \lambda_H^1, \qquad \lambda_N^2 \geq \lambda_N^1 \quad \mbox{  and   } \quad \lambda_W^2 \geq  \lambda_W^1.
$$ 

Since, as we discovered in Chapter~\ref{chap:symmetry}, disease introduction scenarios do not appear to influence the performance of \textit{Wolbachia} in reducing dengue transmission, only one disease introduction scenario is used in this analysis. That is, each dengue serotype is introduced weekly in alternate years. Although the strategy of dengue introductions is different, the total outbreak sizes caused by different serotypes were similar for both scenarios. This means that although $\lambda$ is different, the outbreak size caused by each dengue serotype would be similar for different scenarios of dengue introductions. Therefore, in this analysis, we use the scenario where each dengue serotype is introduced weekly in alternate years.

The outbreak size over two years is used to quantify the proportional reduction in dengue incidence due to \textit{Wolbachia} as it includes the  sizes of the outbreaks caused by both serotypes. The four-year outbreak size used in the preceding chapter is unnecessary here because we are not comparing the performance of \textit{Wolbachia} in reducing dengue under different disease introduction scenarios. The simulation procedures used here are the same as those used in the previous chapters (Chapters~\ref{chap:multipleintro} and~\ref{chap:symmetry}).
\section{Results}
In this section, the results of simulations exploring the effects of asymmetry in the  antibody-dependent enhancement (ADE)  (Section~\ref{AssADE}) and the transmission probabilities (Section~\ref{AssTN}) on \textit{Wolbachia} performance in reducing dengue incidence are presented.  A  discussion and conclusion appears in Section~\ref{AssDiscussion}. 

\subsection{Effect of Asymmetric  Antibody-Dependent Enhancement}\label{AssADE}
In our simulations for this section, the  ADE for serotype~2 is varied and that of serotype~1 is fixed.

\begin{figure}[h!]
\centering
\includegraphics[width=1.05\textwidth]{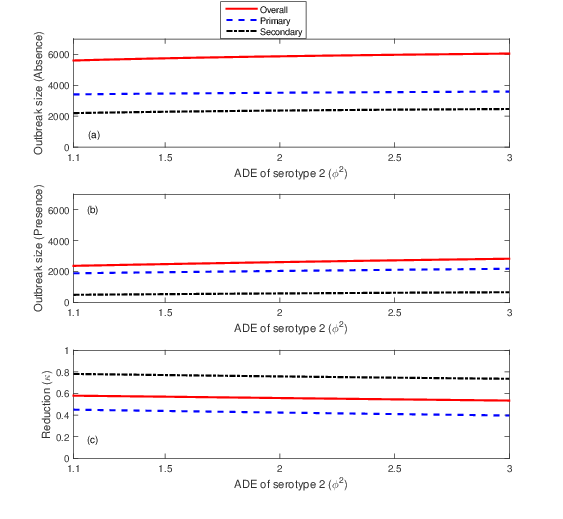}
\caption[The effect of changes in the ADE for serotype 2 on primary and secondary infections]{The effect of changes in the antibody dependent enhancement for serotype 2 ($\phi^2$) on  dengue cases. All plots show overall (solid red lines), primary (dashed blue lines) and secondary (dash-dot black lines) infections. Plots (a) and (b) show outbreak sizes in the absence and presence of \textit{Wolbachia}-carrying mosquitoes, respectively, and plot (c) shows the  proportional reduction in dengue incidence due to \textit{Wolbachia}.}
\label{fig:Thesis_Asymmetry_Phi2}
\end{figure}

When the dengue serotypes have different  ADE, an increase in the  ADE for serotype~2 leads to slight increases in the overall outbreak size and the outbreak size due to primary and secondary infections in the absence and presence of \textit{Wolbachia}-carrying mosquitoes.  The proportional reduction in dengue incidence caused by \textit{Wolbachia} also decreases (Figure~\ref{fig:Thesis_Asymmetry_Phi2}).  The proportional reduction in secondary infections (73--78\%) is higher than that of primary infections (40--45\%), and the overall reduction in dengue incidence varies between 53\% and 58\%. 

\begin{figure}[h!]
\centering
\includegraphics[width=1.05\textwidth]{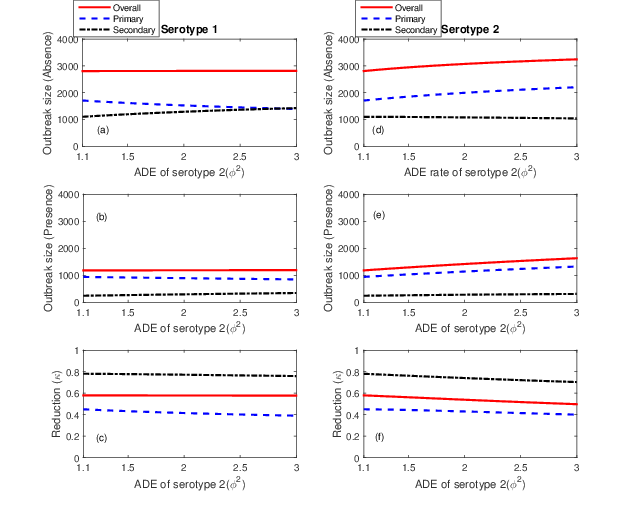}
\caption[The effect of changes in the ADE  for serotype~2 on primary and secondary infections caused by each serotype]{The effect of changes in the  ADE  for serotype 2 when the  ADE  for serotype~1 is fixed ($\phi^1=1.1$). All plots show overall (solid red lines), primary (dashed blue lines) and secondary (dash-dot black lines) infections. Plots at the top ((a) and (d)) and  in the middle ((b) and (e))  show outbreak size in the absence and presence of \textit{Wolbachia}-carrying mosquitoes, respectively.  Plots (c) and (f) show the proportional reduction in dengue
incidence due to \textit{Wolbachia}.  The left hand plots ((a)--(c)) show serotype~1  and the right hand plots ((d)--(f)) show serotype~2.}
\label{fig:Thesis_Asymmetry_Phi2_eachstrain}
\end{figure}

When looking at the effects of each dengue serotype on the performance of \textit{Wolbachia} in reducing dengue incidence, we see that the serotype with the higher  ADE (serotype 2) contributes more to the primary infections, whereas the serotype with the lower ADE (serotype 1) contributes more to the secondary infections (see Figure~\ref{fig:Thesis_Asymmetry_Phi2_eachstrain}). As the  ADE for serotype 2 increases, the overall outbreak size and the outbreak size for primary infections due to  serotype 2 in the absence of \textit{Wolbachia} increases, whereas the outbreak size for secondary infections due to serotype 2 in the absence of \textit{Wolbachia} decreases. The same behaviour is found in the presence of \textit{Wolbachia}-carrying mosquitoes, except that the outbreak size for secondary infections with serotype 2 slightly increases.  On the other hand, as the  ADE for serotype 2  increases, the overall outbreak size for serotype 1 in the absence of \textit{Wolbachia}-carrying mosquitoes remains constant because  the decrease in the number of primary infections is offset by the slight increase in the number of secondary infections. Similar behaviour is  found in the presence of \textit{Wolbachia}-carrying mosquitoes. Moreover, as the  ADE for serotype 2 increases, the performance of \textit{Wolbachia} in reducing the number of dengue cases due to serotype 2 decreases, though the reduction in the secondary infections (70--78\%) is higher than that in the primary infections (40--45\%).  The performance of \textit{Wolbachia} in reducing dengue infections due to serotype~1 is  similar  to its performance against infections caused by serotype~2.

\subsection{Effect of Asymmetric Transmission Probabilities}\label{AssTN}

\begin{figure}[h!]
\centering
\includegraphics[width=1.05\textwidth]{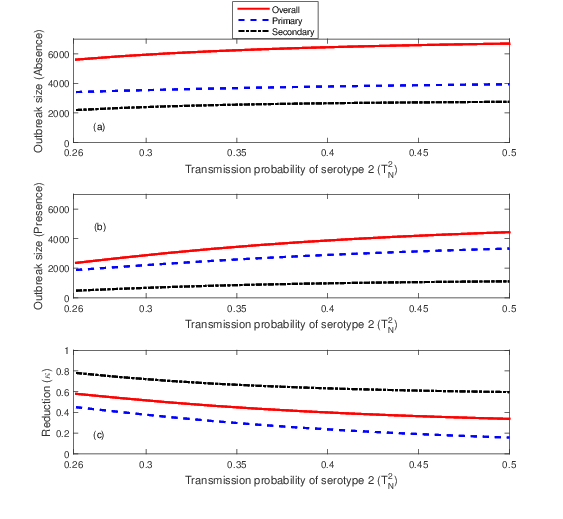}
\caption[The effect of changes in the transmission probability of serotype 2 on primary and secondary infections]{The effect of changes in the transmission probability of serotype 2 in dengue cases, when the transmission probability of serotype 1 is fixed ($T_N^1 = 0.2614$). All plots show the overall (solid red lines), primary (dashed blue lines) and secondary (dash-dot black lines) infections. The plots (a) and (b)  are for outbreak size in the absence and presence of \textit{Wolbachia}-carrying mosquitoes, respectively, and the plot (c) is for proportional reduction in dengue due to \textit{Wolbachia}.}
\label{fig:Thesis_Asymmetry_T2}
\end{figure}

If the transmission probability of serotype 2 is higher than that of serotype 1, the overall outbreak size and the outbreak sizes for primary and secondary infections in the absence and presence of \textit{Wolbachia}-carrying mosquitoes increases, and the proportional reduction in overall, primary and secondary cases due to \textit{Wolbachia} decreases (see Figure~\ref{fig:Thesis_Asymmetry_T2}). The overall reduction varies between 31\% and 58\%, and the reduction in primary and secondary infections decreases from 45\% to 13\% and 78\% to 60\%, respectively.

\begin{figure}[h!]
\centering
\includegraphics[width=0.95\textwidth]{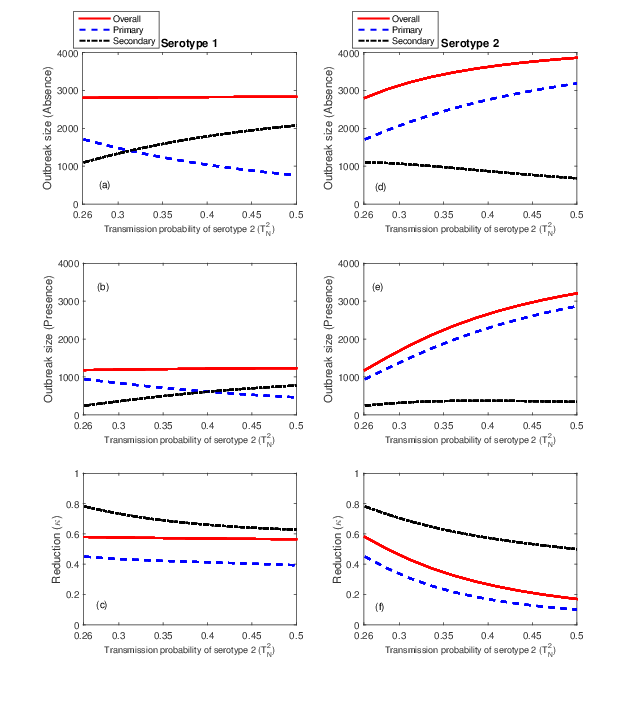}
\caption[The effect of changes in the transmission probability of serotype 2 on primary and secondary infections caused by each serotype]{The effect of changes in the transmission probability of serotype 2 while the transmission probability of serotype 1 is fixed. All plots show the overall (solid red lines), primary (dashed blue lines) and secondary (dash-dot black lines) infections. The plots at the top ((a) and (d)) and in the middle ((b) and (e))  show the  outbreak size in the absence and presence of \textit{Wolbachia}-carrying mosquitoes, respectively, and the plots at the bottom ((c) and (f)) show the proportional reduction in dengue incidence due to \textit{Wolbachia}, with the left hand plots ((a)--(c) showing serotype~1  and the right hand plots ((d)--(f)) showing serotype~2.}
\label{fig:Thesis_Asymmetry_T2_eachstrain}
\end{figure}

A strong effect on the \textit{Wolbachia} performance in reducing dengue caused by differences in the transmission probabilities between dengue serotypes is observed (Figure~\ref{fig:Thesis_Asymmetry_T2_eachstrain}). Generally, the more transmissible serotype (serotype~2) will dominate the primary infections, while the less transmissible serotype (serotype 1) will dominate the secondary infections. The overall outbreak size of dengue caused by serotype 1 remains constant as the decrease in the number of primary infections is balanced by the increase in the number of secondary infections. Interestingly, in the presence of \textit{Wolbachia}-carrying mosquitoes, although the number of primary infections due to serotype 1 decreases, the number of secondary infections due to serotype 2 slightly increases, but the latter is still lower than the former. Furthermore, the results show that the reduction in secondary infections is higher than that in primary infections (see Figure~\ref{fig:Thesis_Asymmetry_T2_eachstrain}).  Moreover, \textit{Wolbachia} can still reduce the number of dengue cases caused by the more transmissible serotype (serotype~2). 

\section{Discussion and Conclusions}\label{AssDiscussion}
In this chapter, we investigated the effect of asymmetric epidemiological characteristics on the performance of \textit{Wolbachia} in reducing dengue incidence. Although ADE influences the performance of \textit{Wolbachia} in reducing dengue incidence, the transmission probability is still the most influential parameter. The results are consistent with our previous findings for  dengue serotypes with symmetric  epidemiological characteristics. In the simulations conducted for this chapter, the reduction in secondary dengue infections due to \textit{Wolbachia} is higher than that for primary infections.

As the ADE of serotype 2 is varied, the overall outbreak sizes in the absence and presence of \textit{Wolbachia}-carrying mosquitoes increases. Furthermore, the reduction in primary and secondary infections due to \textit{Wolbachia} declines slightly as the ADE of serotype 2~($\phi^2$) increases. This implies that ADE  does not have a large impact on the performance of \textit{Wolbachia} in reducing dengue. When looking at the effect on individual serotypes, as the ADE of serotype 2 increases, the outbreak sizes of serotype~2 primary infections increase and the incidence of serotype 1 primary infections decreases. Serotype~2 contributes more to primary infections than serotype~1 as the ADE of serotype~2 increases. Because more individual humans are primarily infected by serotype~2, secondary infections are dominated by serotype~1. If there are two serotypes circulating in the population and  more individuals have primary infections with one of the serotypes, then it is likely that more individuals have secondary infections with the other serotype.

 Interestingly, we find that although the number of primary infections due to serotype~1 decreases in the presence of \textit{Wolbachia}-carrying mosquitoes, the number of secondary infections caused by serotype~2 increases slightly.   However, the  number of secondary infections caused by serotype 2 is still lower than the number of  primary infections caused by serotype 1. This may be for the following reasons.   As the ADE of serotype~2 increases, there are more individuals primarily infected with serotype~2 than serotype~1. The higher number of primary infections with serotype~2 affects the force of infection, resulting in a greater likelihood of secondary infection with serotype 2 for individuals previously infected with serotype~1.    Hence, as the ADE for serotype~2 increases, the  number of secondary infections caused by serotype~2 increases. This implies that there is a complex interaction between variables regulating the force of infections. Note that similar behaviour is also found when varying the transmission probability of serotype~2.

A strong effect on the performance of \textit{Wolbachia} in reducing dengue transmission is found when the transmission probability of one of the serotypes is greatly different from the transmission probability of the other.  When the transmission probability of one serotype is higher than  that of the other serotype, the number of primary infections caused by that serotype (the serotype with higher transmission probability ($T_N^2$)) is increased and secondary infections are more likely to be caused by the other serotype (the serotype with the lower transmission probability ($T_N^1$)). This is biologically realistic because when $T_N^2>T_N^1$, more individuals are primarily infected with serotype~2 and thus they are more likely to be secondarily infected with serotype~1.  Furthermore, \textit{Wolbachia} will be less effective in in reducing the overall number of dengue cases caused by the  more transmissible serotype. Although asymmetry in the transmission probabilities can lead to a higher number of secondary infections due to the less transmissible serotype,  this incidence can be reduced by up to 78\% with the implementation of the \textit{Wolbachia} intervention.


Our findings suggest that if the more transmissible serotype is responsible for more severe illness, then reducing the incidence of primary infections may be of concern.  This is because the more transmissible serotype will cause more primary infections.  Although secondary infections can lead to more severe forms of dengue, paying attention to primary infections is also important given that the level of reduction in dengue primary infections due to the \textit{Wolbachia} intervention is not as great as the reduction in secondary infections. Furthermore, our finding that differences in antibody-dependent enhancement rates between serotypes have a much smaller effect on dengue dynamics than differences in the transmission probabilities indicates that a better understanding of serotype-specific transmission probabilities  may be needed in order to develop strategies to optimize the  delivery of \textit{Wolbachia} interventions.

Further investigation of  the variables that regulate the force of infections is needed to advance our understanding of \textit{Wolbachia} performance in reducing dengue incidence.   A further exploration of serotype-specific effects on the performance of \textit{Wolbachia} in reducing dengue incidence is one possible avenue for future exploration.  Ferguson \textit{et al.}~\cite{Ferguson2015} estimated the basic reproduction numbers for the four dengue serotypes.  If the data becomes available, one could find similar estimates for the parameters that govern the force of infection or the basic reproduction number.   Our theoretical explorations provide insights into the performance of \textit{Wolbachia} when two  dengue serotypes  circulate in the population, and, in particular, when the epidemiological characteristics between serotypes differ.  The results are consistent with our findings for a single serotype presented in part 1 of this thesis.  It may be of future interest to explore \textit{Wolbachia} performance when more than two dengue serotypes circulate.

\section{Summary}
\begin{itemize}[wide=0pt,labelwidth = 1.3333em, labelsep = 0.3333em, leftmargin = \dimexpr\labelwidth + \labelsep\relax ]
\item We investigate the performance of \textit{Wolbachia} in reducing dengue incidence when the level of antibody-dependent enhancement (ADE) and the transmission probability of dengue serotypes are different (Sections~\ref{AssADE} and \ref{AssTN}).
\item When a dengue serotype has higher ADE or transmission probability, that serotype will contribute more to primary infections, whereas the other serotype will contribute more to secondary infections (Sections~\ref{AssADE} and \ref{AssTN}, see Figures~\ref{fig:Thesis_Asymmetry_Phi2_eachstrain} and \ref{fig:Thesis_Asymmetry_T2_eachstrain}).
\item The proportional reduction in secondary infections is higher than that in primary infections when one serotype is more transmissible that the other serotype (around 60--78\% compared to around 13--45\%; see Section~\ref{AssTN}) or has higher ADE than the other serotype (73--78\% compared to 40--45\%; see Section~\ref{AssADE}) .
\end{itemize}


\chapter{Conclusions and Future Directions}\label{chap:Conclusions} 


\ifpdf
    \graphicspath{{9/figures/PNG/}{9/figures/PDF/}{9/figures/}}
\else
    \graphicspath{{9/figures/EPS/}{9/figures/}}
\fi

The central question addressed by this thesis is ``to what extent can \textit{Wolbachia} reduce dengue transmission?''.  We have approached this question by formulating novel mathematical models for populations in which one and two dengue serotypes circulate in the absence and presence of \textit{Wolbachia}-carrying mosquitoes. The details of these mathematical models are given in Chapter~\ref{chap:singlemathmodel} (single serotype dengue models) and Chapter~\ref{chap:modelmultiplestrains} (two-serotype dengue models). As discussed in Section~\ref{ResearchAims}, the aims of this thesis are to

\begin{enumerate}[wide=0pt,labelwidth = 1.3333em, labelsep = 0.3333em, leftmargin = \dimexpr\labelwidth + \labelsep\relax ]
\item Determine the level of reduction in dengue incidence caused by the presence of \textit{Wolbachia}-carrying mosquitoes.
\item \textcolor{blue}{Explore the effects of seasonality and other important parameters on dengue transmission dynamics and the persistence of \textit{Wolbachia}-carrying mosquitoes.}
\item Determine the effects of dengue introduction scenarios on the performance of \textit{Wolbachia} in reducing dengue incidence.
\item Investigate the effects of two dengue serotypes and dengue serotype characteristics on the performance of \textit{Wolbachia} in reducing dengue incidence.
\end{enumerate}

Our original results, presented in Chapters~\ref{chap:singlemathmodel}--\ref{chap:asymmetry}, fulfil the aims of this thesis as listed above. Aim 1 is satisfied in Chapters~\ref{chap:singleintro}, \ref{chap:multipleintro}, \ref{chap:symmetry} and  \ref{chap:asymmetry}.  Aim 2 is satisfied in Chapter~\ref{chap:singleintro}. Aim 3 is satisfied in Chapters~\ref{chap:multipleintro} and \ref{chap:symmetry}, while Aim 4 is satisfied in Chapters~\ref{chap:symmetry}--\ref{chap:asymmetry}.  A summary of the original results contained in this thesis is given below in Section~\ref{SummaryofResults}.

\section{Summary of the Results}\label{SummaryofResults}
The original findings discovered in this thesis can be grouped into two parts:  results which consider the circulation of a single dengue serotype and results which consider the circulation of two dengue serotypes.

Part 1 of this thesis (Chapters~\ref{chap:singlemathmodel}-\ref{chap:multipleintro}), is concerned with the development of novel mathematical models for dengue transmission dynamics where it is assumed that  a single dengue serotype is circulating in the population.  Separate models are developed to consider both the absence and presence of \textit{Wolbachia}-carrying mosquitoes.  These models are then used to analyse the effects of \textit{Wolbachia} on dengue transmission dynamics when dengue cases are introduced once and multiple times.

When dengue cases are introduced into the population once, and \textit{Wolbachia}-carrying mosquitoes persist, a reduction in dengue incidence of up to 90\%  can be obtained.  Furthermore, the choice of \textit{Wolbachia} strain for inoculation of the mosquitoes has a significant effect on the ability for \textit{Wolbachia}-carrying mosquitoes to persist in competition with non-\textit{Wolbachia} mosquitoes.  The \textit{WMel} strain, which reduces the lifespan of the mosquito by at most 10\%, allows the \textit{Wolbachia}-carrying mosquitoes to persist, whereas the \textit{WMelPop} strain is unlikely to allow persistence since it reduces the mosquito lifespan by up to 50\%. 

Cytoplasmic incompatibility, when considered in isolation from all other physiological effects of \textit{Wolbachia} on the mosquitoes, was found to reduce the number of human dengue cases. When exploring the key parameters for our models, we found the range of the ratios of reproductive and death rates for adult non-\textit{Wolbachia} mosquitoes which allowed \textit{Wolbachia}-carrying mosquitoes to persist. Furthermore, the transmission probability, the biting rate and the average death rate were found to be the most influential parameters, the latter having a negative relationship with the cumulative number of infectious individuals. The results for the basic reproduction number, $\mathcal{R}_0$, show that \textit{Wolbachia} reduces the number of days for which $\mathcal{R}_0 >1$. 

When investigating multiple introductions of dengue cases into the population, we simulated weekly introductions of dengue cases into the population throughout the year and over three month periods.  We found that \textit{Wolbachia} reduces the potential length of the season in  which epidemics are likely to occur by up to six weeks, depending on the strength of seasonality.  The largest reduction in dengue incidence found was around 80\%, which is obtained when the strength of the seasonal forcing is low. The efficacy of \textit{Wolbachia} also depends on the transmission rate, with the bacteria most effective in reducing dengue incidence at moderate transmission rates ranging from 0.08--0.12. 
This is consistent with fitted estimates for Cairns, Australia as given in Chapter~\ref{chap:singleintro}.  

In part 2 of the thesis (Chapters~\ref{chap:modelmultiplestrains}--\ref{chap:asymmetry}), the performance of \textit{Wolbachia} in reducing dengue incidence when two serotypes of dengue circulate in the population is investigated.

When considering the case where epidemiological characteristics of these serotypes are the same (the `symmetric case'), we found that varying the disease introduction scenario does not affect dengue transmission dynamics. Although the timing of dengue introductions is irregular and is not easy to  investigate, the results for the disease introduction scenarios that we simulated suggest that the \textit{Wolbachia} intervention can  be implemented, regardless of the disease introduction scenario. Furthermore, the antibody-dependent enhancement rate does not noticeably affect the impact of \textit{Wolbachia} on dengue incidence unless the transmission probability is high. A considerable reduction in secondary infections of 60-80\% is achieved through the \textit{Wolbachia} intervention.

When the epidemiological characteristics of the dengue serotypes differ, there is a shift towards greater numbers of primary infections of the serotype with the higher enhancement rate or higher transmission probability. Where only the antibody-dependent enhancement rate differs between serotypes, we find relatively little decline in the effectiveness of \textit{Wolbachia}. However, if one serotype is more transmissible than the other, \textit{Wolbachia} is less effective in reducing dengue cases caused by the more transmissible serotype. Although the number of secondary infections caused by the less transmissible serotype is higher, a reduction of around 78\% in the number of secondary infections can be obtained by the  implementation of the \textit{Wolbachia} intervention.

\section{Conclusions}
In this thesis we sought to answer the question: ``to what extent can \textit{Wolbachia} reduce dengue transmission?''. 

We found that once mosquitoes carrying the \textit{Wolbachia} bacterium persist, a reduction in dengue incidence of up to 80\% can be obtained in the presence of one or two circulating dengue serotypes. A higher reduction in the incidence of secondary infections can be obtained regardless of the epidemiological characteristics of the serotypes.  Different disease introduction scenarios and levels of antibody-dependent enhancement in secondary cases do not affect the performance of \textit{Wolbachia} in reducing dengue incidence. By contrast, differences in the transmission probabilities of the serotypes have a much greater effect. If the transmission probability is high, \textit{Wolbachia} can reduce the number of dengue cases by only around 20\%-40\%. Where serotypes have different epidemiological characteristics, the more transmissible serotype will dominate primary infections, and the less transmissible serotype will dominate secondary infections. 

Our findings suggest that \textit{Wolbachia} should reduce the number of primary dengue cases in areas with moderate transmission levels, and cause even greater reductions in the incidence of secondary cases.  Given the higher risk of severe outcomes in secondary cases, \textit{Wolbachia} has great potential for improving public health.

\section{Future Directions}\label{FutureDirection}
The work contained in this thesis makes a significant contribution to the current knowledge of dengue transmission dynamics and provides evidence of the effectiveness of the \textit{Wolbachia} intervention in reducing dengue transmission. To advance our understanding of the performance of \textit{Wolbachia} in reducing dengue incidence, several possible directions are open for study.  These may utilise our work as a baseline.

\begin{enumerate}[wide=0pt,labelwidth = 1.3333em, labelsep = 0.3333em, leftmargin = \dimexpr\labelwidth + \labelsep\relax ]
\item In this thesis, deterministic mathematical models are formulated, which are appropriate for large population sizes.  A stochastic approach could be used to further our understanding of the impact of small infected population sizes on dengue dynamics in the presence of \textit{Wolbachia}-carrying mosquitoes. Alternatively, a hybrid modelling approach,  coupling deterministic and stochastic models could be used.  In such an approach, a stochastic model would be used when the infected population becomes small, and a deterministic model when the infected population exceeds a certain threshold value.

\item We found that the biting rate is one of the important parameters that regulates the transmission dynamics of dengue. In this thesis, the biting rate is assumed to be fixed. To further our understanding of the effect of the biting rate on dengue transmission dynamics in the presence of \textit{Wolbachia}-carrying mosquitoes,  it may be of interest to explore  variations in the biting rate of mosquitoes.  Such variations could occur as mosquitoes get older and may have an impact on regulating the transmission dynamics of dengue.  

\item It may be of interest to explore the effect of feminisation of the population of \textit{Wolbachia}-carrying mosquitoes on the performance of \textit{Wolbachia} in reducing dengue transmission.  The release of more males than females seems to have an effect on the mosquito population dynamics~\cite{Hancock2011a, Zheng2014}, and, hence, on dengue transmission dynamics.

\item We define an ``instantaneous basic reproduction number'' using a formula for autonomous systems and replace the equilibrium mosquito population with a time-varying mosquito population. It may be interesting to derive the basic reproduction number for the full seasonally-dependent model, or use the concept proposed by Bacaer~\cite{Bacaer2007} to investigate the epidemic threshold.


\item A different approach such as metapopulation modelling or integro-differential equations may be used to investigate the effects of human movements on dengue transmission dynamics.

\end{enumerate}


\appendix
\backmatter


\chapter{Appendix\\Basic Reproduction Number}\label{app:BasicReproductionNumber} 


\ifpdf
    \graphicspath{{9/figures/PNG/}{9/figures/PDF/}{9/figures/}}
\else
    \graphicspath{{9/figures/EPS/}{9/figures/}}
\fi

We derive the basic reproduction number for dimensionalised models. 
The infected subsystem of the single serotype dengue model in the persence of \textit{Wolbachia}-carrying mosquitoes (Equations~\eqref{nondim10}--\eqref{nondim21}) is given by

\begin{align}
\frac{dE_H}{dt}&=b_NT_{N}LI_NS_H+b_WT_{HW}LI_WS_H-\gamma_H E_H -\mu_H E_H,\label{infecsubs1aps}\\[5mm]
\frac{dI_H}{dt}&=\gamma_HE_H-\sigma I_H -\mu_H I_H,\label{infecsubs2aps}\\[5mm]
\frac{dE_N}{dt}&=b_NT_{N}I_H S_N-\left(\gamma_N+\mu_N(t)\right)E_N,\label{infecsubs3aps}\\[5mm]
\frac{dI_N}{dt}&=\gamma_NE_N-\mu_N(t)I_N,\label{infecsubs4aps}\\[5mm]
\frac{dE_W}{dt}&=b_WT_{N}I_HS_W-\left(\gamma_W+\mu_W(t)\right)E_W,\label{infecsubs5aps}\\[5mm]
\frac{dI_W}{dt}&=\gamma_WE_W-\mu_W(t)I_W . \label{infecsubs6aps}
\end{align}
At the infection-free steady state, $E_H=I_H=E_N=I_N=E_W=I_W=0$, and $S_H=1$. For small $\left(E_H, I_H, E_N, I_N, E_W, I_W\right)$, the linearised infected subsystem is approximated by Equations~\eqref{infecsubs1aps}--\eqref{infecsubs6aps}, with $S_H=1$.


Let $\mathbf{x}=\left(E_H, I_H, E_N, I_N, E_W, I_W\right)'$.   We want to write the linearised infected subsystem in the form
\begin{equation*}
\mathbf{\frac{dx}{dt}} =\left(\mathbf{T} + \pmb{\Sigma}\right)\mathbf{x}\,,
\end{equation*}
where  $\mathbf{T}$ is the transmission matrix, whose entries correspond to transmission events, and $\pmb{\Sigma}$ is the transition matrix, whose entries correspond to movement between the infected compartments including deaths.  

The transmission matrix $\mathbf{T}$ is

$$\mathbf{T} =
 \begin{pmatrix}
  0&0&0&b_{N}LT_{N}&0&b_{W}LT_{HW}\\
  0&0&0&0 &0&0 \\
  0&b_{N}T_{N}S_N(t)&0&0&0&0\\
  0&0&0&0&0&0\\
  0&b_{W}T_{N}S_W(t)&0&0&0&0\\
  0&0&0&0&0&0
 \end{pmatrix}\,,$$
%
%
where $S_N(t)$ and $S_W(t)$ are the number of mosquitoes with and without \textit{Wolbachia}.  As mosquito populations vary seasonally, both of these are functions of time. The transition matrix $\pmb{\Sigma}$ is
  $$\pmb{\Sigma} =
  \begin{pmatrix}
   -\left(\gamma_H+\mu_H\right)&0&0&0&0&0 \\
   \gamma_H&-\left(\sigma+\mu_H\right)&0&0&0&0 \\
   0&0&-\left(\gamma_N+\mu_N(t)\right)&0&0&0 \\
   0&0&\gamma_N&-\mu_N(t)&0&0\\
   0&0&0&0&-\left(\gamma_W+\mu_W(t)\right)&0\\
   0&0&0&0&\gamma_W&-\mu_W(t)
  \end{pmatrix}.$$
   
We next find the inverse of the transition matrix $\pmb{\Sigma}^{-1}$:

 $$\pmb{\Sigma}^{-1} =
  \begin{pmatrix}
   -\frac{1}{\left(\gamma_H+\mu_H\right)}& 0& 0& 0& 0& 0 \\
   -\frac{\gamma_H}{\left(\gamma_H+\mu_H\right)\left(\sigma+\mu_H\right)}&-\frac{1}{\left(\sigma+\mu_H\right)}& 0&0&0&0 \\
   0&0&-\frac{1}{\left(\gamma_N+\mu_N(t)\right)}&0&0&0 \\
   0&0&-\frac{\gamma_N}{\mu_N(t)\left(\gamma_N+\mu_N(t)\right)}&-\frac{1}{\mu_N(t)}&0&0\\
   0&0&0&0&-\frac{1}{\left(\gamma_W+\mu_W(t)\right)}&0\\
   0& 0& 0& 0&-\frac{\gamma_W}{\mu_W(t)\left(\gamma_W+\mu_W(t)\right)}&-\frac{1}{\mu_W(t)}
  \end{pmatrix}. $$

We multiply the transmission matrix and the inverse of transition matrix to obtain

$$-\mathbf{T}\pmb{\Sigma}^{-1} =
 \begin{pmatrix}
  0&0&\frac{b_{N}LT_{N}\gamma_N}{\mu_N\left(\gamma_N+\mu_N\right)}&\frac{b_{N}LT_{N}}{\mu_N}& \frac{b_{W}LT_{HW}\gamma_W}{\mu_W\left(\gamma_W+\mu_W\right)}& \frac{b_{W}LT_{HW}}{\mu_W}\\
  0&0&0&0 &0&0 \\
  0&\frac{b_{N}T_{N} \gamma_H S_N(t)}{\left(\sigma+\mu_H\right)\left(\gamma_H + \mu_H\right)}&\frac{b_{N}T_{N}S_N(t)}{\left(\sigma+\mu_H\right)}&0&0&0\\
  0&0&0&0&0&0\\
  0&\frac{b_{W}T_{N} \gamma_H S_W(t)}{\left(\sigma+\mu_H\right)\left(\gamma_H + \mu_H\right)}&\frac{b_{W}T_{N}S_W(t)}{\left(\sigma+\mu_H\right)}&0&0&0\\
  0&0&0&0&0&0
 \end{pmatrix}.$$

The spectral radius of the matrix $-\mathbf{T}\pmb{\Sigma}^{-1}$ is the basic reproduction number, that is,

\begin{equation}
\resizebox{0.9\hsize}{!}{$\mathcal{R}_P=\sqrt{\frac{b_N^2 L T_N^2 \gamma_N \gamma_H S_N(t)}{\mu_N(t)\left(\gamma_N + \mu_N(t)\right)\left(\sigma +\mu_H\right)\left(\gamma_H + \mu_H\right)} +\frac{b_W^2 L T_{HW} \gamma_W T_N \gamma_H S_W(t)}{\mu_W(t)\left(\gamma_W + \mu_W(t) \right)\left(\sigma + \mu_H\right)
\left(\gamma_H + \mu_H\right)}}$}.
\label{eq:R0apps}
\end{equation}

Equation~\eqref{eq:R0apps} gives the basic reproduction number in the presence of \textit{Wolbachia}. In the absence of \textit{Wolbachia}, after setting all \textit{Wolbachia}-related parameters to zero, Equation~\ref{eq:R0apps} is reduced to

\begin{equation}
\mathcal{R}_A=\sqrt{\frac{b_N^2 L T_N^2 \gamma_N \gamma_H S_N(t)}{\mu_N(t)\left(\gamma_N + \mu_N(t)\right)\left(\sigma +\mu_H\right)\left(\gamma_H + \mu_H\right)}}.
\label{eq:R0Absapps}
\end{equation}


The basic reproduction number found here can be considered to be an instantaneous basic reproduction number which depends on fluctuations in  the mosquito population, $S_N(t)$ and $S_W(t)$. 

 \singlespace
  \addcontentsline{toc}{chapter}{Bibliography}
\bibliography{refs}
  \bibliographystyle{plain}
\end{document}